\documentclass[structabstract]{aa}  
\usepackage{url}
\usepackage{cancel}
\usepackage{graphics}
\usepackage{latexsym}
\usepackage{graphicx}
\usepackage{amsmath}
\usepackage{lscape}
\usepackage{xcolor}
\usepackage{txfonts}
\usepackage{natbib}
\usepackage{chngcntr}
\usepackage{soul}
\usepackage{siunitx}
\usepackage[textwidth=3.5cm]{todonotes}
\usepackage{blindtext}

\renewcommand{\d}[0]{{\rm d}}
\newcommand{\e}[0]{{\rm e}}
\renewcommand{\i}[0]{{\rm i}}
\newcommand{\ave}[1]{\langle #1 \rangle}
\newcommand{\Ave}[1]{\Big\langle #1 \Big\rangle}
\renewcommand{\Ref}[1]{(\ref{#1})}

\newcommand{\mat}[1]{\tens{#1}}

\newcommand{\msol}[0]{M_\odot}

\newcommand{\sfont}[1]{{\scriptscriptstyle\rm #1}}

\begin{document}

\title{KiDS-1000: Detection of deviations from a purely cold
  dark matter power spectrum with tomographic weak gravitational
  lensing}

\author{Patrick Simon$^1$\thanks{\email{psimon@astro.uni-bonn.de}},
  Lucas Porth$^1$, Pierre Burger$^{2,3,1}$, and Konrad Kuijken$^4$}
\institute{$^1$ Universit\"at Bonn, Argelander-Institut f\"ur Astronomie, Auf dem H\"ugel 71, 53121 Bonn, Germany\\
  $^2$ Waterloo Centre for Astrophysics, University of Waterloo, Waterloo, ON N2L 3G1, Canada\\
  $^3$ Department of Physics and Astronomy, University of Waterloo, Waterloo, ON N2L 3G1, Canada\\
  $^4$ Leiden Observatory, Leiden University, Einsteinweg 55, 2333 CC
  Leiden, The Netherlands}

\date{Received \today}

\authorrunning{Simon et al.}  \titlerunning{Detection of deviations
  from a purely $\Lambda$CDM power spectrum in KiDS-1000}

\abstract{Model uncertainties in the non-linear structure growth limit
  current probes of cosmological parameters.  To shed more light on
  the physics of non-linear scales, we reconstructed the finely binned
  three-dimensional power-spectrum from lensing data of the
  Kilo-Degree Survey (KiDS), relying solely on the background
  cosmology, the source redshift distributions, and the intrinsic
  alignment (IA) amplitude of sources (and their uncertainties). The
  adopted Tikhonov regularisation stabilises the deprojection,
  enabling a Bayesian reconstruction in separate $z$-bins.  Following
  a detailed description of the algorithm and performance tests with
  mock data, we present our results for the power spectrum as relative
  deviations from a $\Lambda\rm CDM$ reference spectrum that includes
  only structure growth by cold dark matter. Averaged over the full
  range \mbox{$z\lesssim1$}, a \emph{Planck}-consistent reference then
  requires a significant suppression on non-linear scales,
  $k=0.05$--$10\,h\,\rm Mpc^{-1}$, of up to $20\%$--$30\%$ to match
  KiDS-1000 ($68\%$ credible interval, CI).  Conversely, a reference
  with a lower \mbox{$S_8\approx0.73$} avoids suppression and matches
  the KiDS-1000 spectrum within a $20\%$ tolerance. When resolved into
  three $z$-bins, however, and regardless of the reference, we detect
  structure growth only in the range $z\approx0.4$--$0.13$, but not in
  the range $z\approx0.7$--$0.4$. This could indicate spurious
  systematic errors in KiDS-1000, inaccuracies in the intrinsic
  alignment (IA) model, or potentially a non-standard cosmological
  model with delayed structure growth. In the near future, analysing
  data from Stage IV surveys with our algorithm promises a
  substantially more precise reconstruction of the power spectrum.}


\maketitle

\section{Introduction}

In contrast to the continuous background expansion of the Universe,
the structure growth in the matter density field is less certain, with
uncertainties in theoretical models that vary by $10\%$ or more in the
non-linear regime through their dependence on baryon-galaxy feedback;
possible deviations from purely cold, stable, and interaction-free
dark matter; or, perhaps, modifications in the standard model of
gravity \citep[e.g.,][and references therein]{2024PhRvD.109d3507L,
  2024A&A...683A.152B, 2024A&A...685A.156M, 2023arXiv230911129F,
  2023MNRAS.523.2247S, 2022MNRAS.514.3802S, 2015MNRAS.450.1212H,
  2011PhRvD..84f3507S, 2006ApJ...640L.119J}. Currently the theory
uncertainties are limiting studies of cosmological parameters because
they often rely on measures of the cosmic structure, such as the
matter power spectrum, $P_\delta(k,z)$, at different redshifts,
\mbox{$z=:1/a-1$}, and spatial (comoving) wave number, $k$
\citep[e.g.,][]{2024arXiv241021980P, 2024arXiv240313794G,
  PhysRevD.105.023515, 2023PhRvD.108l3518L, 2021A&A...645A.104A,
  2013MNRAS.432.2433H}. To shed more light on the physics imprinted on
non-linear scales, direct measurements of a model-free $P_\delta(k,z)$
with a minimum of assumptions promise to be a valuable model
test. That this is already feasible for \mbox{$z\lesssim1$} and
\mbox{$10^{-2}\lesssim k/(h\,\rm Mpc^{-1})\lesssim10$} with Stage III
galaxy surveys when exploiting the weak gravitational lensing effect,
such as with KiDS-1000 \citep{2019A&A...625A...2K,
  2015MNRAS.454.3500K}, is shown in this work. This foreshadows
exciting applications to future lensing data, such as those by
\emph{Euclid} \citep{2024arXiv240513491E}, the \emph{Vera C. Rubin}
Observatory Legacy Survey of Space and Time
\citep{2018arXiv180901669T}, or the \emph{Nancy Grace Roman} Space
Telescope \citep{2015arXiv150303757S}.

The coherent distortion of distant galaxy images, just sources
hereafter, by the weak shearing of light bundles passing through
intervening foreground structure probes the matter power spectrum (for
a review, see, e.g., \citealt{2015RPPh...78h6901K},
\citealt{2006glsw.conf..269S}, \citealt{2001PhR...340..291B}). More
specifically, and putting negligible higher-order corrections aside
\citep{2009A&A...499...31H}, the second-order correlations,
$\xi_\pm(\theta)$, at lag $\theta$ in the cosmic shear field are
simply a linear projection of $P_\delta(k,z)$ throughout the look-back
light cone up to the distance of the sources
\citep{2024arXiv240418240P, 2012JCAP...06..005M, 2012A&A...543A...2S,
  2005MNRAS.363..723B, 2003MNRAS.346..994P, 2002PhRvD..66j3508T,
  1998ApJ...506...64S}. Therefore, by using $\xi_\pm^{(ij)}(\theta)$
between tomographic bins, $i$ and $j$, for different characteristic
distances of sources, the correlation data can be deprojected to
recover, within limitations, the original $P_\delta(k,z)$. More
conveniently, as in \citet[S12]{2012A&A...543A...2S}, we focus on a
transfer function,
\mbox{$f_\delta(k,z)=P_\delta(k,z)/P^{\rm fid}_\delta(k,z)$}, with
respect to a reference model, $P^{\rm fid}_\delta(k,z)$, to highlight
deviations from the presumed reference inside an average $z$-bin. For
this reference, we chose a purely cold dark matter (CDM) model to
probe for deviations, \mbox{$f_\delta(k,z)\ne1$}, that may hint at
missing physics in a basic, well-understood dark matter scenario or at
unidentified systematic errors in the data. However, at the end of the
day, the kind of reference is a deliberate choice.

Required for the deprojection, on the other hand, is the knowledge of
the projection (lensing) kernel and the inclusion of the intrinsic
alignment (IA) of sources in the correlation function, an essential
part of modern cosmic-shear analyses \citep{2004PhRvD..70f3526H,
  2001ApJ...559..552C, 2000ApJ...545..561C}. The lensing kernel is
fully described by the radial distribution of sources inside the
tomographic bins; the average matter density in today's Universe,
$\Omega_{\rm m}$, relative to the critical density,
$\rho_{\rm crit}=3H_0^2(8\pi G_{\rm N})^{-1}$; and the expansion rate
\mbox{$E(a):=H_0^{-1}\dot{a}/a$}, where
$H_0=100\,h\,\rm km\,s^{-1}\,\rm Mpc^{-1}$ is the Hubble constant and
$G_{\rm N}$ is Newton's gravitational constant. The kernel parameters
$\Omega_{\rm m}$ and $E(a)$ are confined to percentage precision by
cosmological experiments already
(\mbox{$E^2(a)\approx\Omega_{\rm m}[a^{-3}-1]+1$} for a flat
universe), especially owing to CMB experiments
\citep{2020A&A...641A...6P, 2013ApJS..208...19H}, but also by combining
probes from the closer Universe
\citep[e.g.,][]{2021A&A...646A.140H,2018PhRvD..98d3526A}. Nevertheless,
in the refinement of the method in S12 for KiDS-1000, we marginalised
here over the small background cosmology uncertainties, and
over those in the source distributions
\citep{2021A&A...647A.124H}. Another refinement is the required
inclusion of the IA, and its uncertainty, in the deprojection
procedure by the widely employed non-linear alignment (NLA) model
\citep{2011A&A...527A..26J, 2007NJPh....9..444B}, informed by the IA
constraints in \cite{2021A&A...645A.104A}.

Despite the simplicity of the well-defined projection, the noise level
in $\xi^{(ij)}_\pm(\theta)$ makes the recovery of the matter power
spectrum still challenging because the deprojection has to undo a
convolution in radial and transverse direction, producing strongly
correlated, oscillating noise in an unstable reconstruction. This
complication could be mitigated by data volumes substantially larger
than KiDS-1000, beating down oscillating noise, or by additional
assumptions on the redshift dependence or shape of the power spectrum
in fitting an analytical model with few parameters to the data
\citep{2025arXiv250206687P, 2024arXiv241107082Y, 2024arXiv241018191T,
  2024arXiv240913404B, 2024arXiv240418240P, 2023MNRAS.525.5554P,
  2003MNRAS.346..994P, 1998ApJ...506...64S}. Avoiding analytical
models, we instead propose constraining $f_\delta(k,z)$ averages in
different $k$- and $z$-bins, as in S12 but additionally filtered by a
classic Tikhonov regularisation, for instance as applied in the
mathematically related problem of deconvolving noisy images
\citep[e.g.,][]{2022PASJ...74.1329M}. In contrast to S12, the Tikhonov
regularisation, installed on our Bayesian statistical model as prior,
penalises strongly oscillating power spectra for noisy data, giving an
advantage to solutions of $f_\delta(k,z)$ that, on the one hand, are
smoother in the $k$-direction, but, on the other hand, use no prior
information on their $z$-dependence. We show with verification data
that this regularisation indeed stabilises the reconstruction and
extracts from KiDS-1000 data useful constraints on the transfer
function $f_\delta(k,z)$, either averaged over the full redshift range
or averaged within three separate redshift bins to broadly probe a
redshift evolution. Furthermore, for an efficient sampling of the
posterior constraints of the matter power spectrum, we describe and
test a Markov chain Monte Carlo (MCMC) code tailored to deal with the
numerous $60$ (or more) degrees-of-freedom of the binned, band power
$P_\delta(k,z)$; its practical details are given in Appendix
\ref{ap:mcmc}.

The outline of the paper is as follows. Section \ref{sect:WL} reviews
our weak lensing formalism and shows that the projection of
$P_\delta(k,z)$ into $\xi_\pm^{(ij)}(\theta)$, both binned, can be
reduced to one projection matrix, to be reused as long as the same
lensing kernel and IA parameters, including source redshift
distributions, are employed. Section \ref{sect:statanalysis} defines
the statistical model for our Bayesian analysis, including a prior for
the Tikhonov regularisation. We summarise the KiDS-1000 data for our
cosmic shear analysis in Sect. \ref{sect:data} and report our results
in Sect. \ref{sect:results}. Section 5 also reports, as verification
of our reconstruction method, the results of a mock analysis based on
KiDS-1000-like data produced by ray-tracing $N$-body data. Another
verification test, this time based on the data vector of an analytical
model subject to random noise, is presented in Appendix
\ref{ap:mcmc}. We discuss our results and conclusions on the
reconstructed matter power spectrum in
Sect. \ref{sect:discussion}. Notably, our figures use comoving wave
numbers, $k$, for spatial scales of the matter power spectrum,
emphasised by the `c' in the unit $[k]=h\,\rm cMpc^{-1}$.

\section{Weak lensing formalism}
\label{sect:WL}

This work is an application of the well-established theory of cosmic
shear. For a review and its mathematical foundation, we refer to
\cite{2015RPPh...78h6901K} or \cite{2006glsw.conf..269S}, and only
briefly summarise our formalism here.

\subsection{Cosmic shear}

Weak gravitational lensing by fluctuations in the large-scale
structure of the foreground matter density,
$\delta_{\rm m}=\rho_{\rm m}/\bar{\rho}_{\rm m}-1$, distorts images of
background galaxies. Their average shear distortion in direction
$\vec{\theta}$ of the sky shall be expressed by the complex
$\gamma(\vec{\theta})=\gamma_1(\vec{\theta})+\i\gamma_2(\vec{\theta})$
for an ensemble of galaxies with probability distribution function
(PDF) $p_\chi(\chi)$ in comoving distance, $\chi$. To lowest order,
and sufficiently accurate in practical applications, the Fourier
transform $\tilde{\gamma}(\vec{\ell})$ is related to the linear
projection of fluctuations along $\vec{\theta}$,
\begin{equation}
  \label{eq:kappa}
  \kappa(\vec{\theta})=
  \frac{3H_0^2\,\Omega_{\rm m}}{2c^2}\,
  \int_0^{\chi_{\rm h}}\frac{\d\chi}{a(\chi)}\,
  \overline{W}(\chi)\,f_\sfont{K}(\chi)\,
  \delta_{\rm m}\!\left[f_\sfont{K}(\chi)\vec{\theta},\chi\right]\;,
\end{equation}
through the convolution ($\ell\ne0$)
\begin{equation}
  \label{eq:gammakappa}
  \tilde{\gamma}(\vec{\ell})
  =\int_{\mathbb{R}^2}\d^2\theta\,\gamma(\vec{\theta})\,\exp{(-\i\vec{\theta}\cdot\vec{\ell})}
  =\frac{(\ell_1+\i\ell_2)^2}{\ell^2}\,\tilde{\kappa}(\vec{\ell})\;,
\end{equation}
where $c$ is the vacuum speed of light. In the above equations,
$\tilde{\kappa}(\vec{\ell})$ is the Fourier transform of the
convergence $\kappa(\vec{\theta})$, the scalar $f_\sfont{K}(\chi)$ is
the (comoving) angular diameter distance for the curvature scalar $K$,
the vector $f_\sfont{K}(\chi)\vec{\theta}$ denotes a separation in a
tangential plane on the sky at distance $\chi$, the term
$1+z=a^{-1}(\chi)$ is our relation between redshift, $z$, and scale
factor, $a(\chi)$, at the look back time of $\chi$, and
\begin{equation}
  \label{eq:lenseff}
  \overline{W}(\chi):=
  \int_\chi^{\chi_{\rm h}}\d\chi^\prime\;
  p_\chi(\chi^\prime)\,
  \frac{f_\sfont{K}(\chi^\prime-\chi)}{f_\sfont{K}(\chi^\prime)}  
\end{equation}
is the lensing efficiency, cut off at the size of the observable
Universe, $\chi_{\rm h}$.  As approximation, we assumed a flat sky
with Cartesian coordinates for $\vec{\theta}$ but expect negligible
inaccuracies for angular separations below several degrees.

Practical estimators of $\gamma(\vec{\theta})$ in the direction
$\vec{\theta}$ of a galaxy image use a convenient definition of galaxy
ellipticity, $\epsilon$, that is calibrated to be an unbiased
estimator, $\ave{\epsilon}=\gamma$, for a selected source population
\citep[e.g.,][]{2021A&A...645A.105G}.

\subsection{Second-order statistics}

Our analysis exploits the coherent shear distortion of galaxy images
over the sky to infer the three-dimensional power spectrum,
$P_\delta(k,z)$, of matter density fluctuations inside the light cone,
\begin{equation}
  \ave{\tilde{\delta}_{\rm m}(\vec{k},z)\,\tilde{\delta}^\ast_{\rm m}(\vec{k}^\prime,z)}=:
  (2\pi)^3\,\delta_{\rm D}(\vec{k}-\vec{k}^\prime)\,P_\delta(k,z)\;,    
\end{equation}
for a range of wave numbers and redshifts. The
$\delta_{\rm D}(\vec{x})$ denotes the Dirac delta function, and
$\tilde{\delta}_{\rm m}(\vec{k},z)$ is the Fourier coefficient of a
fluctuation mode at redshift $z$ inside the light cone. To obtain the
power spectrum from $\gamma(\vec{\theta})$, we estimate the two-point
correlation function of cosmic shear,
\begin{equation}
  \xi_\pm(\theta):=
  \Ave{\gamma_{\rm t}(\vec{\theta}_1)\,\gamma_{\rm t}(\vec{\theta}_2)}\pm
  \Ave{\gamma_\times(\vec{\theta}_1)\,\gamma_\times(\vec{\theta}_2)}
\end{equation}
at lag \mbox{$\theta=|\vec{\theta}_2-\vec{\theta}_1|$} from an
ensemble of sources. Here, $\gamma_{\rm t}$ and $\gamma_\times$ denote
the tangential and cross-shear components relative to the orientation
of \mbox{$\vec{\theta}_2-\vec{\theta}_1=:\theta\,\e^{\i\phi}$} on the
sky, defined by
\mbox{$\gamma_{\rm t}+\i\gamma_\times:=-\e^{-2\i\phi}\gamma$}. In
addition, using the hybrid extended Limber approximation in Fourier
space \citep{2017MNRAS.472.2126K,1992ApJ...388..272K}, the correlation
function of cosmic shear is a linear projection of $P_\delta(k,z)$,
\begin{multline}
  \label{eq:xipmauto}
  \xi_\pm(\theta)\\
  =\frac{9H_0^4\Omega_{\rm m}^2}{4c^4} \int_0^{\chi_{\rm h}}
  \!\!\!\!\int_0^\infty\frac{\d\chi\,\d\ell\,\ell}{2\pi}
  \frac{\overline{W}^2(\chi)}{a^2(\chi)}
  J_{0,4}(\ell\theta)\,P_\delta\left(\frac{\ell+1/2}{f_\sfont{K}(\chi)},\chi\right)
\end{multline}
through the lensing kernel
$K(\chi)\propto\Omega_{\rm m}^2\,\overline{W}^2(\chi)\,a^{-2}(\chi)$
-- the minimal ingredient that has to be known to invert the
projection. The expressions $J_n(x)$ denote $n^{\rm th}$-order Bessel
functions of the first kind, of which $J_0(x)$ has to be applied for
$\xi_+$ and $J_4(x)$ for $\xi_-$.

\subsection{Tomographic lensing and intrinsic alignment of sources}

\begin{figure}
  \begin{center}
    \includegraphics[width=90mm]{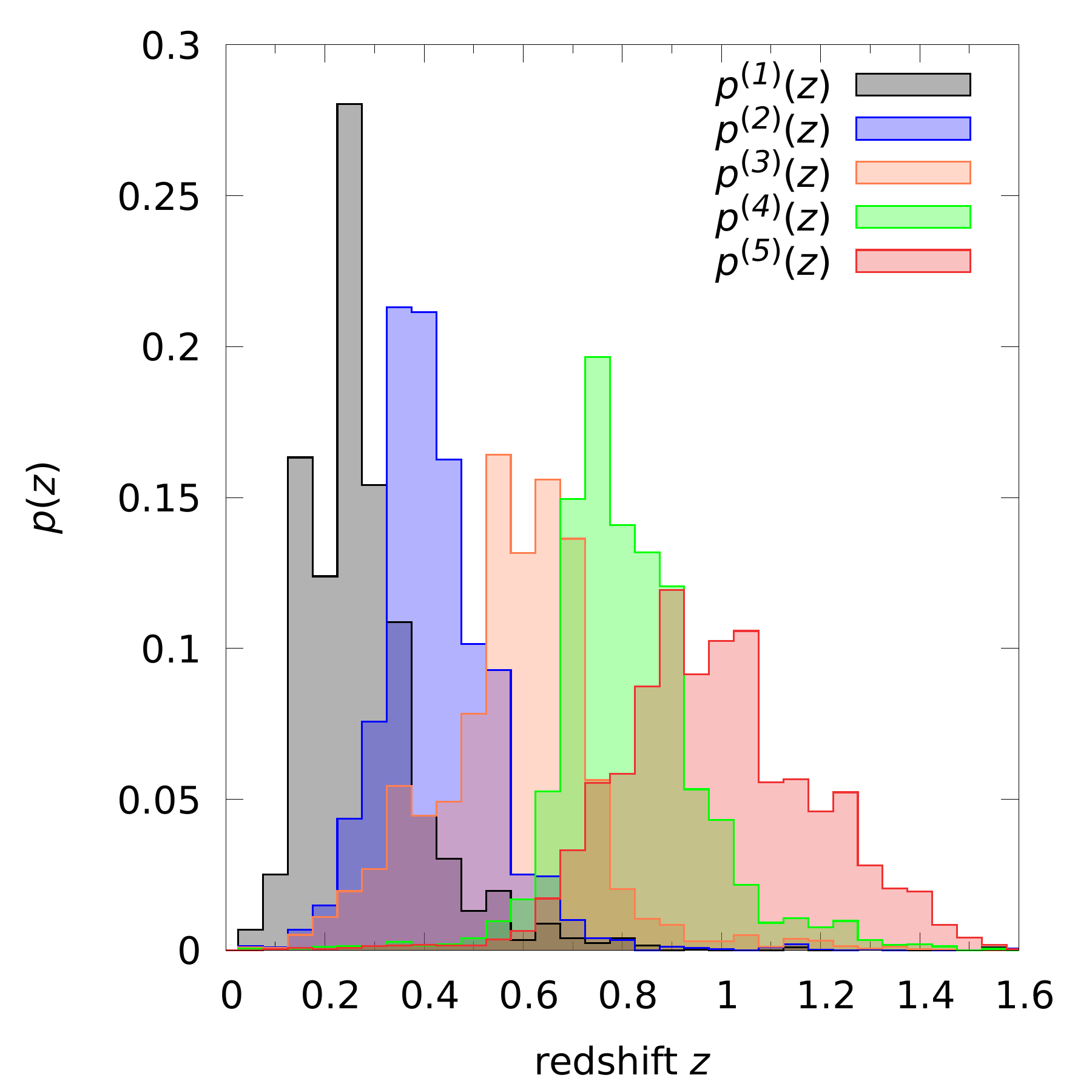}
  \end{center}
  \caption{\label{fig:pofz} Probability density distribution
    functions, $p_z^{(i)}(z)$, of KiDS-1000 source galaxies within the
    five tomographic redshift bins (from $i=1$ to $i=5$): $(0.1,0.3]$,
    $(0.3,0.5]$, $(0.5,0.7]$, $(0.7,0.9]$, and $(0.9,1.2]$. These
    estimates are   from \cite{2021A&A...647A.124H}.}
\end{figure}

Owing to the information loss in the projection Eq. \Ref{eq:xipmauto}
for a single sample of sources, a power spectrum averaged over the
entire light cone could be obtained at best, as in
\cite{2003MNRAS.346..994P} or \cite{2002A&A...396....1S}. More
advanced is splitting the source samples by redshift in a tomographic
analysis for a better statistical precision, reflected by improved
cosmology constraints in tomographic analyses, or to achieve a
(limited) redshift resolution for the recovered
$P_\delta(k,z)$. Therefore, we split our source sample into five
redshift bins, using non-overlapping ranges of photometric redshifts,
namely the $z_{\rm B}$-intervals $(0.1,0.3]$, $(0.3,0.5]$,
$(0.5,0.7]$, $(0.7,0.9]$, and $(0.9,1.2]$, as done already in
\cite{2021A&A...647A.124H}. Figure \ref{fig:pofz} plots estimates of
the resulting distributions $p_z^{(i)}(z)$ for each bin $i$. The
tomographic analysis then correlates the shear signal between bins $i$
and $j$, denoted by the superscript ``$(ij)$'' in
$\xi_\pm^{(ij)}(\theta)$. Compared to Eq. \Ref{eq:xipmauto}, this
increases the data vector size by a factor of $5(5+1)/2=15$ but also
adds information on the $P_\delta(k,z)$ evolution because the shear
tomography probes the same foreground with sources at different $z$.

This tomographic analysis requires special attention with regard to
additional contributions to the shear signal from the intrinsic
alignment (IA) of sources \citep[e.g.,][]{2023arXiv230908605L}. In the
ideal absence of IA, the orientations of intrinsic source
ellipticities are statistically independent among each other and to
the cosmic shear signal. So-called `II' contributions, however,
originate from correlated orientations of physically close
galaxies. In addition, galaxy shapes are aligned to the surrounding
matter density, giving rise to a `GI' signal
\citep{2004PhRvD..70f3526H}. While II contributions could in principle
be reduced by correlating only bin combinations with little radial
overlap, and the GI signal could be suppressed by a nulling technique
\citep{2008A&A...488..829J}, we followed the common approach of
modelling the II and GI signal in $\xi_\pm^{(ij)}(\theta)$ to avoid
information loss,
\begin{equation}
  \label{eq:xipm}
  \xi_\pm^{(ij)}(\theta)=
  {}^0\xi_\pm^{(ij)}(\theta)+{}^\sfont{\rm II}\xi_\pm^{(ij)}(\theta)+
  {}^\sfont{\rm GI}\xi_\pm^{(ij)}(\theta)\;.
\end{equation}
In this equation,
\begin{multline}
  \label{eq:xipm_pure}
  {}^0\xi_\pm^{(ij)}(\theta):=\frac{9H_0^4\Omega_{\rm m}^2}{4c^4}\\
  \times\int_0^{\chi_{\rm h}}
  \!\!\!\!\int_0^\infty\frac{\d\chi\,\d\ell\,\ell}{2\pi}
  \frac{\overline{W}^{(i)}(\chi)\overline{W}^{(j)}(\chi)}{a^2(\chi)}
  J_{0,4}(\ell\theta)\,P_\delta\left(\frac{\ell+1/2}{f_\sfont{K}(\chi)},\chi\right)
\end{multline}
is the ideal shear signal without IA.  Covering both auto-
(\mbox{$i=j$}) and cross-correlations (\mbox{$i\ne j$}) of source
samples, this expression is more general than
Eq. \Ref{eq:xipmauto}. Herein, the lensing efficiency with
superscript, $\overline{W}^{(i)}(\chi)$, refers to
Eq. \Ref{eq:lenseff} but for the PDF $p_\chi^{(i)}(\chi)$ of the $i$th
bin. Notably, the $p_\chi^{(i)}(\chi)$ are related to the
$p_z^{(i)}(z)$ in Fig. \ref{fig:pofz} by
$p_\chi^{(i)}(\chi)=p_z^{(i)}[z(\chi)]\,|\d z/\d\chi|$ for
\mbox{$\d z/\d\chi=H_0\,c^{-1}\,E[(1+z)^{-1}]$}.

For the II and GI terms, we employed the NLA model by
\cite{2011A&A...527A..26J}, derived from the linear alignment model by
\cite{2007NJPh....9..444B} but replacing the linear matter
power-spectrum by the non-linear one.  Clearly just tweaking the
original linear IA model, the NLA is nevertheless sufficiently
accurate to model contemporary lensing data
\citep{2022MNRAS.509.3868H}, predicting
\begin{multline}
  {}^\sfont{\rm II}\xi_\pm^{(ij)}(\theta):=\int_0^{\chi_{\rm h}}
  \!\!\!\!\int_0^\infty\frac{\d\chi\d\ell\,\ell}{2\pi}
  F^2(\chi)\\
  \times\frac{p_\chi^{(i)}(\chi)\,p_\chi^{(j)}(\chi)}{f^2_\sfont{K}(\chi)}
  J_{0,4}(\ell\theta)\,P_\delta\left(\frac{\ell+1/2}{f_\sfont{K}(\chi)},\chi\right)
\end{multline}
for II correlations and for the GI term 
\begin{multline}
  \label{eq:xipm_gi}
  {}^\sfont{\rm GI}\xi_\pm^{(ij)}(\theta):=
  \frac{3H_0^2\Omega_{\rm m}}{2c^2}\int_0^{\chi_{\rm h}}
  \!\!\!\!\int_0^\infty\frac{\d\chi\d\ell\,\ell}{2\pi}\,F(\chi)\,\\
  \times 
  \frac{\overline{W}^{(i)}(\chi)\,p_\chi^{(j)}(\chi)+\overline{W}^{(j)}(\chi)\,p_\chi^{(i)}(\chi)}
  {a(\chi)\,f_\sfont{K}(\chi)}
  J_{0,4}(\ell\theta)\,P_\delta\left(\frac{\ell+1/2}{f_\sfont{K}(\chi)},\chi\right)\;.
\end{multline}
The amplitude of the IA signal scales in this model with
\begin{multline}
  F(\chi)=
  -A_\sfont{IA}\,C_1\,\rho_{\rm crit} \frac{\Omega_{\rm
      m}}{D_+(\chi)}
  \\\approx-2.4\times10^{-2}\,
  \left(\frac{A_\sfont{IA}}{3.0}\right)\,
  \left(\frac{\Omega_{\rm m}}{0.3}\right)\,
  \left(\frac{D_+(\chi)}{0.5}\right)^{-1}
\end{multline}
and depends on distance $\chi$ (or redshift) only through the linear
growth factor, $D_+(\chi)$ (by definition \mbox{$D_+\equiv1$} at
\mbox{$\chi=0$}). For our main result, we neglected further
dependencies on redshift or the evolution of the average galaxy
luminosity with redshift, similar to \cite{2021A&A...645A.104A} who
find in their cosmological analysis of KiDS-1000 little evidence for a
more complex $F(\chi)$.

With respect to future applications, the work by
\cite{2021MNRAS.501.2983F}, and, more recently, by
\cite{2024arXiv240418240P} observe that the NLA model, even if overly
simplistic, could be accurate enough to model IA in Stage IV survey
data if a redshift dependence of the NLA parameters is accounted
for. We briefly return to this topic in our discussion on conceivable
systematic uncertainties in our analysis in
Sect. \ref{sect:discussion}.

\subsection{Projection kernels of the matter power spectrum}

The shear correlation function with IA terms, Eq. \Ref{eq:xipm}, is
still linear in the matter power spectrum, hence a deprojection of
$\xi_\pm^{(ij)}(\theta)$ into $P_\delta(k,z)$ is, as in S12,
principally possible through linear minimum-variance estimators that
depend only on the background fiducial cosmology, source
distributions, and IA parameters.  On the practical side, however, the
deprojection is hampered by broad, partly similar lensing kernels, the
relatively low signal-to-noise ratio, and a confined $\theta$-range,
rendering the estimators ill-conditioned (or singular). They are also
biased when ignoring the $\theta$-confinement which amounts to setting
$\xi^{(ij)}(\theta)\equiv0$ outside the $\theta$ range (Section 7 in
\citealt{2002A&A...396....1S}). Similar to S12, we addressed these
practicalities by boundary conditions $P_\delta(k,z)\ge0$ through
priors in the framework of a Bayesian analysis, at most a couple of
redshift bins for $P_\delta(k,z)$, and by varying $P_\delta(k,z)$
within a confined region in $(k,z)$-space only, assuming a reference
power, $P^{\rm fid}_\delta(k,z)$, otherwise. Additionally, it is
sensible to constrain the relative deviations
$f_\delta(k,z):=P_\delta(k,z)/P^{\rm fid}_\delta(k,z)$ instead of
$P_\delta(k,z)$ directly: Most of the evolution is probably already
accounted for in $P^{\rm fid}_\delta(k,z)$, and a slowly changing
$f_\delta(k,z)$ within a broad $z$-bin is a reasonable quantity to be
averaged. We describe the implementation details below.

In our set-up, the $f_\delta(k,z)$ shall be constant within cells of a
regular mesh of $(N_k+1)\times(N_z+1)$ mesh points $(k_m,\chi_n)$,
where \mbox{$k_m<k_{m+1}$}, \mbox{$z_n<z_{n+1}$}, and
$\chi_n:=\chi(z_n)$; the intervals $[z_1,z_{N_z+1})$ and
$[k_1,k_{N_k+1})$ define the $z$- and $k$-ranges where the power
spectrum may differ from $P_\delta^{\rm fid}(k,z)$, here for
\mbox{$k/(h\,\rm Mpc^{-1})\in[10^{-2},20)$} and
\mbox{$z\in[0,2)$}. The matter power spectrum for Eq. \Ref{eq:xipm}
thus equals
\begin{eqnarray}
  P_\delta(k,\chi)&=&
  P^{\rm fid}_\delta(k,\chi)\left(
    1+\sum_{n,m=1}^{N_z,N_k}
    H_{mn}(k,\chi)\left[f_{\delta,mn}-1\right]\right)\;,
\end{eqnarray}
where
\begin{equation}
  H_{mn}(k,\chi):=\left\{
    \begin{array}{ll}
      1 & ,~{\rm if}~k\in[k_m,k_{m+1})~{\rm
          and}~\chi\in[\chi_n,\chi_{n+1})\\
      0 & ,~{\rm otherwise}
    \end{array}\right.
\end{equation}
is a two-dimensional top-hat function. The average of $f_\delta(k,z)$
within a cell (or band) is denoted by the coefficient
$f_{\delta,mn}$. The mesh in $k$-direction is equi-spaced on a
log-scale, $\Delta k=k\,N_k^{-1}\,\ln{(k_{N_k+1}/k_1)}$, while the
$z$-direction uses for one variant \mbox{$z_n\in\{0,0.3,0.6,2\}$}, for
$N_z=3$, and $z_n\in\{0,2\}$ in other variant with one wide bin,
$N_z=1$, when averaging $f_\delta(k,z)$ over the entire redshift
range. The number of $k$-bins is always \mbox{$N_k=20$}.

Using this $f_{\delta,mn}$-representation of $P_\delta(k,z)$,
Eq. \Ref{eq:xipm} is now recast into
\begin{equation}
  \label{eq:model}
  \xi_\pm^{(ij)}(\theta)=
  \sum_{n,m=1}^{N_z,N_k}
  X^{(ij)}_\pm(\theta;m,n)\,f_{\delta,mn}+
  \xi_{\pm,\rm fid}^{(ij)}(\theta)\;,
\end{equation}
for the projection matrix
\begin{multline}
  \label{eq:projmatrix}
  X^{(ij)}_\pm(\theta;m,n)\\
   :=\frac{1}{2\pi\,\theta^2}\,
  \int_{\chi_n}^{\chi_{n+1}}\d\chi\;
  \left({}^0{\cal K}^{(ij)}_{\gamma\gamma}(\chi)+
    {}^\sfont{II}{\cal K}^{(ij)}_{\gamma\gamma}(\chi)+
    {}^\sfont{GI}{\cal K}^{(ij)}_{\gamma\gamma}(\chi)\right)\\
  \times\int\limits_{k_m f_\sfont{K}(\chi)\theta}^{k_{m+1}f_\sfont{K}(\chi)\theta}\d s\,s\,
  J_{0,4}(s)\,P^{\rm fid}_\delta\!\left(k[s,\chi,\theta],\chi\right)\;,
\end{multline}
the integral kernels
\begin{eqnarray}
  {}^0{\cal K}^{(ij)}_{\gamma\gamma}(\chi)&:=&
  \frac{9H_0^4\Omega_{\rm m}^2}{4c^4}\,
  \frac{\overline{W}^{(i)}(\chi)\overline{W}^{(j)}(\chi)}{a^2(\chi)}\;;
  \\
  {}^\sfont{II}{\cal K}^{(ij)}_{\gamma\gamma}(\chi)&:=&              
  F^2(\chi)\,\frac{p_\chi^{(i)}(\chi)\,p_\chi^{(j)}(\chi)}{f^2_\sfont{K}(\chi)}\;;
  \\
  \nonumber
  {}^\sfont{GI}{\cal K}^{(ij)}_{\gamma\gamma}(\chi)&:=&
                                                        \frac{3H_0^2\Omega_{\rm m}}{2c^2}\, F(\chi)\\
  &&
  \times\frac{\overline{W}^{(i)}(\chi)\,p_\chi^{(j)}(\chi)+\overline{W}^{(j)}(\chi)\,p_\chi^{(i)}(\chi)}
  {a(\chi)\,f_\sfont{K}(\chi)}\;,
\end{eqnarray}
and a constant offset
\begin{multline}
  \xi_{\pm,\rm fid}^{(ij)}(\theta)\\
    :=\frac{1}{2\pi\,\theta^2}\,\int_{0}^{\chi_{\rm h}}\d\chi
    \left({}^0{\cal K}^{(ij)}_{\gamma\gamma}(\chi)+
     {}^\sfont{II}{\cal K}^{(ij)}_{\gamma\gamma}(\chi)+
     {}^\sfont{GI}{\cal K}^{(ij)}_{\gamma\gamma}(\chi)\right)\\
  \times\int\limits_{0}^{\infty}\d s\,s\,
  J_{0,4}(s)\,
  \cancel{P}^{\rm fid}_\delta\left(k[s,\chi,\theta],\chi\right)
\end{multline}
from $(k,z)$ regions that are unaffected by the choice of
$f_{\delta,mn}$. The offset employs the definition
\begin{equation}
\cancel{P}^{\rm fid}_\delta\left(k,\chi\right):=
\left\{
\begin{array}{ll}
  P^{\rm fid}_\delta(k,\chi) & ,~{\rm if}~\sum\limits_{m,n=1}^{N_z,N_k}H_{mn}(k,\chi)=0\\
  0 & ,~{\rm otherwise}
\end{array}
\right.\;,
\end{equation}
and the foregoing equations abbreviate
$k[s,\chi,\theta]:=\frac{s+\theta/2}{f_\sfont{K}(\chi)\theta}$.

Our code computed the projection matrix $X_\pm^{(ij)}(\theta;m,n)$
once for a series of $\theta$-bins, enabling a quick prediction of
$\xi_\pm^{(ij)}(\theta)$ when Monte Carlo sampling the posterior PDF
of $f_{\delta,mn}$.  An efficient way to numerically calculate
$X_\pm^{(ij)}(\theta;m,n)$ and $\xi_{\pm,\rm fid}^{(ij)}(\theta)$ is
given in Appendix A of S12, albeit with the little modification
$(2f_\sfont{K}[\chi])^{-1}$ in what is now
$\bar{k}_j:=(2f_\sfont{K}[\chi])^{-1}+(\hat{k}_j+\hat{k}_{j+1})/2$ due
to the extended hybrid Limber approximation adopted here. As further
practical footnote on code implementation, we performed the above
integrals with integration variable $a$ instead of $\chi$, as well as
$p_z(z)$ instead of $p_\chi(\chi)$, for which the following
transformations are notable, given a general function $g(\chi)$ and
$H(a)=:H_0\,E(a)$,
\begin{equation}
  \int_{\chi_1}^{\chi_2}\d\chi\;g(\chi)=
  \frac{c}{H_0}\int_{a_2}^{a_1}\frac{\d a}{a^2\,E(a)}\;g[\chi(a)]
\end{equation}
since $|\d\chi/\d a|=c/(a^2H[a])$, 
\begin{equation}
  \int_{\chi_1}^{\chi_2}\d\chi\;p^{(i)}_\chi(\chi)\,g(\chi)=
  \int_{a_2}^{a_1}\frac{\d a}{a^2}\;p_z^{(i)}[z(a)]\,g[\chi(a)]
\end{equation}
since $|\d z/\d\chi|=H(z)/c$, and hence
\begin{multline}
  \int_{\chi_1}^{\chi_2}\d\chi\;p^{(i)}_\chi(\chi)\,p^{(j)}_\chi(\chi)\,g(\chi)\\
  =\frac{H_0}{c}\,
  \int_{a_2}^{a_1}\frac{\d a\;E(a)}{a^2}\,p_z^{(i)}[z(a)]\,p_z^{(j)}[z(a)]\,g[\chi(a)]\;.
\end{multline}
When integrating over $p_z^{(i)}(z)$, we adopted histograms with
top-hat functions, without interpolation, exactly as indicated in
Fig. \ref{fig:pofz}. The very first bin, left-aligned towards $z=0$,
has $p_z^{(i)}(z)=0$. Probably obvious through these integrals and the
pre-factors in Eqs. \Ref{eq:xipm_pure}--\Ref{eq:xipm_gi}, the
projections of $P_\delta(k,z)$ into $\xi^{(ij)}(\theta)$ are
independent of $H_0$ as long as $[k]=h\,\rm Mpc^{-1}$ (comoving),
whereas the $\xi_\pm^{(ij)}(\theta)$ scale with $\Omega_{\rm m}^2$
since $F(\chi)\propto\Omega_{\rm m}$.

\begin{table}
  \begin{center}
    \caption{\label{tab:fiducialmodel} Parameters of the reference
      power spectrum, $P^{\rm fid}_\delta(k,z)$, and the projection
      parameters (lensing kernel, IA) used to infer
      $f_\delta(k,z)=P_\delta(k,z)/P^{\rm fid}_\delta(k,z)$.  For
      $P^{\rm fid}_\delta(k,z)$ in this $\Lambda\rm CDM$ setting, we
      employed an updated version of \texttt{halofit}
      (\citealt{2003MNRAS.341.1311S,2012ApJ...761..152T}). The
      reference assumes a flat cosmology, this means
      $\Omega_\Lambda=1-\Omega_{\rm m}$.}
    \begin{tabular}{cccc}
      $A_\sfont{IA}$ & $\Omega_{\rm m}$ & $\Omega_{\rm b}$ & $n_{\rm s}$\\
      \hline\\
      $+1.070$ & $0.305$ & $0.047$ & $0.901$
      \\\\
      $h$ & $\sigma_8$ & $\Gamma$ & $w_0$\\
      \hline\\
      $0.695$ & $0.720^\dagger$ & $0.167$ & $-1.000$                                                        
    \end{tabular}
  \end{center}
  \tablefoot{$A_\sfont{IA}$: IA amplitude of the NLA model;
    $\Omega_{\rm m}$: (total) matter density parameter;
    $\Omega_{\rm b}$: baryon density parameter; $n_{\rm s}$: shape
    parameter of the primordial power spectrum; $h$: Hubble constant
    in units of $100\,\rm km\,s^{-1}\,Mpc^{-1}$; $\sigma_8$:
    normalisation of the linear power spectrum at $z=0$; $\Gamma$:
    shape parameter according to \cite{1995ApJS..100..281S}; $w_0$:
    equation-of-state parameter for dark energy; ${}^\dagger$ this
    value has been lowered from the $0.76$ in
    \cite{2021A&A...646A.140H}, Table C.1, to obtain an average of
    $\bar{f}_\delta\approx1$ over all $k$ and $z$
    (Sect. \ref{sect:refpower})}
\end{table}

\section{Bayesian inference of the three-dimensional matter
  power-spectrum}
\label{sect:statanalysis}

With the linear relation between the three-dimensional (3D) power
spectrum and the two-point shear correlation functions described, we
now outline the statistical model and the numerical sampler for the
posterior probability density of $f_{\delta,mn}$ within a Bayesian
framework \citep[e.g.,][for a review]{gelman2003bayesian}. In
comparison to S12, our updated methodology introduces the Tikhonov
regularisation as Bayesian prior, in effect inferring a $k$-smoothed
power spectrum, and a Hamiltonian MCMC for improved computational
speed for $N_z\times N_k\sim60$ model variables.

\subsection{Statistical model}
\label{sect:statmodel}

The statistical model of $f_{\delta,mn}$ employs a compact notation
for the tomographic data, model variables and parameters, all
presented here. The vector of model variables
\mbox{$\vec{\pi}=\left(f_{\delta,mn}|1\le n\le N_z, 1\le m\le
    N_k\right)$} compiles the bin-averaged $f_\delta(k,z)$ on a
$(N_k+1)\times(N_z+1)$ mesh, and the data vector $\vec{d}$ contains
the binned $\xi_\pm^{(ij)}(\theta_k)$ for a sequence of
\mbox{$N_\theta=9$} angular separations, $\theta_k$. Equation
\Ref{eq:model} predicts the components of $\vec{d}$ for a given (flat)
background cosmology, IA parameters, and source redshift
distributions, altogether compressed into
\mbox{$\vec{q}=\{\Omega_{\rm
    m},A_\sfont{IA},p_z^{(1)}(z),\ldots,p_z^{(5)}(z)\}$}
hereafter. The prediction is denoted by the model vector
$\vec{m}(\vec{\pi},\vec{q})=\mat{X}_{\vec{q}}\,\vec{\pi}+\vec{\xi}^{\rm
  fid}_{\vec{q}}$ with elements arranged in the same order as those in
$\vec{d}$, using the projection matrix $\mat{X}_{\vec{q}}$ which
contains the coefficients $X^{(ij)}_\pm(\theta;m,n)$ for fixed
$\vec{q}$. The elements of $\vec{\xi}^{\rm fid}_{\vec{q}}$ are the
offsets $\xi_{\pm,\rm fid}^{(ij)}(\theta_k)$, also for fixed
projection parameters $\vec{q}$. The statistical information on
$\vec{\pi}$ is expressed by the Bayesian posterior PDF of $\vec{\pi}$
conditional on $\vec{d}$,
\begin{equation}
  \label{eqn:posterior}
  P(\vec{\pi}|\vec{d})=
  {\cal E}^{-1}(\vec{d})\,{\cal L}_{\vec{q}}(\vec{d}|\vec{\pi})\,P_{\rm hat}(\vec{\pi})\,P_\tau(\vec{\pi})\;,
\end{equation}
adopting a Gaussian likelihood
\begin{equation}
  \label{eq:likelihood}
  -2\,\ln{{\cal L}_{\vec{q}}(\vec{d}|\vec{\pi})}=
  \left[\vec{d}-\vec{m}(\vec{\pi},\vec{q})\right]^{\rm T}\mat{C}^{-1}
  \left[\vec{d}-\vec{m}(\vec{\pi},\vec{q})\right]
\end{equation}
with error covariance $\mat{C}$ of the measurement $\vec{d}$, and two
prior probability densities, $P_{\rm hat}(\vec{\pi})$ and
$P_\tau(\vec{\pi})$, to regularise the sampling in
$f_\delta$-parameter space. The normalisation,
${\cal E}(\vec{d})=\int\d\vec{\pi}\,{\cal
  L}_{\vec{q}}(\vec{d}|\vec{\pi})\,P_{\rm
  hat}(\vec{\pi})\,P_\tau(\vec{\pi})\ne0$, is not of interest for the
Monte Carlo sampler, and hence we set ${\cal E}(\vec{d})\equiv1$
without impacting the results. We describe the two prior PDFs in the
following.

\subsection{Tikhonov regularisation prior}
\label{sect:tikhonov}

\begin{figure*}
  \begin{center}
    \includegraphics[width=185mm]{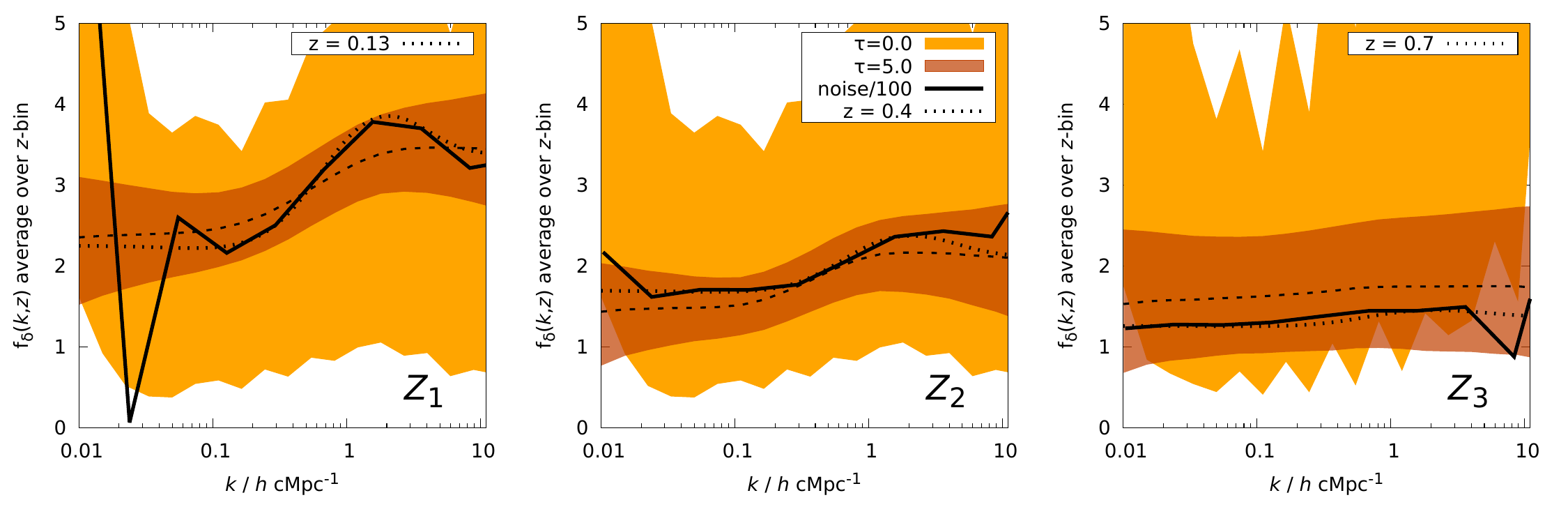}
  \end{center}
  \caption{\label{fig:tikhonov} Impact of the Tikhonov regularisation
    used to suppress oscillating solutions $f_\delta(k,z)$. Shown are,
    for a noise-free mock data vector and the KiDS-1000 error
    covariance, the posterior constraints ($68\%$ credible regions) on
    $f_\delta$ averaged over the redshift bin $Z_1=[0,0.3]$
    (\emph{left}), $Z_2=[0.3,0.6]$ (\emph{middle}), and $Z_3=[0.6,2]$
    (\emph{right}) with and without regularisation (dark orange
    $\tau=5.0$ with median as dashed line or light orange
    $\tau=0$). To boost scale-dependence and evolution,
    $f_\delta(k,z)$ is here defined relative to the power spectrum at
    fixed redshift, probing the relative structure growth since
    $z=1$. The solid line is the median posterior $f_\delta$ for
    $100\times$ reduced measurement errors, providing a nearly
    noise-free reference (noise/100) that averages the growth over the
    redshift bin while still exhibiting artefacts near the edges. The
    dotted lines are the theoretical
    $P_\delta^{\rm fid}(k,\bar{z})/P_\delta^{\rm fid}(k,z=1)$ for one
    specific $\bar{z}$ chosen to most closely match the solid lines,
    indicating the redshift with highest weight in the average.}
\end{figure*}

The Tikhonov regularisation prior, $P_\tau(\vec{\pi})$, greatly
improves the constraints, pruning implausible solutions for
$f_{\delta,mn}$. This is necessary because the combination of heavy
smoothing and added noise in $\xi_\pm^{(ij)}(\theta)$ has the
unpleasant side-effect of producing strong oscillations, correlated
errors, in $f_{\delta}(k,z)$ when inverting the tomographic
signal. Similar to 3D-lensing mass reconstructions
\citep{2002PhRvD..66f3506H}, the oscillations are reduced by applying
regularisation conditions. To this effect, the Tikhonov regularisation
down-weights oscillating $f_\delta(k,z)$, preferring solutions of
$f_{\delta,mn}$ smoothed in $k$-direction,
\begin{equation}
  -\frac{1}{\tau}\,\ln{P_\tau(\vec{\pi})}=
  \sum_{n=1}^{N_z}\sum_{m=1}^{N_k-1}
  \left(f_{\delta,mn}-f_{\delta,(m+1)n}\right)^2\;.
\end{equation}
At the same time, no prior constraints are imposed on the overall
amplitude of $f_{\delta,mn}$ or the difference signal between $z$-bins
because only signal differences from the same $z$-bin, $n$, appear
inside the prior density. The weight of the prior relative to the data
likelihood ${\cal L}_{\vec{q}}(\vec{d}|\vec{\pi})$ is controlled by
the Tikhonov parameter $\tau$, demonstrated in Fig. \ref{fig:tikhonov}
for a simulated analysis with three redshift bins, \mbox{$N_z=3$}.

This intentionally extreme scenario in Fig. \ref{fig:tikhonov}
highlights the impact of the regularisation with a truly $z$-evolving
and scale-dependent $f_{\delta,mn}$ by choosing a constant reference
$P_\delta^{\rm fid}(k,z)$ for all $z$-bins: a theory power spectrum at
\mbox{$z=1$}. Unlike the actual KiDS analysis in
Sect. \ref{sect:results}, the $f_{\delta,mn}$ therefore now quantify
the structure growth relative to \mbox{$z=1$}. To assess the quality
of the reconstruction and our ability to detect the $z$- and
$k$-dependence of $f_{\delta,mn}$, we analysed a mock vector
$\vec{d}=\vec{m}(\vec{\pi},\vec{q})$ with parameters in Table
\ref{tab:fiducialmodel} and KiDS-1000-like noise, $\mat{C}$, yielding
posterior $68\%$ credible regions with regularisation
(\mbox{$\tau=5$}, dark orange) and without (\mbox{$\tau=0$}, light
orange). The figure compares these credible regions to virtually
noise-free data, unaffected by the Tikhonov regularisation, depicted
as solid lines (``noise/100'').  The solid lines are, to account for
the $z$-weighting of $f_\delta(k,z)$ in a reconstruction, the
posterior medians for a data vector with unrealistically high
signal-to-noise ratio (S/N) of $\mat{C}^\prime=10^{-4}\times\mat{C}$
as error covariance and no Tikhonov regularisation. But even for such
high S/N data, artefacts appear towards large scales, close to
\mbox{$k\sim0.02\,h\,\rm Mpc^{-1}$}, and at the high-$k$ end, near
\mbox{$k\sim10\,h\,\rm Mpc^{-1}$}, thus towards $k$ where the
tomographic data poorly constrain the power spectrum. To mitigate
these artefacts, the additional dotted lines depict for each $z$-bin a
theoretical
\mbox{$P_\delta^{\rm fid}(k,\bar{z})/P_\delta^{\rm fid}(k,z=1)$} at
one specific redshift, $\bar{z}$, that matches the solid lines most
closely, and which shall be the `true' average $f_{\delta,mn}$ for
this experiment. Undoubtedly, a Tikhonov prior, \mbox{$\tau=5$},
substantially reduces the statistical uncertainty in the
reconstruction (sizes of credible regions for \mbox{$\tau=0$} versus
\mbox{$\tau=5$}). On the downside, however, the regularisation might
be overly restrictive by straightening out oscillations actually
present in the data, as already partly present for \mbox{$\tau=5$}
(dotted lines versus dashed lines, the posterior median). By running
more tests with different $\tau$, not shown here, we consider
\mbox{$\tau=5$} a good compromise that catches gradual trends with $k$
while improving the precision in the reconstruction, as in the left
panel. We reiterate that the degree of evolution and scale-dependence
in this test, due to the static reference power spectrum, is extreme
compared to what is expected in the KiDS-1000 analysis. If the
reference is close to the true matter power at all $z$,
$f_{\delta,mn}$ will be close to unity.

A cruder alternative to Tikhonov filtering ($\tau=0$) for smoothing
the reconstruction involves drastically reducing the number of
$k$-bins, such as to $N_k=5$. This approach, however, significantly
blurs $f_\delta(k,z)$ and their scale-dependent features, effectively
smoothing it to the fixed size of the larger $k$-bins, resulting in a
distinctly lower resolution compared to Tikhonov filtering. In
contrast, Tikhonov filtering applies adaptive smoothing, using smaller
smoothing kernels where the S/N is higher and broader kernels
elsewhere. This method's resolution is limited only by the size of the
numerous, smaller $k$-bins, for which we chose $N_k=20$.

\subsection{Positivity priors}

Already applied for Fig. \ref{fig:tikhonov}, we restricted valid
solutions to positive $f_{\delta,mn}$ (since
\mbox{$P_\delta(k,z)\ge0$}), by using uniform, top-hat prior PDFs for
\mbox{$f_{\delta,mn}\in[0,f_{\delta,\rm max}]$},
\begin{multline}
  -\sigma_{\rm f}^2\,\ln{P_{\rm hat}(\vec{\pi})}=
  \sum_{n,m=1}^{N_z,N_k}\left(
    f_{\delta,mn}^2\,\mathrm{H}\left[-f_{\delta,mn}\right]\right.\\
  \left. +\left[f_{\delta,mn}-f_{\delta,\rm
        max}\right]^2\,\mathrm{H}\left[f_{\delta,mn}-f_{\delta,\rm
        max}\right] \right)\;,
\end{multline}
where $\mathrm{H}(x)$ is the Heaviside step function. Enforcing
positive solutions for the matter power spectrum is beneficial in
reducing both the statistical errors and oscillations in the
reconstruction, as already reported in S12. The positivity priors
adopt wide intervals, \mbox{$f_{\delta,\rm max}=100$}, compared to
expected values of \mbox{$f_{\delta,mn}\sim1$}. In addition, for
numerical convenience, the top-hat priors have soft edges,
\mbox{$\sigma_{\rm f}=10^{-2}$}, to relax issues with low acceptance
rates and undefined gradients in the Monte Carlo sampler of
$P(\vec{\pi}|\vec{d})$ near the edges.

\subsection{Monte Carlo sampler}

For sampling the posterior PDF, Eq. \Ref{eqn:posterior}, we applied a
MCMC technique that represents $P(\vec{\pi}|\vec{d})$ by
\mbox{$n_{\rm mcmc}=10^4$} points $(\vec{\pi}_i,w_i)$ with statistical
weights, $w_i$, for given projection parameters $\vec{q}$. The MCMC
sampler is a Hamiltonian Monte Carlo algorithm, known to be efficient
even for high-dimensional models ($N_z\times N_k\sim60$) by proposing
new, quickly decorrelating sampling points with high acceptance rate
\citep{gelman2003bayesian}. The sampler, however, requires as input
the gradient
$\nabla_{\vec{\pi}}\ln{{\cal L}_{\vec{q}}(\vec{d}|\vec{\pi})}$ and the
gradient of the logarithmic prior densities -- all easily available in
our case, foremost because of the linear projection
$\vec{m}(\vec{\pi},\vec{q})=\mat{X}_{\vec{q}}\vec{\pi}+\vec{\xi}^{{\rm
    fid}}_{\vec{q}}$. We demonstrate the appropriate sampler
performance in Appendix \ref{ap:mcmc} and provide more implementation
details there. Due to the efficient decorrelation of the MCMCs, we
shortened the burn-in phase for the KiDS-1000 analysis,
\mbox{$n_{\rm burnin}=5\times10^3$}, in comparison to Appendix
\ref{ap:mcmc}, and used shorter chain lengths because of the
subsequent merger of many chains when marginalising over $\vec{q}$,
outlined in the next section.

\subsection{Marginal errors including uncertainties in IA and lensing
  kernel parameters}
\label{sect:errormodel}

The parameters for the lensing kernel and IA are neither exactly known
nor intended to be constrained by our shear data. Therefore, to
account for their uncertainties in $f_{\delta,mn}$, the analysis
combines many chains with \mbox{$n_{\rm runs}=500$} realisations of
$\vec{q}$ drawn from a baseline error model. The error baseline
adopted a flat cosmology, $\Omega_\Lambda=1-\Omega_{\rm m}$, and a
random sample of the posterior joint distribution of
$(\Omega_{\rm m},A_\sfont{IA})$ from the recent
\mbox{$3\times2\,\rm pt$} cosmological analysis in
\citet{2021A&A...646A.140H}; their marginal errors are
\mbox{$\Omega_{\rm m}=0.305_{-0.012}^{+0.012}$} and
\mbox{$A_\sfont{IA}=+1.04_{-0.30}^{+0.28}$} ($68\%$ CI). The
\mbox{$3\times2\,\rm pt$} experiment combines the KiDS-1000 cosmic
shear constraints with those from galaxy-galaxy lensing and,
importantly, the galaxy clustering in the partly overlapping,
spectroscopic surveys of both the 2-degree Field Lensing Survey and
the Baryon Oscillation Spectroscopic Survey
\citep{2016MNRAS.462.4240B,2015ApJS..219...12A} to break the
$\sigma_8-\Omega_{\rm m}$ degeneracy in the shear data. This
substantially reduces the uncertainty of $\Omega_{\rm m}$ in the
lensing kernel while, at the same time, using only \mbox{$z\lesssim1$}
data. Furthermore, for error realisations of the source distributions,
we shift the histograms in Fig. \ref{fig:pofz} according to
$p^{(i)}_z(z)\to p^{(i)}_z(z+\delta^i_z)$, for random draws of
$\delta^i_z$ confined by the error model in
\cite{2021A&A...647A.124H}, their Figure 6 and Table 3; the
root-mean-square (RMS) error of $\delta_z^i$ is
\mbox{$\sigma^i_z\approx1.3\times10^{-2}$} for all tomographic bins,
except for $i=3$ where \mbox{$\sigma_z^i=2\times10^{-2}$}. Within this
Monte Carlo process, we fixed the reference $P_\delta^{\rm fid}(k,z)$
for all chains to that in Table \ref{tab:fiducialmodel} to avoid
conflicting definitions of $f_{\delta,mn}$ while varying
$\Omega_{\rm m}$.

After all MCMCs with randomised lensing kernels and IA parameters were
available, we combined the $n_{\rm runs}$ runs by selecting from each
chain $j$ a number of \mbox{$n_{\rm merge}=10^3$} sampling points
$\vec{\pi}_{ij}$ with probability \mbox{$w_{ij}/\sum_i
  w_{ij}$}. Selected points were put back into the sample $j$, to be
possibly selected again. By doing so, the selected $n_{\rm merge}$
points sample the posterior of chain $j$ when equally weighted. Mixing
the MCMC points from all $n_{\rm run}$ chains for the final
Monte Carlo sample then contains
\mbox{$n_{\rm merge}\times n_{\rm runs}=5\times10^5$}, equally
weighted sampling points $\vec{\pi}_{ij}$ of the marginalised
posterior distribution of $f_{\delta,mn}$.

\begin{table}
\centering
\caption{\label{tab:gold}Overview of the KiDS-1000 `gold sample'}
\begin{tabular}{c c c c c}
\hline 
 \hline 
 \noalign{\smallskip}
Bin & $z_{\rm B}$ range & $n_{\rm eff}/\rm arcmin^{-2}$ & $\sigma_{\epsilon}$ & $\delta^i_z$ 
 \\
 \noalign{\smallskip}
 \hline 
 \noalign{\smallskip}
 1
 & $(0.1,0.3]$
 & 0.62
 & 0.270
 & $+0.000 \pm 0.0106$
 \\
 \noalign{\smallskip}
 2
 & $(0.3,0.5]$
 & 1.18
 & 0.258
 & $+0.002 \pm 0.0113$
 \\
 \noalign{\smallskip}
 3
 & $(0.5,0.7]$
 & 1.85
 & 0.273
 & $+0.013 \pm 0.0118$
 \\
 \noalign{\smallskip}
 4
 & $(0.7,0.9]$
 & 1.26
 & 0.254
 & $+0.011 \pm 0.0087$
 \\
 \noalign{\smallskip}
 5
 & $(0.9,1.2]$
 & 1.31
 & 0.270
 & $-0.006 \pm 0.0097$ 
 \\
 \noalign{\smallskip}
 \hline \hline
 \vspace{0.1cm}
 \end{tabular}
 \tablefoot{The data in this table are compiled from
   \cite{2021A&A...647A.124H} and \cite{2021A&A...645A.105G}. The
   effective number density, $n_{\rm eff}$, accounts for the
   \emph{lens}fit weights, see \cite{2021A&A...646A.129J} for
   details. The fourth column displays the measured ellipticity
   dispersion per component. Our analysis takes into account the
   correlation of the redshift bias uncertainties; $z_{\rm B}$ denotes
   the photometric redshift.}
\label{tab:KiDS1000}
\end {table}

\section{Data}
\label{sect:data}

We used the public data from the fourth Data Release (DR4) of the
Kilo-Degree Survey \citep{2019A&A...625A...2K}, commonly referred to
as the KiDS-1000 data.\footnote{The KiDS data products are public and
  available through \url{http://kids.strw.leidenuniv.nl/DR4}} The
images underlying the data were taken by the high-quality VST-OmegaCAM
\citep{2011Msngr.146....8K}, covering a total area of about
$1006 \, \deg^2$ in four optical filters ($ugri$). After masking, the
effective area of the KiDS-1000 data is $777.4 \, \deg^2$. Its overlap
with the VIKING survey \citep[Vista Kilo-degree Infrared Galaxy
survey, ][]{2013Msngr.154...32E}, which observes in five near-infrared
bands, $ZYJHK_s$, allows a better control over redshift uncertainties
\citep{2021A&A...647A.124H}.

\subsection{Source catalogue}

The following summarises the production of the shear catalogue,
carried out by the KiDS-1000 Consortium. The images were processed
using the {\tiny{THELI}} \citep{2005AN....326..432E} and
A{\tiny{STRO}}-WISE \citep{2013ExA....35....1B} pipelines, source
ellipticities were estimated by \emph{lens}fit
\citep{2013MNRAS.429.2858M,2017MNRAS.467.1627F}. Based on the
nine-band catalogue, individual photometric redshift estimates,
$z_{\rm B}$, were obtained using \texttt{BPZ}
\citep{2000ApJ...536..571B}, dividing the sources into five
tomographic bins (Fig. \ref{fig:pofz}).

To calibrate the redshift distribution of each tomographic bin,
\cite{2020A&A...637A.100W} construct a self-organising map (SOM)
linking the nine-band photometry of the sources to a representative
spectroscopic sample. Galaxies without matching spectroscopic
counterparts, or for which the $z_{\rm B}$ are catastrophically
different from the redshift of the matched spectroscopic sample, are
expunged from the catalogue \citep{2020A&A...640L..14W}. The mean and
the scatter, $\delta^i_z$, on the bias of the means of the
$p_z^{(i)}(z)$, estimated by the SOM, are numerically estimated using
a suite of KiDS-like mock observations
\citep{2021A&A...647A.124H}.\footnote{Sample variance of the
  large-scale structure introduces a correlation between the scatter
  of $\delta^i_z$ across different tomographic redshift bins, see
  Figure 2 in \cite{2021A&A...647A.124H} for the explicit values.}
The resulting `gold sample' contains $\sim2.1\times10^7$ sources with
well-calibrated photometric redshift distributions;
\cite{2021A&A...645A.105G} gives more details. Table
\ref{tab:KiDS1000} in the Appendix summarises the main parameters
relevant for our analysis. Cosmological parameter constraints from the
KiDS-1000 tomographic data, with the methodology in
\cite{2021A&A...646A.129J}, are presented in
\cite{2021A&A...645A.104A}, exclusively using cosmic shear, in
\cite{2021A&A...646A.140H}, using 3$\times$2pt data, and in
\cite{2021A&A...649A..88T}, for a beyond $\Lambda$CDM analysis using
3$\times$2pt data.

\subsection{Data vector and covariance of measurement noise}

Based on the tomographic shear data, the KiDS-1000 DR4 provides
estimates of the binned shear correlation function,
$\xi_\pm^{(ij)}(\theta)$, for every combination of tomographic bins,
$(ij)$, employing the popular computer software \texttt{TreeCorr}
\citep{2004MNRAS.352..338J,2015ascl.soft08007J}. We used these data
without modification. For the $\xi_\pm$ estimator details, we refer to
\cite{2021A&A...645A.104A}, Section 2.1 and Appendix C, and only note
that we analysed the $\xi^{(ij)}_\pm(\theta)$ binned into
\mbox{$N_\theta=9$} angular log-bins for the range
\mbox{$\theta=0\farcm5$--$\ang{;300;}$}. Another notable technical
detail, the data vector for the DR4 was obtained by rebinning a
previously finely binned vector, instead of rerunning
\texttt{TreeCorr}. Further, \cite{2021A&A...645A.104A} exclude scales
smaller than $\ang{;4;}$ for $\xi_-$ due to high systematic
uncertainties in the theoretical matter power-spectrum on small
scales, whereas our model-free measurement of $P_\delta(k,z)$ included
scales down to $0\farcm5$.

The covariance matrix of noise, $\mat{C}$ in Eq. \Ref{eq:likelihood},
for the estimator of $\xi_\pm^{(ij)}(\theta)$ used the analytical
model by \cite{2021A&A...646A.129J}, Appendix E. This model accounts
for the intrinsic shape scatter of sources, for both Gaussian and
non-Gaussian cosmic variance, mostly relevant towards larger angular
scales, for the super-sample covariance due to fluctuation modes
greater than the survey footprint, and for the calibration uncertainty
in the multiplicative shear bias. As for $\xi_\pm^{(ij)}(\theta)$,
the covariance matrix is part of the package of additional DR4 data
products.

\subsection{Systematic shear error}
\label{sect:bmode}

In addition to intrinsic alignments, a residual systematic error in
the shear signal through the data reduction or the instrument may also
be potentially present in $\xi_\pm^{(ij)}$. Detecting $B$-modes in the
shear field, specifically
\mbox{$\tilde{\kappa}(\vec{\ell})\ne\tilde{\kappa}^\ast(-\vec{\ell})$}
as defined in Eq. \Ref{eq:gammakappa}, is considered a strong
indicator of systematic errors, because $B$-modes generated by weak
gravitational lensing are negligible. Clearly separating $B$-modes
from $E$-modes in a $\xi_\pm^{(ij)}$ with finite support is, however,
unfeasible, but other related statistics are known to separate both
modes (and ambiguous modes), most prominently COSEBIs
\citep{2010A&A...520A.116S}.  For KiDS-1000, logarithmic COSEBIs
capturing scales \mbox{$\ell\lesssim10^3$} do not indicate $B$-modes
on a $95\%$ confidence level \citep{2021A&A...645A.104A,
  2021A&A...645A.105G}, several other tests for systematic errors are
negative, and the cosmological analysis with the $E$-mode COSEBIs
yields results consistent with a $\xi_\pm^{(ij)}$ analysis. Therefore,
a significant bias of parameters in a cosmological analysis due to a
systematic shear error is unlikely.

Notably, however, the COSEBIs cosmological analysis alone is
insensitive to small angular scales of $\ell\sim10^4$, included in our
analysis through $\theta\lesssim\ang{;4;}$ in $\xi^{(ij)}_-$. In
addition, even a small systematic shear error in $\xi_\pm^{(ij)}$
might cause artificial deviations of $P_\delta(k,z)$ from the
reference $P_\delta^{\rm fid}(k,z)$, i.e., \mbox{$f_{\delta,mn}\ne1$},
if the error is of the order of the \emph{difference signal} between
the measured $\xi^{(ij)}_\pm$ and the expected signal in
Eq. \Ref{eq:model} for \mbox{$f_{\delta,mn}\equiv1$}. The
$1\sigma$--$2\sigma$ window left open for $B$-modes by the logarithmic
COSEBIs therefore leaves the possibility that deviations in KiDS-1000
might partly be related to a systematic shear error in the data. We
return to this caveat in Sect. \ref{sect:discussion}.

\section{Results}
\label{sect:results}

This section reports our results for the relative power spectrum,
$f_{\delta,mn}$, in two variants: an entirely average, non-evolving
scenario (one broad redshift bin) and an evolving scenario with three
independent redshift bins. In addition, we verified our pipeline with
results from a mock analysis based on ray-traced $N$-body data in an
entirely $\Lambda$CDM framework.

\subsection{Reference $\Lambda$CDM matter power-spectrum}
\label{sect:refpower}

For the KiDS-1000 analysis presented here, the reference power
spectrum, $P^{\rm fid}_\delta(k,z)$, is a revised \texttt{halofit}
model with the parameters in Table \ref{tab:fiducialmodel}
\citep{2012ApJ...761..152T, 2003MNRAS.341.1311S}. These parameters are
maximum-posterior values taken from Table C.1
(\mbox{$3\times2\,\rm pt$}, joint) in \cite{2021A&A...646A.140H} with
one modification: the value for $\sigma_8$ was lowered from $0.76$ to
$0.72$ to achieve the best fit to KiDS-1000 data that averages out to
\mbox{$\ave{f_{\delta,mn}}_{mn}\approx1$} over all $m$ and $n$ (the
next Sect. \ref{sect:devidence} provides more details). This
$\sigma_8$ reduction is presumably required because the
$P_\delta(k,z)$ model in \citet{2021A&A...646A.140H} includes neutrino
suppression and baryonic feedback, missing here, and is, even without
neutrinos and baryons, already systematically lower by roughly
$5\%$--$10\%$ within $k\approx0.1$--$10\,h\,\rm Mpc^{-1}$ compared to
\texttt{halofit}. We refer to Section 4 in \cite{2015MNRAS.454.1958M}
for a discussion on the latter aspect, and reiterate that our
$P_\delta^{\rm fid}(k,z)$ choice is arbitrary -- in our analysis, a
purely $\Lambda$CDM model that matches the KiDS-1000 data. The results
for $f_\delta(k,z)$ in the following sections consequently probe the
deviations of the true $P_\delta(k,z)$ relative to the best-fitting
$\Lambda$CDM reference.

\begin{figure}
  \begin{center}
    \includegraphics[width=75mm]{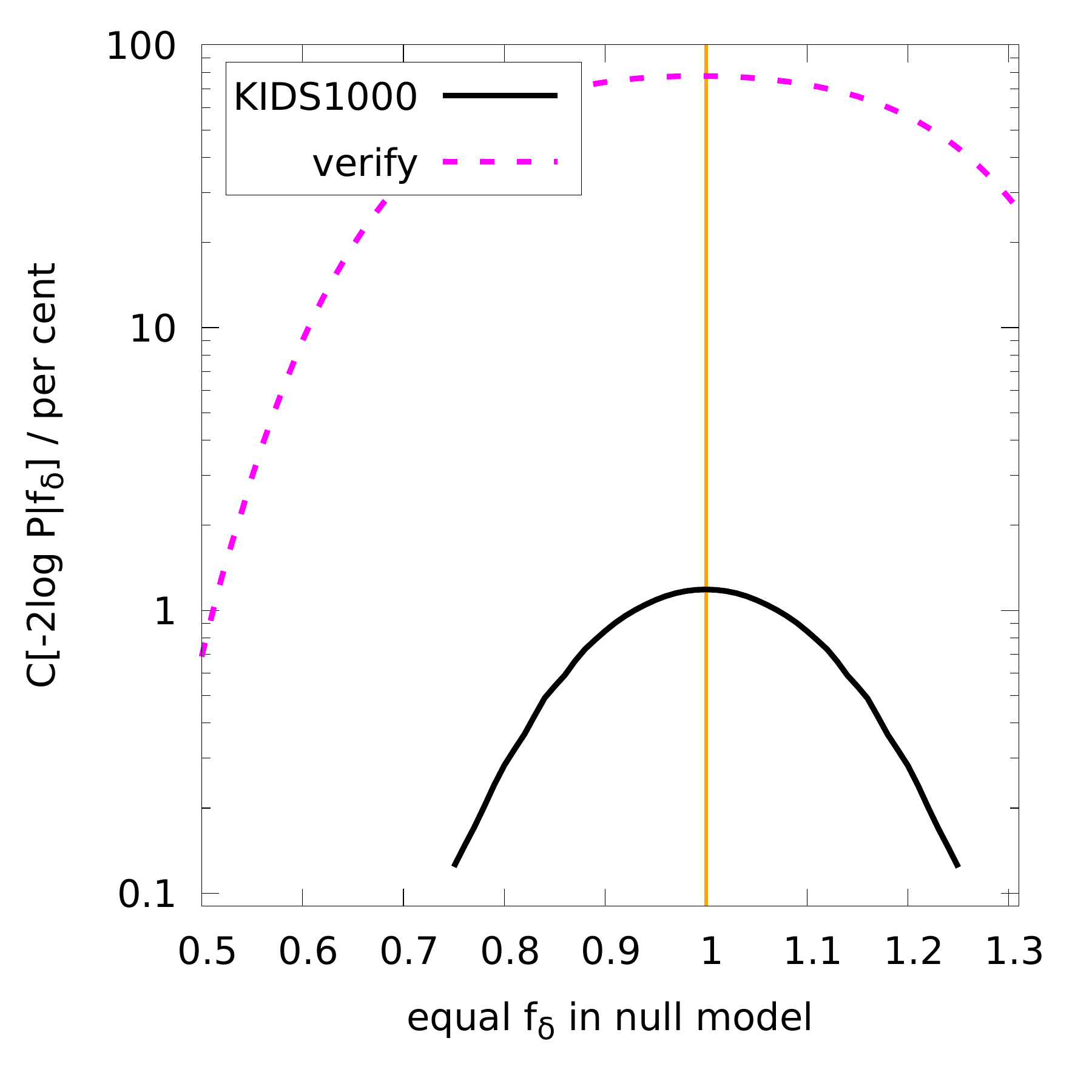}
    \caption{\label{fig:nulltest} Test of data against a null model
      with identical \mbox{$f_{\delta,mn}=\bar{f}_0$} for all $m$ and
      $n$ ($N_z=3$). Shown on the $y$-axis is the probability of
      $-2\ln{P(f_{\delta,mn}=\bar{f}_0|\vec{d})}$ being greater than
      in the null model. The dotted line is the result for one noisy
      verification data vector that has \mbox{$\bar{f}_0=1$}; the
      solid line is for the KiDS-1000 data. The set-up in Table
      \ref{tab:fiducialmodel} and the source distributions in Fig.
      \ref{fig:pofz} without errors in the projection parameters are
      assumed.}
  \end{center}
\end{figure}

\subsection{Evidence for deviation from reference}
\label{sect:devidence}

A hypothesis test against null models with constant
\mbox{$f_{\delta,mn}=\bar{f}_0$}, for all $m$ and $n$, and fixed
projection parameters $\vec{q}$ shows that the KiDS-1000 data support
deviations from the reference power spectrum in either $k$ or $z$,
shown by the results in Fig. \ref{fig:nulltest}. For each $\bar{f}_0$
on the $x$-axis in this figure, $10^5$ noise realisations of
$\vec{d}_{\rm null}$ were used in the null model to predict the
cumulative distribution $C[t_{\rm null}]$ of the test statistic
\mbox{$t_{\rm null}:=-2\ln{P(f_{\delta,mn}=\bar{f}_0|\vec{d}_{\rm
      null})}$}. The $y$-axis shows the probability,
\mbox{$p:=C[t_d]$}, that the null model exceeds the
\mbox{$t_d:=-2\ln{P(f_{\delta,mn}=\bar{f}_0|\vec{d})}$} in the actual
data (solid line) or, for verification, in random null-model data
(dashed line). Starting with the reference power spectrum
(\mbox{$\bar{f}_0=1$}), as indicated by the yellow vertical line, the
KiDS-1000 data have \mbox{$p=0.012$} and the verification data
\mbox{$p=0.78$}. Therefore, while the verification data are well
within the expectation of the null hypothesis, the KiDS-1000 data are
inconsistent with reference model on a \mbox{$1-p=98.8\%$} confidence
level. Varying $\bar{f}_0$, however, would not improve the test result
because the KiDS-1000 curve already peaks at
\mbox{$\bar{f}_0\approx1$}: the reference power spectrum,
$P_\delta^{\rm fid}(k,z)$, is the best fit out of a family of models
with constant $f_{\delta,mn}$. The only way to improve the fit would
be to make $f_{\delta,mn}$ varying with either scale, $k$ (this means
$m$), or redshift, $z$ (this means $n$).

In fact, the peak at \mbox{$\bar{f}_0\approx1$} was our deliberate
choice for KiDS-1000, achieved by lowering $\sigma_8$ for
$P^{\rm fid}_\delta(k,z)$ to $0.72$ compared to the best-fitting value
of $0.78$ in \cite{2021A&A...646A.140H} in order to move the peak from
\mbox{$\bar{f}_0\approx0.9$} to its final location
\mbox{$\bar{f}_0\approx1$}. This way, the $f_{\delta,mn}$ in our data
are normalised to the average $\ave{f_{\delta,mn}}_{mn}\approx1$. The
failed null test still stands, however, and is evidence for the
existence of a better fit by introducing either a $k$- or
$z$-dependence of $f_{\delta,mn}$, or both. This is explored in the
following two more complex scenarios.

\begin{figure*}
  \begin{center}
    \includegraphics[width=60mm]{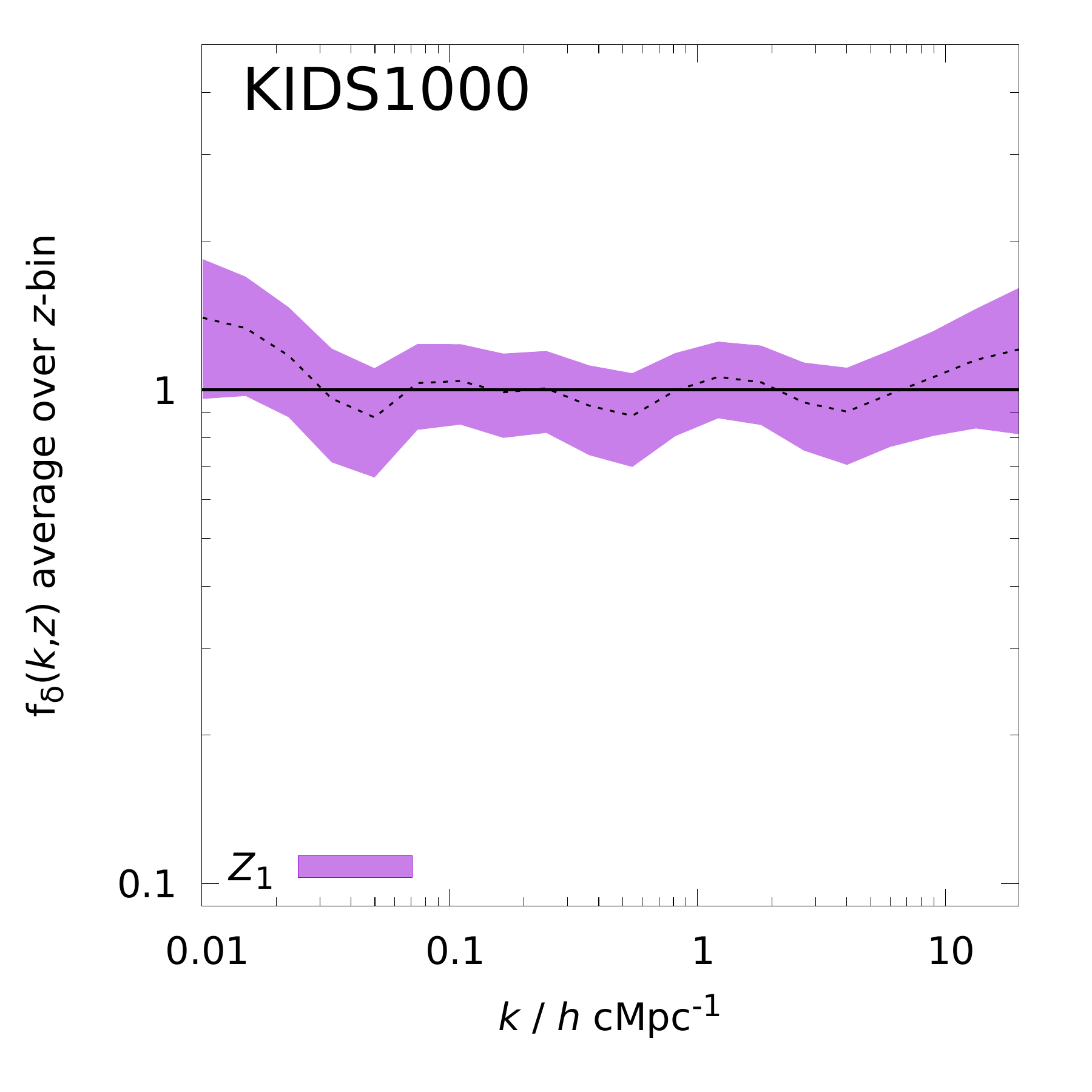}
    \includegraphics[width=60mm]{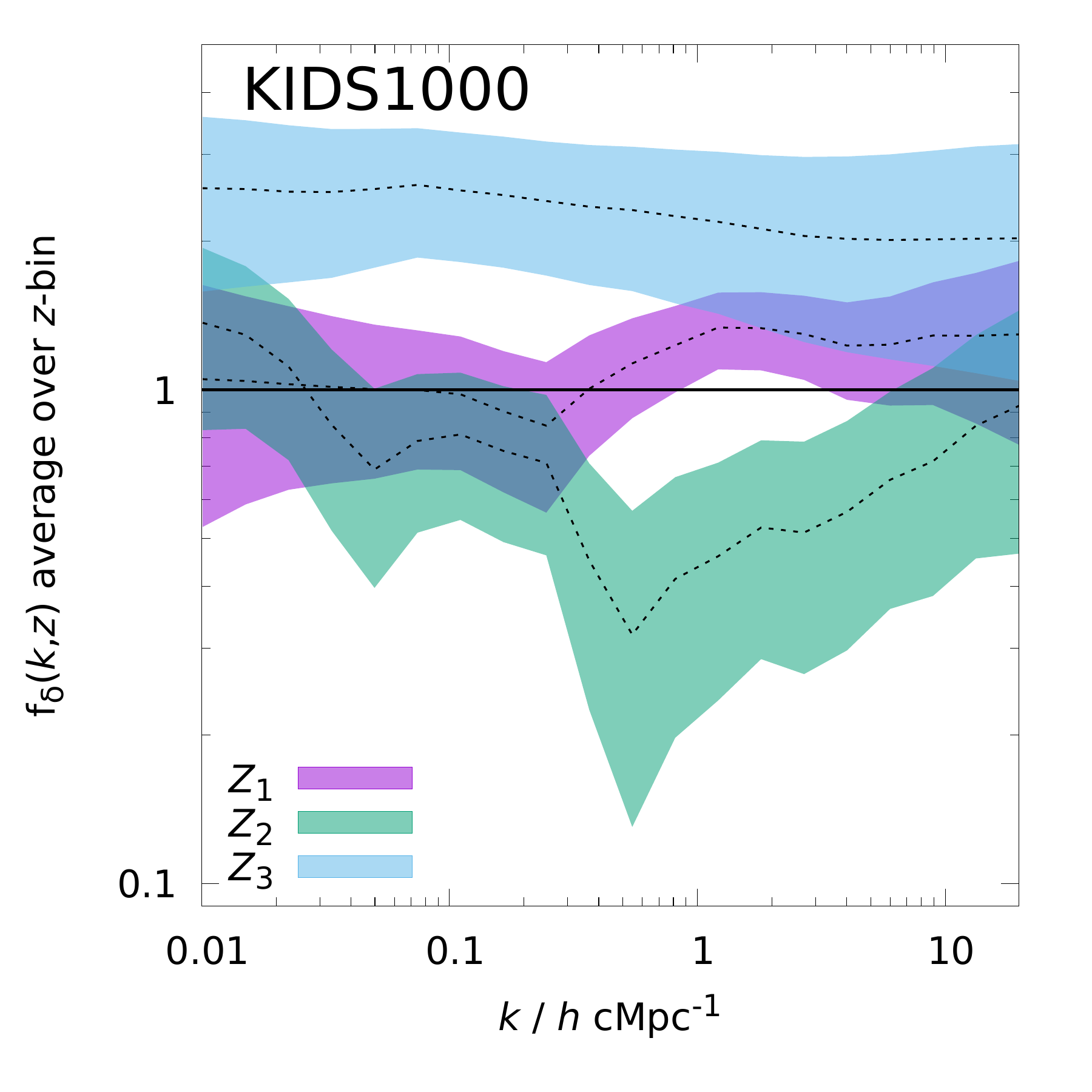}
    \includegraphics[width=60mm]{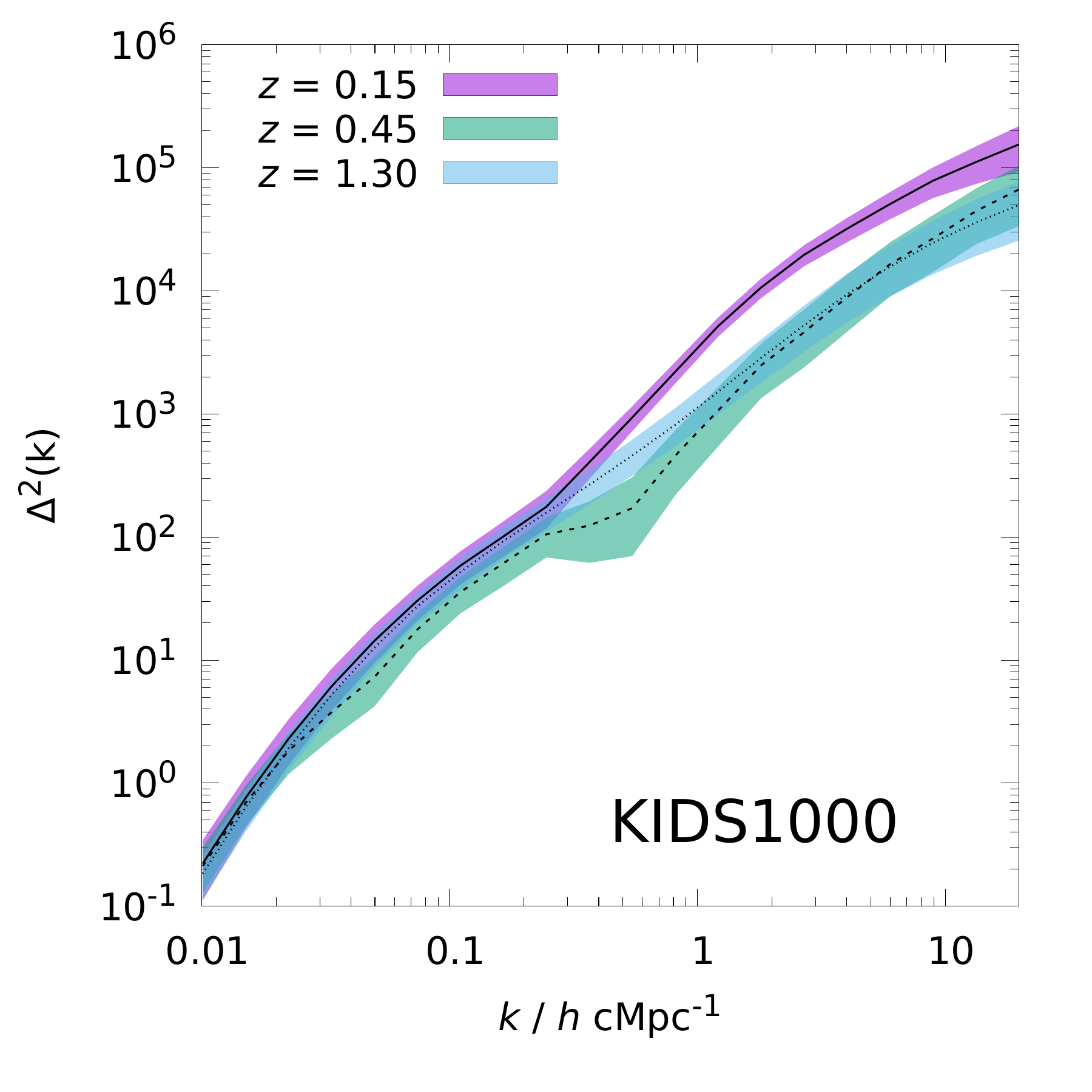}
    \caption{\label{fig:kidsrelpower} Reconstructed matter power
      spectrum in KiDS-1000 for $\tau=5.0$ and $N_k=20$ in three
      variants. The errors marginalise over uncertainties in the lensing
      kernel and IA. Shown are the posterior $68\%$ CI about the
      median (lines). \emph{Left panel:} Transfer function
      $f_\delta(k,z)$ averaged over the entire redshift range
      $Z_1=[0,2]$ and relative to the reference,
      $P_\delta^{\rm fid}(k,z)$, with parameters listed in Table
      \ref{tab:fiducialmodel}. \emph{Middle panel:} Same as the left panel,
      but  for three independent transfer functions averaged
      separately within the ranges $Z_1=[0,0.3]$, $Z_2=[0.3,0.6]$, and
      $Z_3=[0.6,2]$. \emph{Right panel:} Dimensionless power spectrum,
      $\Delta^2(k,z)=4\pi\,k^3P_\delta(k,z)=4\pi\,k^3\,f_\delta(k,z)\,P^{\rm
        fid}_\delta(k,z)$, interpolated to the centres of three
      redshift bins $Z_1$, $Z_2$, and $Z_3$.}
    \end{center}
\end{figure*}

\subsection{Average deviations over full redshift range}

The first scenario introduces a $k$-dependence for $f_{\delta,mn}$
without redshift evolution (\mbox{$N_z=1$}), reported with $68\%$ CI
about the median in the left panel of Fig. \ref{fig:kidsrelpower}. In
other words, the redshift evolution of $P_\delta(k,z)$ is assumed to
be that of the reference power spectrum. Towards non-linear scales
\mbox{$k\gtrsim0.1\,h\,\rm Mpc^{-1}$}, there is no indication for a
deviation from the reference exceeding the $68\%$ CI, in particular
the suppression of power does not fall below
\mbox{$f_\delta\approx0.8$}, nor do the data support a boost of power
beyond \mbox{$f_\delta\approx1.2$}. Towards linear scales, below
\mbox{$k\approx0.1\,h\,\rm Mpc^{-1}$}, is a slight, yet insignificant,
rise towards \mbox{$f_\delta\approx1.5$}. This might be a sign that
\texttt{halofit} with higher $\sigma_8$ provides a slightly better fit
on linear scales. Yet, on average, variations relative to
$P_\delta^{\rm fid}(k,z)$ stay within about $20\%$ ($68\%$ CI) down to
\mbox{$k\sim10\,h\,\rm Mpc^{-1}$}.

\subsection{Deviations for three separate redshift bins}
\label{sect:threezbins}

For more insight in a possible $z$-evolving $f_\delta$ in the second
scenario, we probed the average $f_\delta(k,z)$ inside three separate
redshift bins, \mbox{$Z_1=[0,0.3]$}, \mbox{$Z_2=[0.3,0.6]$}, and
\mbox{$Z_3=[0.6,2]$}. Their marginalised posterior constraints ($68\%$
CI about the median) are plotted as three credible regions in the
middle panel of Fig. \ref{fig:kidsrelpower}, one region for each
redshift bin.  The lowest bin, $Z_1$, is fully consistent with
\mbox{$f_\delta=1$}, on $k$-average \mbox{$f_\delta=1.15\pm0.28$},
while the middle bin, $Z_2$, on average \mbox{$f_\delta=0.57\pm0.27$},
indicates a power suppression, reaching the depth
\mbox{$f_\delta=0.3\pm0.2$} at \mbox{$k\sim0.5\,h\,\rm Mpc^{-1}$}, and
the highest bin, $Z_3$, overall prefers a boosted power of
\mbox{$f_\delta=2.22\pm0.81$}. The signal suppression in the middle
redshift bin and the boost in the highest bin cancel each other when
averaging $f_\delta$ over the whole redshift range, resulting in the
\mbox{$f_\delta=0.99\pm0.20$} in the left panel.

The close description of the $\xi^{(ij)}_\pm(\theta)$ data points by
the flexible $f_\delta$ model is illustrated by the posterior
predictive of the model in Fig. \ref{fig:kidspostpred}
(Appendix). Shown in the various panels are our KiDS-1000 data points
(black with $1\sigma$ errors bars) for $\theta\,\xi^{(ij)}_-(\theta)$
(lower left panels) and $\theta\,\xi^{(ij)}_+(\theta)$ (upper right
panels) in comparison to the posterior model constraints ($68\%$ and
$95\%$ CIs in dark and light blue).  The CIs in this plot do not
marginalise over $\vec{q}$; however, this has an effect of less than
$10\%$. The red line corresponds to the $\Lambda\rm CDM$ reference
model, this means \mbox{$f_{\delta,mn}\equiv1$}, obtained from a best
fit of $\sigma_8$ to the data points (Sect. \ref{sect:refpower}).
Allowing for deviations \mbox{$f_{\delta,mn}\ne1$} in the three
separate redshift bins $Z_1$ to $Z_3$ moves the CIs relative to the
reference, thereby improving the fit, although the reference still
stays within the $95\%$ CI, as, for instance, for
$\xi_\pm^{(55)}(\theta)$. Whilst an overall good match to the
(correlated) data points, a conflict with the model is probably
present for $\xi_+^{(22)}(\theta)$, and perhaps in
$\xi_+^{(12)}(\theta)$ to $\xi_+^{(15)}(\theta)$, where the data
prefer a higher amplitude than the model fitting all tomographic
bins. Figure \ref{fig:postpredVer} shows a similar plot with
verification data, based on the reference model, as comparison. A
higher value of
\mbox{$S_8:=\sqrt{\Omega_{\rm m}/0.3}\,\sigma_8\approx0.791$}, as
indicated by the green line for the $N$-body verification data
(without IA, Sect. \ref{sect:tks17}), would account for the higher
$\xi^{(1j)}_+$-amplitude but, on the other hand, is rejected by the
data due to mismatches at higher redshift (e.g., $\xi_+^{(35)}$ or
$\xi_+^{(45)}$). This might indicate spurious systematic errors in the
low-$z$ bins, discussed in Sect. \ref{sect:discussion}.

Expressed in terms of the actual power spectrum,
\mbox{$\Delta^2(k,\bar{z}):=4\pi\,k^3\,f_\delta(k,\bar{z})\,P_\delta^{\rm
    fid}(k,\bar{z})$}, interpolated to the $z$-bin centres, $\bar{z}$,
the diverse redshift evolution of $f_\delta(k,z)$ in the KiDS-1000
data translates into a suppression of non-linear structure growth
between \mbox{$0.3\lesssim z\lesssim2$}. Namely, the
\mbox{$\bar{z}=0.15$} matter power spectrum in the right panel of
Fig. \ref{fig:kidsrelpower} (magenta) has clearly more power than the
two higher redshift bins (green for \mbox{$\bar{z}=0.45$} and cyan for
\mbox{$\bar{z}=1.3$}) for \mbox{$k\gtrsim0.1\,h\,\rm Mpc^{-1}$}, while
the two other power spectra are consistent with each other, maybe with
the green rising over cyan near \mbox{$k=10\,h\,\rm
  Mpc^{-1}$}. Consequently, with cyan and green being moved closer to
each other, there is for \mbox{$k\gtrsim0.1\,h\,\rm Mpc^{-1}$} less
structure growth detected between $\bar{z}=0.45$--$1.3$ than in the
reference $\Lambda$CDM cosmology but, conversely, more between
$\bar{z}=0.15$--$0.45$, pulling back the low-$z$ bin to the reference
power spectrum.

\begin{figure}
  \begin{center}
    \includegraphics[width=65mm]{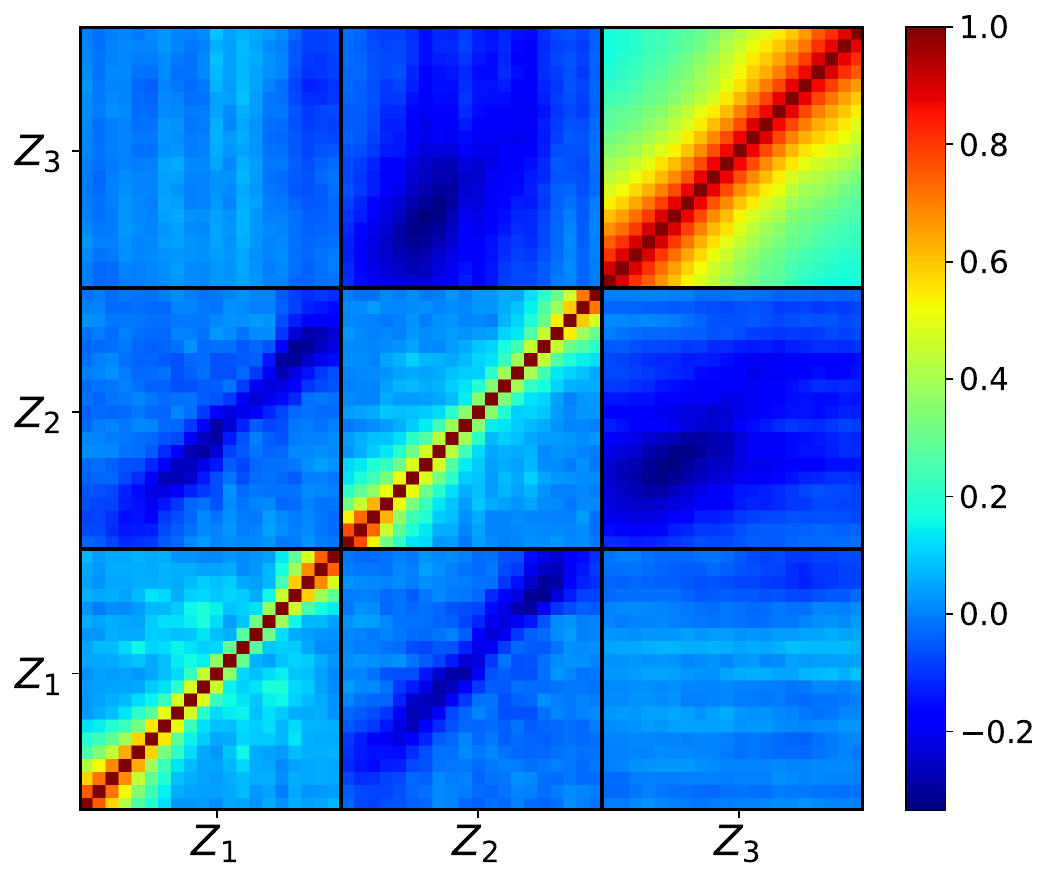}
  \end{center}
  \caption{\label{fig:corrmatrix} Correlation matrix of
    $f_{\delta,mn}$ errors for the KiDS-1000 analysis with three
    redshift bins, corresponding to the middle and right panels in
    Fig. \ref{fig:kidsrelpower}. The three blocks on the diagonal from
    bottom left to top right are the correlation matrices for the
    $z$-bins \mbox{$Z_1=[0,0.3]$}, \mbox{$Z_2=[0.3,0.6]$},
    \mbox{$Z_3=[0.6,2]$}, respectively; off-diagonal blocks show
    correlations between distinct $z$-bins; pixels within the
    $20\times20$ blocks are correlations between the $k$-bins,
    ascending from left to right inside a block.}
\end{figure}

It is noteworthy that the statistical errors of $f_{\delta,mn}$ and
hence $\Delta^2(k,z)$ are correlated, best visualised by the
correlation matrix in Fig. \ref{fig:corrmatrix}. Each pixel in the
correlation matrix encodes the Pearson correlation coefficient, $r$,
between two $k$-bins, either from same or from different $z$-bins. The
matrix is organised in $N_k\times N_k$ blocks, where $k$ increases
from left to right and top to bottom. The three blocks on the
diagonal, bottom left to top right (lowest $z$ to highest $z$), are
correlations within the same redshift bin, and the six off-diagonal
blocks are cross-correlations between distinct $z$-bins. The
enhancement of correlations within the blocks, making them stick out
as tiles in a $3\times3$ mosaic, is a side-effect of the Tikhonov
regularisation smoothing the $f_{\delta,mn}$ along the
$k$-axis. Correlations are high for adjacent $k$-bins in the same
$z$-bin, $r\approx0.4$--$0.8$, especially in the highest $z$-bin (top
right), but quickly drop off with distance on the $k$-scale. Adjacent
$k$-bins are also statistically dependent when from adjacent $z$-bins,
albeit now anti-correlated with typically \mbox{$r\approx-0.4$} to
$-0.1$. Not shown is the correlation matrix for a \mbox{$N_z=1$}
reconstruction, left panel in Fig. \ref{fig:kidsrelpower}, to save
space; it is essentially a diagonal matrix where adjacent $k$-bins
have \mbox{$|r|\lesssim0.1$}, except at the lower and upper boundary
of the $k$-range where \mbox{$r\approx0.5$--$0.6$}.

\subsection{Verification test with ray-traced mock data}
\label{sect:tks17}

To add more realism to the code verification than in Appendix
\ref{ap:mcmc}, especially with respect to the $f_\delta$ result for
\mbox{$N_z=3$}, we performed a more stringent test with $N$-body
ray-traced data \citep{2017ApJ...850...24T}. As in
\cite{2024A&A...683A.103B}, we subdivided the $108$ full-sky shear
fields from the simulated data according to the KiDS-1000 sky area
and, for each tomographic bin, combined $30$ shear grids for ascending
source redshifts according to the redshift distributions in
Fig. \ref{fig:pofz}. This yields an ensemble of \mbox{$n=1944$} mock
KiDS-1000 surveys. From these, source positions were picked to match
the angular positions, shear weights, and shape noise to the measured
counterparts in the public KiDS-1000 ellipticity catalogue
\citep{2021A&A...645A.105G}. In contrast to Appendix \ref{ap:mcmc},
however, intrinsic alignments of sources are not included here, thus
\mbox{$A_\sfont{IA}=0$}. Similar to the actual KiDS-1000 data vector,
we used \texttt{TreeCorr} to compute $\xi^{(ij)}_\pm(\theta)$ from the
mock shear catalogues.

\begin{figure}
  \begin{center}
    \includegraphics[width=75mm]{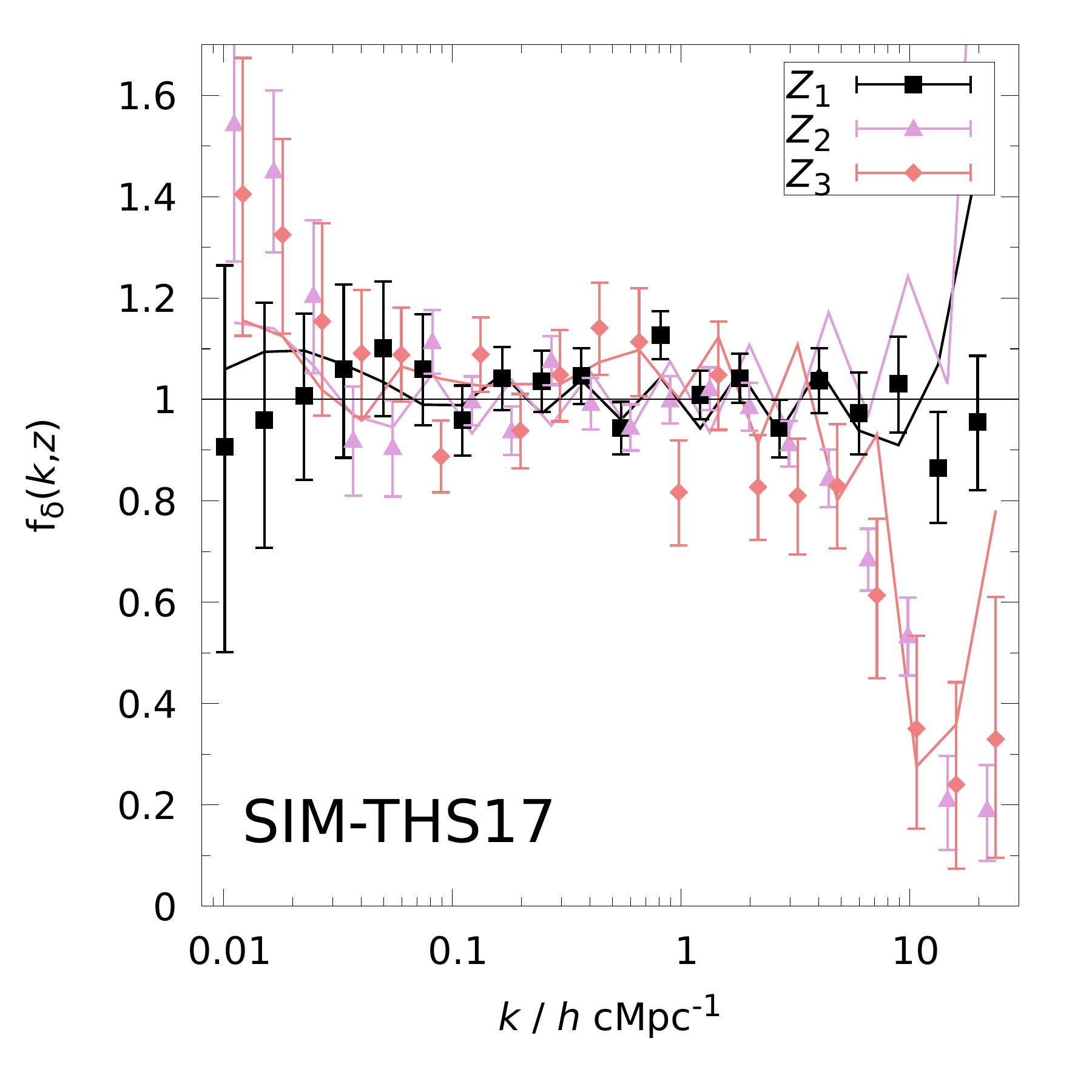} \\
    \includegraphics[width=75mm]{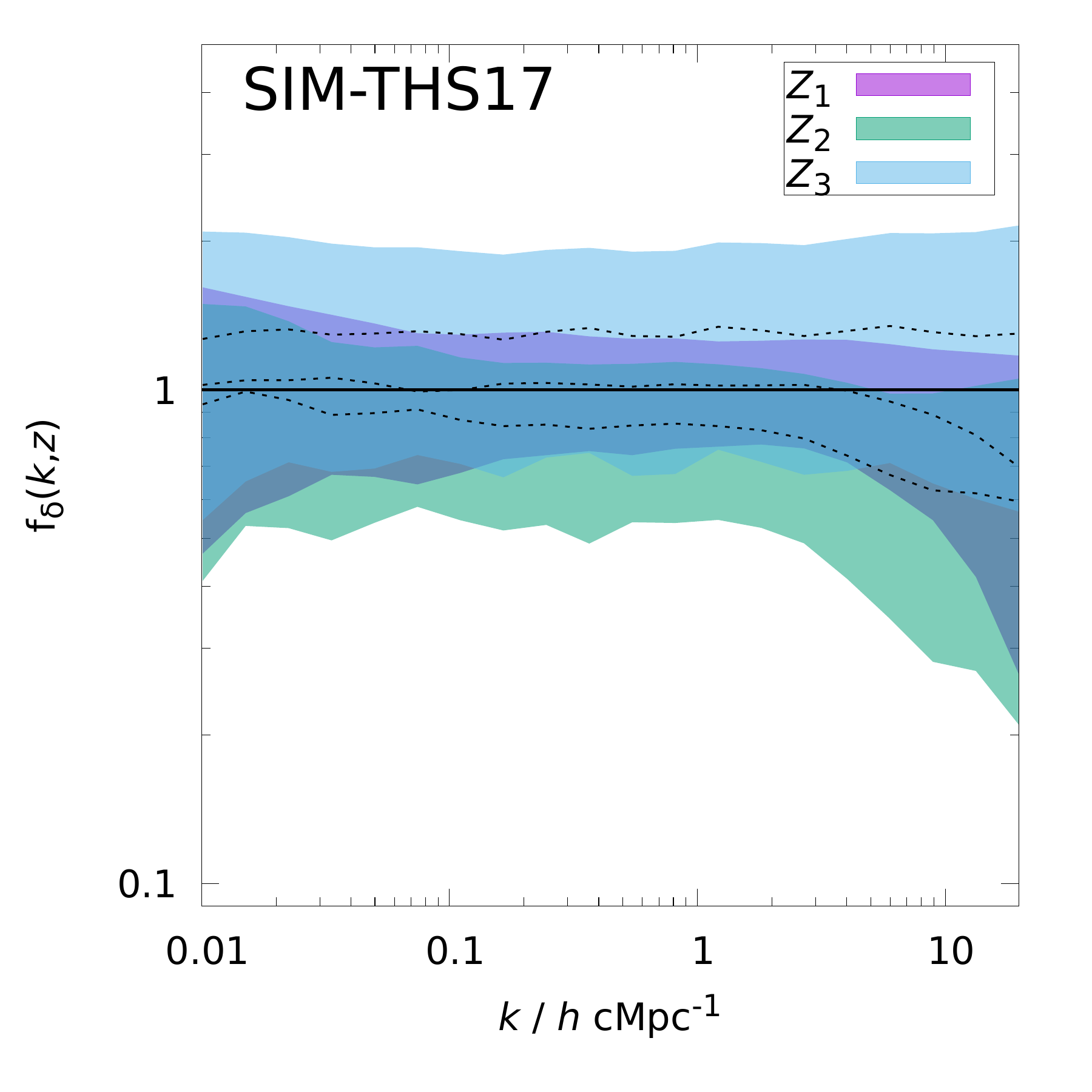}
    \caption{\label{fig:ths17relpower} Verification test of the
      analysis code for $f_\delta(k,z)$ using: an $N$-body simulated
      data vector, averaged over \mbox{$n=1944$} realisations of the
      KiDS-1000 data without IA; the code and data set-up for the
      KiDS-1000 analysis has $N_k=20$ $k$-bins and three redshift bins
      $Z_1=[0,0.3]$, $Z_2=[0.3,0.6]$, and $Z_3=[0.6,2]$.  The
      reference power, $P^{\rm fid}_\delta(k,z)$, is here identical to
      the cosmology of the simulation \citep{2017ApJ...850...24T}.
      The errors ($68\%$ CI) do not marginalise over a projection
      kernel uncertainty. \emph{Top panel:} Simulated
      covariance of the average data, comprising $n$ times the area of
      KiDS-1000.  Deviations from the expected
      \mbox{$f_\delta(k,z)=1$} towards higher $z$ and
      $k\gtrsim1\,h\,\rm Mpc^{-1}$ are related to the finite grid
      pixel size in the simulation (pixelation bias). The lines are
      additional reconstructions using noise-free data, based on
      Eq. \Ref{eq:xipm_pure} and \texttt{halofit} without pixelation
      bias. \emph{Bottom panel:} Same as top panel, but with statistical errors as in a
      single KiDS-1000 survey.}
    \end{center}
\end{figure}

Driven first by the question whether our code can recover
\mbox{$f_\delta(k,z)=1$} from a simulated (unrealistically) high S/N
data vector when using exactly the same cosmological parameters as in
the simulation, we averaged all $n$ realisations of the data vector,
denoted by $\bar{\vec{d}}^{\rm sim}$, and used the individual
realisations, $\vec{d}_i^{\rm sim}$, to estimate the (standard) error
covariance of $\bar{\vec{d}}^{\rm sim}$,
\begin{equation}
  \mat{C}^{\rm sim}=\frac{1}{n(n-1)}
  \sum_{i=1}^n
  \left(\vec{d}^{\rm sim}_i-\bar{\vec{d}}^{\rm sim}\right)
  \left(\vec{d}^{\rm sim}_i-\bar{\vec{d}}^{\rm sim}\right)^{\rm T}\;.
\end{equation}
This covariance matrix was turned into an unbiased estimator of the
inverse covariance using \cite{2007A&A...464..399H}. The average
$\bar{\vec{d}}^{\rm sim}$ and $\mat{C}^{\rm sim}$ are basically a
single measurement in a hypothetical survey with $n$ times the angular
area than KiDS-1000, albeit underestimating the cosmic variance error.
For an initial inspection, the solid green lines SIM-THS17 in
Fig. \ref{fig:kidspostpred} (Appendix) compares
$\bar{\vec{d}}^{\rm sim}$ to the real KiDS-1000 data (black data
points with error bars): The simulated tomography correlations are
mostly within the $1\sigma$ noise scatter of the KiDS-1000 data
points, yet exhibit systematically more signal compared to the
reference model (red lines) by occasionally exceeding the $95\%$ CIs,
probably due to the higher fluctuation amplitude,
\mbox{$S_8\approx0.791$} in the simulation compared to
\mbox{$S_8\approx0.765$} in KiDS-1000, and, less relevantly, the
missing IA terms in the simulation. Now, using
$\bar{\vec{d}}^{\rm sim}$ and $\mat{C}^{\rm sim}$ as input, the data
points in the top panel of Fig. \ref{fig:ths17relpower} show the code
results for $f_\delta(k,z)$ in three redshift bins. To test the code's
ability to recover the power spectrum in the simulation,
\mbox{$f_\delta(k,z)=1$}, the reference $P^{\rm fid}_\delta(k,z)$ was
set here to the same cosmological parameters as in the $N$-body data,
namely $\Omega_{\rm m}=0.279$, $\Omega_\Lambda=1-\Omega_{\rm m}$,
$\Omega_{\rm b}=0.046$, $h=0.7$, $n_{\rm s}=0.97$, and
$\sigma_8=0.82$. Since the lensing kernel is exactly known in this
experiment, no marginalisation of $\vec{q}$ was done. We find the
following.

The reconstructed power spectrum in Fig. \ref{fig:ths17relpower}, top
panel, for the lowest redshift bin (filled squares) is overall
consistent with \mbox{$f_{\delta,mn}=1$} for the range
\mbox{$k=0.01\mbox{--}20\,h\,\rm Mpc^{-1}$} within the $68\%$ CI,
indicated by the error bars. But two significant deviations emerge:
first, in the higher redshift bins beyond
\mbox{$k\approx 5\,h\,\rm Mpc^{-1}$}, and, second, smaller deviations
around $0.01\,h\,\rm Mpc^{-1}$. The high-$k$ suppression of the signal
is a real feature in the simulation data due to the pixelated shear
grids of the simulation (pixel size $0\farcm4$), significantly biasing
low the values for $\xi_-^{(ij)}$ for
\mbox{$\theta\lesssim\ang{;5;}$}. The lowest redshift bin is
presumably also affected, but scales of a few arcmins correspond here
to a characteristic $k$ too high to be visible in the plot: the
typical wave number,
\mbox{$k_{\rm eff}=2\pi/(f_\sfont{K}(\chi_{\rm
    d})\,\theta_{1^\prime})$}, of an angular scale
$\theta_{1^\prime}=1^\prime$ on a lens plane at distance
$\chi_{\rm d}$ corresponds to
$k_{\rm eff}\approx33.5,12.0,5.2\,h\,\rm Mpc^{-1}$ for lens planes
located at the centres of the bins $Z_1$, $Z_2$, and $Z_3$,
respectively. The additional lines in the plot support the idea of a
pixelation bias: they are reconstructions of noise-free data, based on
the simple Limber-Kaiser projection in Eq. \Ref{eq:xipm_pure} --
crucially, not showing the systematic signal drop for $Z_2$, or for
$Z_3$ around $k=1\,h\,\rm Mpc^{-1}$. Instead, however, the lines
exhibit artefacts as zig-zag oscillations of amplitude
\mbox{$\Delta f_\delta\approx0.1$}, strongly increasing towards the
edges of the plotted $k$-range and probably related to the limitations
of the deprojection method (Sect. \ref{sect:statmodel}). The artefacts
are partly mirrored in the data point scatter and overlap with the
pixelation bias for $Z_3$ (yet not for $Z_2$) at large $k$.

The second, weaker \mbox{$\sim1.5\sigma$--$2\sigma$} deviations appear
around \mbox{$k=0.01\,h\,\rm Mpc^{-1}$} for $Z_2$ and $Z_3$.  They,
too, are probably to some extend present in the $N$-body data: Figure
19 in \cite{2017ApJ...850...24T} report a $10\%$ deviation from
\texttt{halofit} near \mbox{$k=0.01\,h\,\rm Mpc^{-1}$}. Nevertheless,
we doubt that this explains the full effect and point out that the
plots compare an average data vector in a simulation of finite volume,
still subject to noise, to the cosmic average. Especially on large
scales, cosmic variance noise is relevant and under-estimated in
$\mat{C}^{\rm sim}$. In addition, the noise-free data (solid lines)
suggest some excess signal on large scales due to edge artefacts. In
summary, the bottom panel of Fig. \ref{fig:ths17relpower} is an
excellent reconstruction of the relative power with
\mbox{$\Delta f_\delta\approx0.1$} accuracy, or better, for
\mbox{$k\lesssim10\,h\,\rm Mpc^{-1}$} and $z\lesssim1$.

Another inference of $f_\delta(k,z)$ with the simulated data is shown
in the bottom panel of Fig. \ref{fig:ths17relpower} -- this time for a
KiDS-1000-like error covariance, $n\times\mat{C}^{\rm sim}$. By using
the average $\bar{\vec{d}}^{\rm sim}$ as input, the (essentially
linear) reconstruction in the figure represents the ensemble mean of
$f_\delta(k,z)$ posteriors in three redshift bins. In this average
reconstruction, the posterior errors are similar to the ones in the
KiDS-1000 reconstruction, middle panel of Fig. \ref{fig:kidsrelpower},
increasing in size for higher $z$, although now the $f_{\delta,mn}$
are consistent with \mbox{$f_\delta=1$} throughout ($68\%$ CI). The
suppression of power by pixelation bias is here fully contained within
the statistical errors, only vaguely hinted at in the green $Z_2$
posterior region of Fig. \ref{fig:ths17relpower}, bottom panel, which
falls just below \mbox{$f_\delta=1$} around
\mbox{$k=10\,h\,\rm Mpc^{-1}$}. Therefore, also for KiDS-1000-like
noise levels, the code recovers the correct power spectrum in the
$N$-body data. The highest precision of \mbox{$f_\delta=1.00\pm0.27$}
is achieved for $Z_1$, the middle bin $Z_2$ has
\mbox{$f_\delta=0.81\pm0.31$}, and $Z_3$ \mbox{$f_\delta=1.29\pm0.63$}
(medians and $68\%$ CIs, no $\vec{q}$ errors).

For comparison, Fig. \ref{fig:ths17relpower3} (Appendix) shows an
ensemble of reconstructions with individual, $16$ randomly chosen
$\vec{d}^{\rm sim}_i$, illustrating the possible variations in the
inferred $f_\delta(k,z)$. Each panel is a prediction of a KiDS-1000
reconstruction in a $\Lambda\rm CDM$ scenario. Here, we observe a
small tendency of the posterior median in $Z_3$ to be above
\mbox{$f_\delta=1$} and the median in $Z_2$ to be below
\mbox{$f_\delta=1$}, despite both having \mbox{$f_\delta\approx1$} for
\mbox{$k\lesssim5\,h\,\rm Mpc^{-1}$}. This tendency is also visible in
Fig. \ref{fig:ths17relpower} of average reconstructions, bottom panel,
with the posterior median always being above \mbox{$f_\delta=1$} for
$Z_3$, or below for $Z_2$, and probably has two reasons.  First, the
$Z_3$ bin is poorly constrained, giving a broad marginalised posterior
PDF skewed towards higher $f_\delta$ values, shifting up the centres
(posterior medians) of our CIs; this is illustrated by
Fig. \ref{fig:posteriorShape} in the Appendix. Second, errors in $Z_2$
and $Z_3$ are anti-correlated, preferring a lower value for $Z_2$, if
the $f_\delta$ at similar $k$ in $Z_3$ is scattered upwards. But, on a
broad note, one should not expect the maximum of the marginalised PDF
to coincide with the maximum of the full posterior (for the
non-marginalised parameters) because this requires that the full
posterior PDF must obey certain symmetries about its maximum, as, for
instance, present in a multivariate Gaussian posterior -- excluded
here because of our skewed, non-Gaussian marginalised posterior
distributions. For increased S/N in the data, however, the long tails
of skewed posteriors are suppressed, and the median shift is no longer
discernible in the top panel of Fig. \ref{fig:ths17relpower}.

\begin{figure}
  \begin{center}
    \includegraphics[width=90mm]{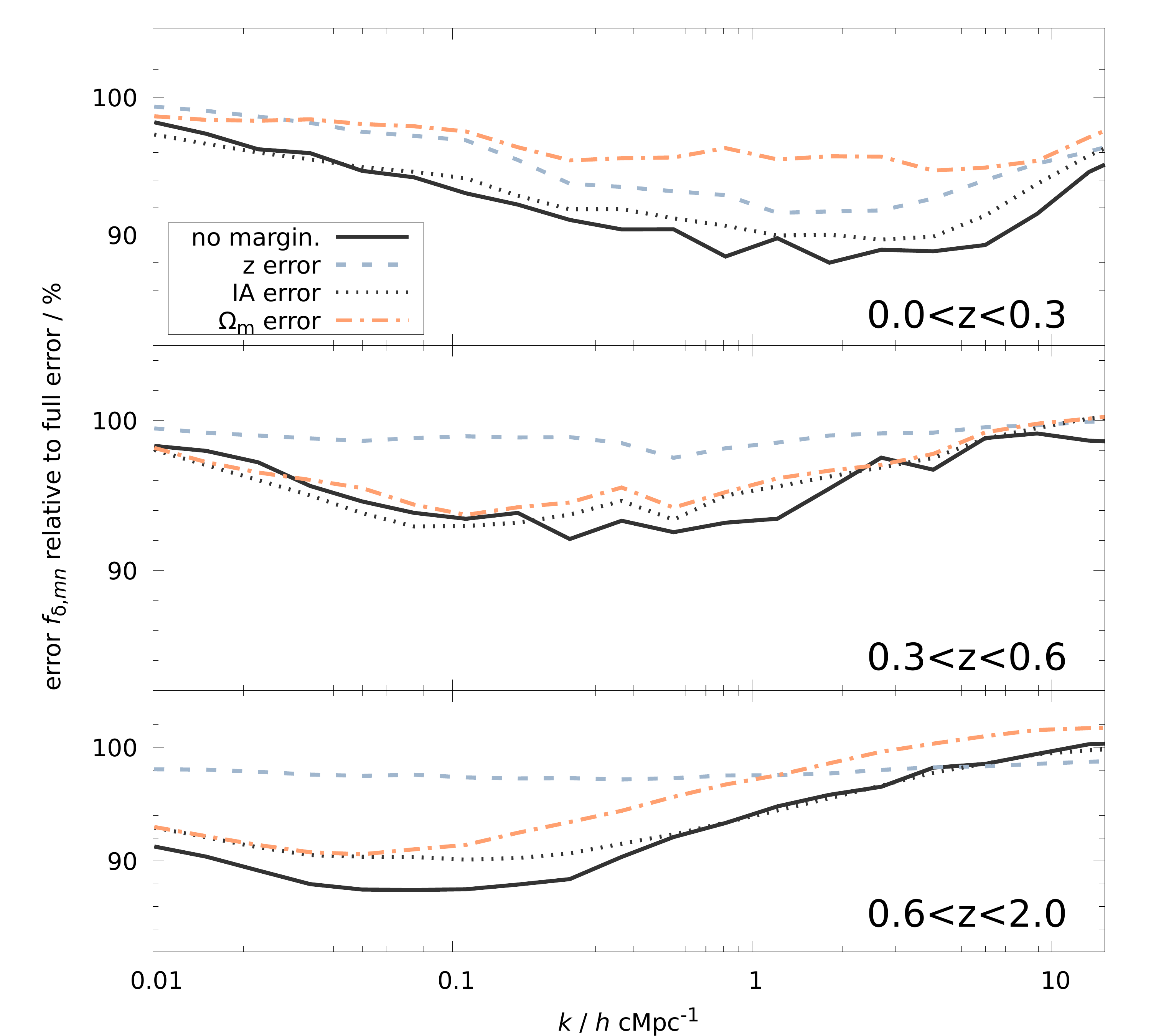}
    \caption{\label{fig:marginerror} Percentage fraction in total
      marginal posterior error of $f_{\delta,mn}$ due to uncertainties
      in the lensing kernel and IA parameter. Shown is the statistical
      error relative to the full marginal error (RMS variance of
      posterior) as a function of $k$ in three redshift bins without
      marginalisation (solid lines), with marginalisation over errors
      in the source redshift distributions only (dashed lines),
      $\Omega_{\rm m}$ in the lensing kernel only (dash-dotted
      lines), and for the $A_\sfont{IA}$ error only (dotted lines).}
  \end{center}
\end{figure}

\subsection{Impact of uncertainty in IA and lensing kernel}

The individual impact by uncertainties in the projection parameters
$\vec{q}$ on the reconstructed $f_{\delta,mn}$ is of interest for
improving future analyses or to gauge the relevance of errors in the
projection kernel. We therefore compare the marginalised
$f_{\delta,mn}$ errors in the middle panel of
Fig. \ref{fig:kidsrelpower} to the statistical errors obtained without
marginalisation, or to the errors where all parameters $\vec{q}$ are
fixed but for one that is marginalised over.

Figure \ref{fig:marginerror} shows the results in these different
scenarios as function of scale $k$ for three $z$-bins, indicated in
the lower right of each panel; a $100\%$ error is as large as in a
scenario where all $\vec{q}$ uncertainties are marginalised over. The
solid black lines use a fixed $\vec{q}$ without any marginalisation,
setting a lower limit for all scenarios (but is also subject to
numerical noise owing to the limited number of realisations,
\mbox{$n_{\rm merge}=500$}). A lower limit of $90\%$ to the black line
for all $k$ means that uncertainties due to source shape noise, cosmic
variance, and calibrated shear bias, all accounted for in the noise
covariance $\mat{C}$, are responsible for at least $90\%$ of the total
error in the reconstruction, while the addition of lensing kernel
uncertainties and IA errors amounts to $10\%$ or less of the
total. This addition is explored in more detail by the dashed lines
that marginalise over redshift errors only, the dotted-dashed lines
for only $\Omega_{\rm m}$ uncertainties in the lensing kernel, and the
dotted lines for $A_\sfont{IA}$ errors only.  In summary, IA errors
have a subdominant impact, about several per cent, reflected by the
closeness of the dotted line to the solid line. The largest impact of
the $\Omega_{\rm m}$ uncertainty is in the lowest $z$-bin, whereas the
uncertainty in the source $z$-distributions is the dominating
contributor in the middle and highest $z$-bin for
$k\lesssim1\,h\,\rm Mpc^{-1}$. Therefore, better calibrating photo-$z$
errors in future analyses yields improvements of up to $10\%$ in the
total statistical error.

\section{Discussion}
\label{sect:discussion}

In this study we presented, verified, and applied to KiDS-1000 data a
revised method for reconstructing the three-dimensional (band) matter
power spectrum with tomographic shear data. Our approach relies on the
specification of a projection kernel only, and is hence agnostic about
an analytical model for $P_\delta(k,z)$, while being applicable for
all cosmologies where the $\xi^{(ij)}_\pm(\theta)$ are projections of
$P_\delta(k,z)$ along the line of sight. Numerically convenient, we
express $P_\delta(k,z)=:f_\delta(k,z)\,P^{\rm fid}_\delta(k,z)$ in
terms of a reference and probe deviations,
\mbox{$f_{\delta,mn}\in[0,100]$}, relative to it, putting the free
variables $f_{\delta,mn}$ into the same dynamical range. Not required,
however, is a reference close to the true power spectrum; for
instance, in Fig. \ref{fig:tikhonov} we use a reference fixed in
cosmic time to highlight the structure growth, or, on the extreme end,
a constant \mbox{$P^{\rm fid}_\delta(k,z)\equiv P_0$} would yield a
$f_{\delta,mn}$ that directly represents the power spectrum. The
rescaling of $P^{\rm fid}_\delta(k,z)$ by a constant inside any
$z$-bin, whilst redefining $f_{\delta,mn}$, does not affect the
inferred
\mbox{$\Delta^2(k,z)\propto f_\delta(k,z)\,P^{\rm fid}_\delta(k,z)$};
the reconstructed matter power spectrum is robust against the
reference used. Choosing a reference towards a slowly evolving
$f_\delta(k,z)$ is nevertheless preferable because the $z$-bin-average
that the $f_{\delta,mn}$ represent depends on the specific noise
properties and source $z$-distributions in the tomographic survey; a
comparison to theoretical models will have to replicate the
data-specific $z$-weighting more closely for a significantly varying
$f_\delta(k,z)$ inside a wide $z$-bin. This is obvious in
Fig. \ref{fig:tikhonov} for a $z$-independent reference, where more
weight is given closer to the lower boundary of each $z$-bin: the
average $f_{\delta,mn}$ is close to the true $f_\delta(k,z)$ near
$z_{\rm max}=0.13$, $0.4$, $0.7$ for $Z_1$, $Z_2$, and $Z_3$; yet the
bin centres are at $z_{\rm mid}=0.15$, $0.45$, $1.30$.  Although a
best-fitting $\Lambda\rm CDM$ model without baryonic feedback or
neutrinos as reference is a good fit to the data already, as shown by
the red lines relative to the black data points in
Fig. \ref{fig:kidspostpred}, the $f_{\delta,mn}$ values indicate where
additional changes might be needed or how much deviation from the
reference is acceptable until we conflict with the data. We proceed
here with two variants of the analysis, one for an
$f_{\delta,mn}$-average over the full redshift range and one to probe
the $z$-evolution of $f_\delta(k,z)$.

\begin{figure}
  \begin{center}
    \includegraphics[width=85mm]{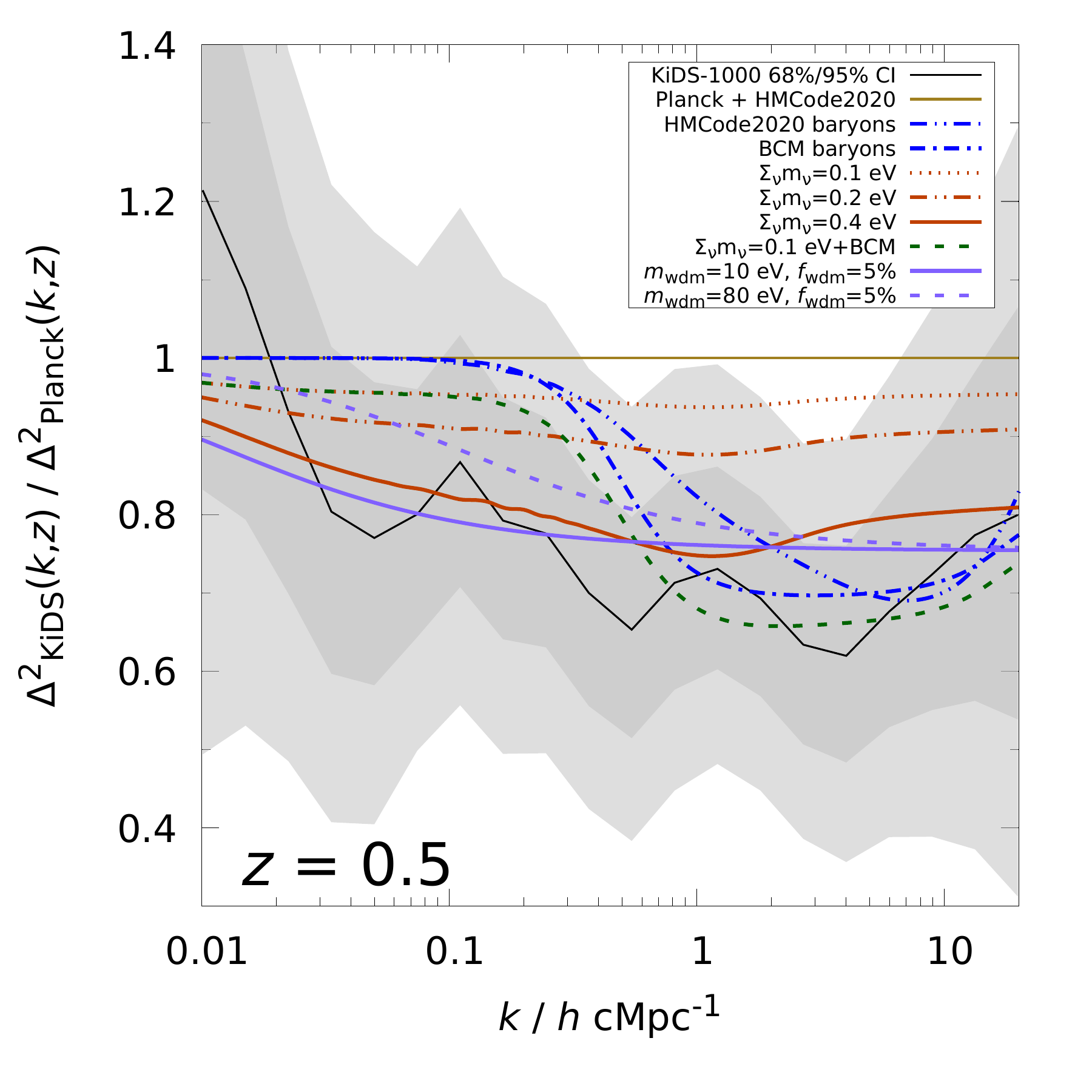}
    \caption{\label{fig:kids100vsplanck} Power spectrum ratio
      \mbox{$\Delta^2_{\rm KiDS}(k,z)/\Delta^2_{\rm Planck}(k,z)$} at
      \mbox{$z=0.5$}.  Shown in grey are the $68\%$ and $95\%$ CI from
      our analysis.  The KiDS-1000 spectrum $\Delta^2_{\rm KiDS}(k,z)$
      uses the $f_{\delta,mn}$ in Fig. \ref{fig:kidsrelpower} averaged
      within one broad redshift bin (left panel). The \emph{Planck}
      model for $\Delta^2_{\rm Planck}(k,z)$ is the \mbox{TT, TE,
        EE+lowE} best fit of a standard $\Lambda\rm CDM$
      \citep{2020A&A...641A...6P} extrapolated to \mbox{$z=0.5$} with
      \texttt{HMCode2020} by \cite{2021MNRAS.502.1401M}. The lines are
      model transfer functions to $\Delta^2_{\rm Planck}(k,z)$
      invoking flavours of extreme baryonic feedback
      (\texttt{HMCode2020}:
      \mbox{$\log_{10}{(T_{\rm AGN}/{\rm K})}=8.8$},
      \mbox{$A_{\rm baryon}=3.13$}, \mbox{$\eta_{\rm baryon}=0.603$};
      \texttt{BCM} by \citealt{2015JCAP...12..049S}:
      \mbox{$\log_{10}{(M_{\rm c}/h^{-1}\msol)}=15.0792$},
      \mbox{$\eta_{\rm b}=0.5$}, \mbox{$k_{\rm s}=55.0$}), neutrino
      suppression for cumulative neutrino masses $\sum_\nu m_\nu$
      (\texttt{BCM}), or a mixture of CDM and WDM
      (\citealt{2016PhRvD..94b3522K}, $m_{\rm wdm}$: WDM particle
      mass, $f_{\rm wdm}$: WDM mass fraction).}
\end{center}
\end{figure}

Our first variant in the left panel of Fig. \ref{fig:kidsrelpower}
averages $f_\delta(k,z)$, denoted $\bar{f}_\delta(k)$ hereafter, over
the entire $z$-range covered by KiDS-1000, with the most signal from
essentially \mbox{$z\lesssim1$}. This analysis shows that in the
non-linear regime, \mbox{$k\approx0.05$--$10\,h\,\rm Mpc^{-1}$}, the
KiDS-1000 matter power-spectrum is consistent with the
$\Lambda\rm CDM$ reference in Table \ref{tab:fiducialmodel} within
$20\%$ ($68\%$ CI). Therefore, there is no evidence for a variation,
suppression or boost, exceeding $20\%$ relative to the reference
scenario with cold dark matter (excluding neutrino suppression and
baryon feedback). This is also within the range reported by other
lensing studies, discussed below, although we caution that the results
are not always one-to-one comparable to ours because they are partly
based on the premise that the lensing $P_\delta(k,z)$ has the
\emph{Planck} $S_8$ \citep{2020A&A...641A...6P}, but is attenuated in
the non-linear regime to fit the lensing measurements of significantly
lower $S_8$ to explain the so-called $S_8$ tension. Such an
attenuation, if present, would be addressed in our reference by a
systematically lower $S_8$ instead of fully showing up in
$\bar{f}_\delta(k)$, especially because our data lack constraining
power in the unattenuated linear regime,
\mbox{$k\lesssim0.01\,h\,\rm Mpc^{-1}$}. To level the field with
analyses using the \emph{Planck} $S_8$ as the premise, we changed the
reference to a power spectrum with \emph{Planck} cosmology,
$\Delta^2_{\rm Planck}(k,z)$, and recast Fig. \ref{fig:kidsrelpower},
left panel, into Fig. \ref{fig:kids100vsplanck} for the ratio
$\Delta^2(k,\bar{z})/\Delta^2_{\rm Planck}(k,\bar{z})$ at
$\bar{z}=0.5$, a redshift where the lensing method is roughly the most
sensitive. Only from this \emph{Planck} perspective do our results now
clearly require a suppression, $1-f_{\delta,mn}$, of the $\Lambda$CDM
reference at \mbox{$k\approx3\,h\,\rm Mpc^{-1}$} of up to
$35\%\pm15\%\pm25\%$ ($68\%$ and $95\%$ CI). As indicated by the lines
in the plot, it is conceivable to explain the tension with
\emph{Planck} by a variety of hypothetical mechanisms involving, among
other things, extreme baryonic feedback, (sterile) neutrinos, or
alternatives to non-interacting cold dark matter.

Therefore, whereas a CDM power spectrum without suppression and low
$S_8$ is on average a good description of the KiDS-1000 shear
tomography, small modifications to CDM would be insufficient to
reconcile KiDS-1000 with \emph{Planck}: mechanisms with a substantial
suppression (more than $20\%$) have to be invoked, in agreement with
other recent studies analysing KiDS-1000 lensing data. For instance,
the recent \citet[S+22]{2022MNRAS.514.3802S} combines KiDS-1000 data
with observations of X-ray emission and the kinematic
Sunyaev--Zeldovich (kSZ) effect in galaxy clusters to inform baryonic
feedback parameters incorporated in their $\Lambda\rm CDM$-flavoured
$P_\delta(k,z)$. By initially using the KiDS-1000 data alone, the
deviations from their model without baryonic feedback are inconclusive
but remain, as in our measurement in the left panel of
Fig. \ref{fig:kidsrelpower}, within $10$--$20\%$ for
$k=1$--$10\,h\,\rm Mpc^{-1}$ ($68\%$ CI). Only when adding the X-ray
and kSZ data do their constraints strongly favour a power suppression
of up to $30\%$ ($68\%$ CI) at \mbox{$k\approx5\,h\,\rm Mpc^{-1}$},
somewhat more than our data support. Building only on lensing
information, \citet[PEG23]{2023MNRAS.525.5554P} analyse KiDS-1000 and
data from the third-year Dark Energy Survey (DES Y3) to extend a
$\Lambda\rm CDM$ model by one modification parameter, $A_{\rm mod}$,
pushing the non-linear $P_\delta(k,z)$, equivalent to
\mbox{$A_{\rm mod}=1$}, towards the linear power spectrum,
\mbox{$A_{\rm mod}=0$}. Their KiDS-1000 fit with uninformative priors
on cosmological parameters and $A_{\rm mod}\in[-5,5]$ is, as in
Fig. \ref{fig:kidsrelpower}, inconclusive,
\mbox{$A_{\rm mod}\approx0.9\pm0.2$} (their Figure 2, left
panel). However, when assuming a \emph{Planck} cosmology and the
suppression-inclined prior \mbox{$A_{\rm mod}\in[0.5,1]$}, PEG23 claim
strong evidence for a suppression of \mbox{$\approx25\%\pm6\%$} at
\mbox{$k\approx3\,h\,\rm Mpc^{-1}$} ($68\%$ CI), more significant yet
fully consistent with our Fig. \ref{fig:kids100vsplanck} (we reiterate
our relaxed prior \mbox{$f_{\delta,mn}\in[0,10^2]$}). Furthermore,
\citet[BK24]{2024arXiv240913404B} fit a double power law
\mbox{$P_\delta(k,z)\propto k^p\,(1+z)^{-m_\sfont{\rm BK}}$} to
$\xi^{(ij)}_\pm(\theta)$ for KiDS-1000, DES Y3, and the third year
Hyper Suprime-Cam Survey (HSC Y3), yielding a reasonable
$\Lambda\rm CDM$ match without suppression in the range
$k=0.01$--$5\,h\,\rm Mpc^{-1}$ ($68\%$ CI) and a $2\sigma$ agreement
among the three Stage III surveys. Compared to our results for
\mbox{$N_z=1$}, similarly averaging over the full redshift range,
their constraints are weaker in the non-linear regime
($f_\delta\approx0.8^{+0.5}_{-0.2}$), despite the same data, probably
because of our tighter $\Omega_{\rm m}$ prior for the projection
kernel: the $P_\delta(k,z)$ amplitude in $\xi_\pm^{(ij)}(\theta)$ is
degenerate with the pre-factor $\Omega^2_{\rm m}$ in the
Eqs. \Ref{eq:xipm_pure}--\Ref{eq:xipm_gi}. Finally, similar results
are reported by authors that use only DES Y3 lensing data although the
required suppression in the non-linear regime relative to
\emph{Planck} tends to be somewhat lower than for KiDS-1000 by a
factor between one and two (\citealt{2024MNRAS.534..655B}, PEG23,
\citealt{2023arXiv230911129F}. \citealt{2023A&A...678A.109A}). Utilising
lensing probes alone, \cite{2025arXiv250206687P} demonstrates
consistency with the \emph{Planck} cosmology on linear scales, yet
reveals a $20$--$25\%$ suppression in the non-linear regime, by
combining DES Y3 and CMB lensing data. In conclusion, other works
support our finding of an average matter power-spectrum that at
$k\gtrsim0.05\,h\,\rm Mpc^{-1}$ is either, within $20\%$ tolerance, a
purely $\Lambda\rm CDM$ spectrum of low \mbox{$S_8\approx0.73$}, or a
spectrum with both higher \emph{Planck} \mbox{$S_8\approx0.83$} and
significant suppression of up to $20\%$--$30\%$ in the non-linear
regime relative to CDM.

Surprisingly, our KiDS-1000 analysis becomes more complex and
discrepant with our best-fitting $\Lambda$CDM reference of low $S_8$
if the $f_{\delta,mn}$, unlike in previous studies, are free to vary
with redshift, achieved in our second analysis by increasing the
number of $z$-bins to \mbox{$N_z=3$} broad $z$-bins. This is only
practical by employing Tikhonov regularisation, or something else to
this effect, as clearly illustrated by the dramatically shrinking
sizes of the credible regions from \mbox{$\tau=0$} to \mbox{$\tau=5$}
in Fig. \ref{fig:tikhonov}. Applied to KiDS-1000, the $68\%$ CI
constraints in the middle panel of Fig. \ref{fig:kidsrelpower} then
depict a remarkably diverse picture: only the $f_{\delta,mn}$ in the
low-$z$ bin, $Z_1=[0,0.3]$, are now consistent with the reference
matter power spectrum, averaged for all $k$-bins to
\mbox{$\bar{f}_\delta=1.15\pm0.28$}, whereas the power spectrum in
$Z_2=[0.3,0.6]$ is overall suppressed,
\mbox{$\bar{f}_\delta=0.57\pm0.27$}, and boosted for $Z_3=[0.6,2]$,
where \mbox{$\bar{f}_\delta=2.22\pm0.81$}. The verification tests in
Appendix \ref{ap:mcmc}, for analytic data (noisy) vectors, and in
Sect.  \ref{sect:tks17}, for mock $N$-body shear catalogues, assure us
of an $f_{\delta,mn}$ accuracy of $10\%$ or better within
\mbox{$k\approx0.01$--$10\,h\,\rm Mpc^{-1}$}. The observed tendency of
shifting the median posterior $f_{\delta,mn}$ of our $68\%$ CIs
towards \mbox{$\bar{f}_\delta\approx1.3$} for $Z_3$ and towards
\mbox{$\bar{f}_\delta\approx0.8$--$0.9$} for $Z_2$, due to the
broadened skewed posterior PDFs in these $z$-bins, actually goes in
the direction of the KiDS-1000 result. It is nevertheless too small
for the pronounced split between $Z_2$ and $Z_3$ credible regions in
the real data, which also are, unlike the verification results,
blurred by the IA and lensing kernel uncertainties (an extra
$\sim10\%$ error, Fig. \ref{fig:marginerror}). For the baseline
\mbox{$f_\delta(k,z)\equiv1$} in Fig. \ref{fig:ths17relpower3}, the
split in the mock data realisations is not nearly as pronounced as for
KiDS-1000, making the KiDS-1000 result very roughly a
\mbox{$1/16\approx6\%$} event, or less, for the baseline scenario.

\begin{figure}
  \begin{center}
    \includegraphics[width=85mm]{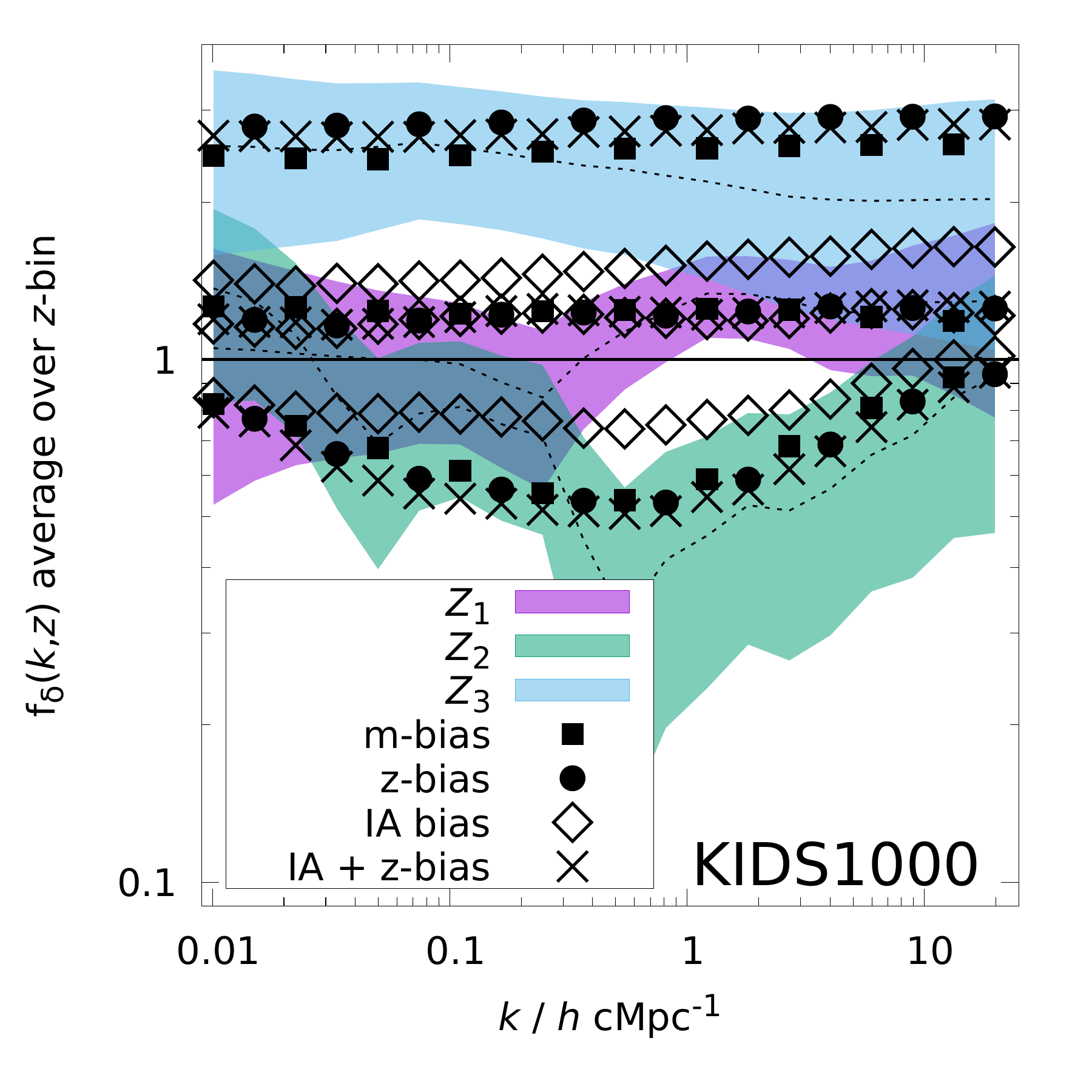}
    \caption{\label{fig:kids1000bias} KiDS-1000 results for
      $f_{\delta,mn}$, as in the middle panel of
      Fig. \ref{fig:kidsrelpower}, in relation to alternative
      hypotheses using the reference cosmology
      $P^{\rm fid}_\delta(k,z)$ and best-fitting bias parameters to
      match the KiDS-1000 data vector, $\xi^{(ij)}_\pm(\theta)$. The
      filled squares assume shear $m$-bias factors
      $\{m_i\}=\{+1.00,+0.48,-0.023,-0.028,+0.035\}$, the filled circles
      assume photo-$z$ biases
      $\{\delta^i_z\}=\{-0.075,-0.079,+0.009,+0.018,-0.052\}$, and
      the open diamonds employ the eNLA model with
      $\{A_\sfont{IA}^\prime,z_{\rm
        piv},\eta\}=\{0.3,0.19,1.5\}$. The crosses assume a hybrid model
      of eNLA and a smaller $z$-bias:
      $\{A_\sfont{IA}^\prime,z_{\rm piv},\eta\}=\{0.45,0.25,2.3\}$ and
      $\{\delta^i_z\}=\{-0.017,-0.058,+0.027,+0.025,-0.043\}$.  The
      data points indicate the posterior median (CIs are similar to
      those of the shaded regions); the filled data points use alternating
      $k$-bins to avoid overlap.}
  \end{center}
\end{figure}

To explore the possibility of spurious systematic errors possibly
mimicking the evolving $f_\delta(k,z)$, we estimated in three
scenarios the amplitude of bias needed to replicate our KiDS-1000
results if the $\Lambda\rm CDM$ reference were the true matter power
spectrum. First, a spurious shear $m$-bias in the data is assumed to
be the dominating source of systematic error. The required bias is
then obtained through fitting
$(1+m_i)\,(1+m_j)\,\xi^{(ij),\rm fid}_\pm(\theta)$, predicted by the
reference model, to the KiDS-1000 data vector by maximising the data
likelihood, Eq. \Ref{eq:likelihood}, with respect to the bias
parameters $m_i$ in the tomographic bins $i$. The corresponding
(posterior median) $f_{\delta,mn}$ of the best-fitting model are
plotted as data points with filled squares in
Fig. \ref{fig:kids1000bias} in comparison to KiDS-1000 (shaded
regions). Evidently, the $m$-bias model closely traces the KiDS-1000
results, but, according to its best-fit parameters
$\{m_i\}=\{+1.00,+0.48,-0.023,-0.028,+0.035\}$, it needs an
unrealistic $m$-bias in the first two source bins that exceeds the
reported values in \cite{2021A&A...645A.104A} by more than $30\sigma$.
The second scenario assumes spurious photo-$z$ errors in
$p_z^{(i)}(z)$ as dominating source of bias. Again, we fit a
$\xi^{(ij),\rm fid}_\pm(\theta)$ to the KiDS-1000 data vector, now for
the $p_z^{(i)}(z+\delta^i_z)$ shifted by the bias parameters
$\delta^i_z$, while the reconstruction uses the original
$p_z^{(i)}(z)$. The best-fit values are
$\{\delta^i_z\}=\{-0.075,-0.079,+0.009,+0.018,-0.052\}$. As in the
previous $m$-bias scenario, this model fits the KiDS-1000 results
equally well (filled circles in Fig. \ref{fig:kids1000bias}), but
requires a high bias (photo-$z$s are systematically low) in the first
two and the last source $z$-bins. This bias level exceeds the
estimated tolerances of photo-$z$ errors in the KiDS-1000 gold sample
by more than $5\sigma$ (Table \ref{tab:gold}). A third conceivable
source of systematic error is an IA model bias. Too simplistic
physically, the approximations in the NLA model are nevertheless
flexible enough to fit measurements of IA as a means for calibrating
the IA distortion of $\xi_\pm^{(ij)}(\theta)$
\citep{2021MNRAS.501.2983F, 2024arXiv240418240P}. It may be, however,
that $z$-binning of $\Delta^2(k,z)$ makes us sensitive to a
$z$-dependent IA amplitude, missing in the NLA model and typically
irrelevant in Stage III cosmological analyses. To test if a
$z$-dependent IA amplitude
$A_\sfont{IA}(z):=A_\sfont{IA}^\prime\,(1+z)^\eta\,(1+z_{\rm
  piv})^{-\eta}$, as in the eNLA model, potentially explains the
$f_{\delta,mn}$ data after all, we fit
$\xi^{(ij),\rm fid}_\pm(\theta)$ to KiDS-1000 by varying
$\{A_\sfont{IA}^\prime,z_{\rm piv},\eta\}$ only, yielding the
parameters $\{0.3,0.19,1.5\}$ for the open diamond data points in
Fig. \ref{fig:kids1000bias}. Crucially, the reconstruction algorithm
is ignorant about $A_\sfont{IA}(z)$ and assumes a constant
$A_\sfont{IA}$. Indeed, then some of the split between $Z_2$ and $Z_3$
(especially the drop in $Z_2$) could be explained by the IA model
bias. And, since \mbox{$z_{\rm piv}=0.3$} and \mbox{$\eta=1.5$} are
deemed consistent with KiDS-1000 uncertainties in
\cite{2021A&A...645A.104A}, this hypothesis is a plausible explanation
for at least part of the $\bar{f}_\delta$ redshift variation. More
plausible still is a combination of $A_\sfont{IA}(z)$ and a moderate
photo-$z$ bias, $2\sigma$--$4\sigma$, indicated by the crosses in
Fig. \ref{fig:kids1000bias}. Lastly, recalling the
$1\sigma$--$2\sigma$ evidence for $B$-modes in Sect. \ref{sect:bmode},
a systematic shear error might also be relevant here. However, we lack
a KiDS-1000 model of these for $\xi^{(ij)}_\pm$, and it is unclear if
they plausibly contribute to the \mbox{$\bar{f}_\delta\ne1$}
result. In summary, whereas (especially) $m$-biases and photo-$z$
biases alone are unlikely sources of systematic errors that shape the
result of evolving $\bar{f}_\delta$, the negligence of a $z$-dependent
IA amplitude, $A_\sfont{IA}$, as in our reconstruction set-up, in
combination with moderate photo-$z$ biases might explain our
\mbox{$N_z=3$} result. At this point, we are unable to exclude either
possibility, or that of small systematic errors related to $B$-modes,
but note that the discussed scenarios mainly target the first two
tomographic bins, $z_{\rm B}\le0.5$, where the posterior predictive of
the model also indicates some conflict with the data
(Fig. \ref{fig:kidspostpred}). Future reconstructions of the matter
power spectrum inside separate redshift bins should also pay close
attention to the IA modelling.

A truly evolving $f_\delta(k,z)$, on the other hand, would offer
insights into the physics of non-linear structure growth at
\mbox{$k\gtrsim0.1\,h\,\rm Mpc^{-1}$} and \mbox{$z\lesssim1$}. In
brief, the KiDS-1000 data indicate too little power for the epoch
$z\approx0.3$--$0.6$ and too much power for $z\approx0.6$--$2$,
compared to the purely-$\Lambda\rm CDM$ reference model that actually
matches the $f_{\delta,mn}$ averaged in a single $z$-bin ($68\%$
CI). Phrasing this a bit more carefully, this statement applies to
weighted averages inside three wide $z$-bins where most weight is
given to $\Delta^2(k,z)$ around \mbox{$z\approx0.13, 0.4, 0.7$} for
$Z_1$, $Z_2$, and $Z_3$, similar to Fig. \ref{fig:tikhonov}. Even so,
the average deviations in the $Z_2$ and $Z_3$ are, in fact, such that
they make $\Delta^2(k,z)$ inside the bins statistically consistent
(right panel of Fig. \ref{fig:kidsrelpower}), independent of the
chosen reference -- thus there is no significant detection of any
growth from about \mbox{$z=0.7$} to $0.4$. The only increase in power
is detected towards lower \mbox{$z\approx0.13$}. This qualitatively
matches the $S_8$-tension: cosmological probes that give more weight
to \mbox{$z\gtrsim0.7$}, such as CMB experiments, would observe a
$\Delta^2(k,z)$ with amplitude higher than in our reference, hence a
higher $S_8$, whereas probes more sensitive to \mbox{$z\sim0.4$} or
lower, such as lensing experiments, would observe a $\Delta^2(k,z)$
with $S_8$ equal to our reference or even lower. We are not aware of a
similar conclusion in the lensing literature, although the recent BK24
report a vanishing growth rate at $1\sigma$, their growth index
$m_\sfont{\rm BK}$, for KiDS-1000 and HSC Y3 (yet not for DES
Y3). That finding for $m_\sfont{\rm BK}$ might be related to our
result, depending on how the $Z_1$--$Z_3$ are effectively weighted in
the BK24 analysis. Possible physical causes of the peculiar structure
growth in KiDS-1000 compared to the reference model are, however,
presumably non-trivial and beyond the scope of this paper. For example,
a baryonic feedback model would have to be strongly evolving with
redshift below \mbox{$z\sim1$} (and non-monotonic) to match the
KiDS-1000 $f_{\delta,mn}$, typically not found in state-of-the art
models \citep[e.g.,][and references therein]{2024arXiv241017109S}.

Finally, the close connection of the $S_8$-tension to the KiDS-1000
result for $\Delta^2(k,z)$ in \mbox{$N_z=3$} $z$-bins makes this an
interesting case, either as evidence in favour of modifications
towards the structure growth in the $\Lambda\rm CDM$ reference, or in
favour of spurious systematic errors in a shear tomography analysis
mimicking the tension. This evidence appears to be unnoticed in
previous Stage III analyses -- a full analytic model of
$\Delta^2(k,z)$ fitted to the shear tomography can give an
unsuspecting match similar to our \mbox{$N_z=1$} analysis -- so that a
confirmation of the KiDS-1000 result with other lensing data is
important. Furthermore, looking forward not too far into the future,
Stage IV data will be available soon with roughly $20$ times the
survey area of KiDS-1000 and also deeper observations, with surveys
such as the space-based missions \emph{Euclid} and \emph{Roman}, or
the ground-based survey by \emph{Rubin}. Of these the Euclid Wide
Survey will provide tomographic measurements within $13$ source
$z$-bins up to \mbox{$z\sim2.5$}, statistical errors reduced by about
one order of magnitude compared to Stage III, and an excellent control
of systematic errors for shear and photo-$z$.  We see no restriction
to applying our reconstruction technique to Stage IV data, merely
assuming a lensing kernel and an IA model (possibly updated and
carefully calibrated). This promises measurements of the matter power
spectrum at different redshifts to unprecedented precision, perhaps up
to $z\sim2$ with \mbox{$N_z=4$} or more bins, critically testing
models of the non-linear structure growth.

\begin{acknowledgements}
  This paper went through the KiDS review process. We also want to
  thank Jeger Broxtermann for useful comments.  LP acknowledges
  support from the DLR grant 50QE2002. PB acknowledges financial
  support from the Canadian Space Agency (Grant No. 23EXPROSS1) and
  the Waterloo Centre for Astrophysics. Some results in this paper are
  based on observations made with ESO Telescopes at the La Silla
  Paranal Observatory under programme IDs 177.A-3016, 177.A- 3017,
  177.A-3018 and 179.A-2004, and on data products produced by the KiDS
  consortium. The KiDS production team acknowledges support from:
  Deutsche Forschungsgemeinschaft, ERC, NOVA and NWO-M grants; Target;
  the University of Padova, and the University Federico II (Naples).
\end{acknowledgements}

\bibliographystyle{aa}
\bibliography{aa54017-25}

\begin{thebibliography}{90}
\expandafter\ifx\csname natexlab\endcsname\relax\def\natexlab#1{#1}\fi

\bibitem[{{Abbott} {et~al.}(2018){Abbott}, {Abdalla}, {Alarcon}, {Aleksi{\'c}},
  {Allam}, {Allen}, {Amara}, {Annis}, {Asorey}, {Avila}, {Bacon}, {Balbinot},
  {Banerji}, {Banik}, {Barkhouse}, {Baumer}, {Baxter}, {Bechtol}, {Becker},
  {Benoit-L{\'e}vy}, {Benson}, {Bernstein}, {Bertin}, {Blazek}, {Bridle},
  {Brooks}, {Brout}, {Buckley-Geer}, {Burke}, {Busha}, {Campos}, {Capozzi},
  {Carnero Rosell}, {Carrasco Kind}, {Carretero}, {Castander}, {Cawthon},
  {Chang}, {Chen}, {Childress}, {Choi}, {Conselice}, {Crittenden}, {Crocce},
  {Cunha}, {D'Andrea}, {da Costa}, {Das}, {Davis}, {Davis}, {De Vicente},
  {DePoy}, {DeRose}, {Desai}, {Diehl}, {Dietrich}, {Dodelson}, {Doel},
  {Drlica-Wagner}, {Eifler}, {Elliott}, {Elsner}, {Elvin-Poole}, {Estrada},
  {Evrard}, {Fang}, {Fernandez}, {Fert{\'e}}, {Finley}, {Flaugher}, {Fosalba},
  {Friedrich}, {Frieman}, {Garc{\'\i}a-Bellido}, {Garcia-Fernandez}, {Gatti},
  {Gaztanaga}, {Gerdes}, {Giannantonio}, {Gill}, {Glazebrook}, {Goldstein},
  {Gruen}, {Gruendl}, {Gschwend}, {Gutierrez}, {Hamilton}, {Hartley}, {Hinton},
  {Honscheid}, {Hoyle}, {Huterer}, {Jain}, {James}, {Jarvis}, {Jeltema},
  {Johnson}, {Johnson}, {Kacprzak}, {Kent}, {Kim}, {King}, {Kirk}, {Kokron},
  {Kovacs}, {Krause}, {Krawiec}, {Kremin}, {Kuehn}, {Kuhlmann}, {Kuropatkin},
  {Lacasa}, {Lahav}, {Li}, {Liddle}, {Lidman}, {Lima}, {Lin}, {MacCrann},
  {Maia}, {Makler}, {Manera}, {March}, {Marshall}, {Martini}, {McMahon},
  {Melchior}, {Menanteau}, {Miquel}, {Miranda}, {Mudd}, {Muir}, {M{\"o}ller},
  {Neilsen}, {Nichol}, {Nord}, {Nugent}, {Ogando}, {Palmese}, {Peacock},
  {Peiris}, {Peoples}, {Percival}, {Petravick}, {Plazas}, {Porredon}, {Prat},
  {Pujol}, {Rau}, {Refregier}, {Ricker}, {Roe}, {Rollins}, {Romer}, {Roodman},
  {Rosenfeld}, {Ross}, {Rozo}, {Rykoff}, {Sako}, {Salvador}, {Samuroff},
  {S{\'a}nchez}, {Sanchez}, {Santiago}, {Scarpine}, {Schindler}, {Scolnic},
  {Secco}, {Serrano}, {Sevilla-Noarbe}, {Sheldon}, {Smith}, {Smith}, {Smith},
  {Soares-Santos}, {Sobreira}, {Suchyta}, {Tarle}, {Thomas}, {Troxel},
  {Tucker}, {Tucker}, {Uddin}, {Varga}, {Vielzeuf}, {Vikram}, {Vivas},
  {Walker}, {Wang}, {Wechsler}, {Weller}, {Wester}, {Wolf}, {Yanny}, {Yuan},
  {Zenteno}, {Zhang}, {Zhang}, {Zuntz}, \& {Dark Energy Survey
  Collaboration}}]{2018PhRvD..98d3526A}
{Abbott}, T.~M.~C., {Abdalla}, F.~B., {Alarcon}, A., {et~al.} 2018, \prd, 98,
  043526

\bibitem[{{Alam} {et~al.}(2015){Alam}, {Albareti}, {Allende Prieto}, {Anders},
  {Anderson}, {Anderton}, {Andrews}, {Armengaud}, {Aubourg}, {Bailey}, {Basu},
  {Bautista}, {Beaton}, {Beers}, {Bender}, {Berlind}, {Beutler}, {Bhardwaj},
  {Bird}, {Bizyaev}, {Blake}, {Blanton}, {Blomqvist}, {Bochanski}, {Bolton},
  {Bovy}, {Shelden Bradley}, {Brandt}, {Brauer}, {Brinkmann}, {Brown},
  {Brownstein}, {Burden}, {Burtin}, {Busca}, {Cai}, {Capozzi}, {Carnero
  Rosell}, {Carr}, {Carrera}, {Chambers}, {Chaplin}, {Chen}, {Chiappini},
  {Chojnowski}, {Chuang}, {Clerc}, {Comparat}, {Covey}, {Croft}, {Cuesta},
  {Cunha}, {da Costa}, {Da Rio}, {Davenport}, {Dawson}, {De Lee}, {Delubac},
  {Deshpande}, {Dhital}, {Dutra-Ferreira}, {Dwelly}, {Ealet}, {Ebelke},
  {Edmondson}, {Eisenstein}, {Ellsworth}, {Elsworth}, {Epstein}, {Eracleous},
  {Escoffier}, {Esposito}, {Evans}, {Fan}, {Fern{\'a}ndez-Alvar}, {Feuillet},
  {Filiz Ak}, {Finley}, {Finoguenov}, {Flaherty}, {Fleming}, {Font-Ribera},
  {Foster}, {Frinchaboy}, {Galbraith-Frew}, {Garc{\'\i}a},
  {Garc{\'\i}a-Hern{\'a}ndez}, {Garc{\'\i}a P{\'e}rez}, {Gaulme}, {Ge},
  {G{\'e}nova-Santos}, {Georgakakis}, {Ghezzi}, {Gillespie}, {Girardi},
  {Goddard}, {Gontcho}, {Gonz{\'a}lez Hern{\'a}ndez}, {Grebel}, {Green},
  {Grieb}, {Grieves}, {Gunn}, {Guo}, {Harding}, {Hasselquist}, {Hawley},
  {Hayden}, {Hearty}, {Hekker}, {Ho}, {Hogg}, {Holley-Bockelmann}, {Holtzman},
  {Honscheid}, {Huber}, {Huehnerhoff}, {Ivans}, {Jiang}, {Johnson},
  {Kinemuchi}, {Kirkby}, {Kitaura}, {Klaene}, {Knapp}, {Kneib}, {Koenig},
  {Lam}, {Lan}, {Lang}, {Laurent}, {Le Goff}, {Leauthaud}, {Lee}, {Lee},
  {Licquia}, {Liu}, {Long}, {L{\'o}pez-Corredoira}, {Lorenzo-Oliveira},
  {Lucatello}, {Lundgren}, {Lupton}, {Mack}, {Mahadevan}, {Maia}, {Majewski},
  {Malanushenko}, {Malanushenko}, {Manchado}, {Manera}, {Mao}, {Maraston},
  {Marchwinski}, {Margala}, {Martell}, {Martig}, {Masters}, {Mathur},
  {McBride}, {McGehee}, {McGreer}, {McMahon}, {M{\'e}nard}, {Menzel},
  {Merloni}, {M{\'e}sz{\'a}ros}, {Miller}, {Miralda-Escud{\'e}}, {Miyatake},
  {Montero-Dorta}, {More}, {Morganson}, {Morice-Atkinson}, {Morrison},
  {Mosser}, {Muna}, {Myers}, {Nandra}, {Newman}, {Neyrinck}, {Nguyen},
  {Nichol}, {Nidever}, {Noterdaeme}, {Nuza}, {O'Connell}, {O'Connell},
  {O'Connell}, {Ogando}, {Olmstead}, {Oravetz}, {Oravetz}, {Osumi}, {Owen},
  {Padgett}, {Padmanabhan}, {Paegert}, {Palanque-Delabrouille}, {Pan},
  {Parejko}, {P{\^a}ris}, {Park}, {Pattarakijwanich}, {Pellejero-Ibanez},
  {Pepper}, {Percival}, {P{\'e}rez-Fournon}, {P{\'e}rez-R{\`a}fols},
  {Petitjean}, {Pieri}, {Pinsonneault}, {Porto de Mello}, {Prada}, {Prakash},
  {Price-Whelan}, {Protopapas}, {Raddick}, {Rahman}, {Reid}, {Rich}, {Rix},
  {Robin}, {Rockosi}, {Rodrigues}, {Rodr{\'\i}guez-Torres}, {Roe}, {Ross},
  {Ross}, {Rossi}, {Ruan}, {Rubi{\~n}o-Mart{\'\i}n}, {Rykoff},
  {Salazar-Albornoz}, {Salvato}, {Samushia}, {S{\'a}nchez}, {Santiago},
  {Sayres}, {Schiavon}, {Schlegel}, {Schmidt}, {Schneider}, {Schultheis},
  {Schwope}, {Sc{\'o}ccola}, {Scott}, {Sellgren}, {Seo}, {Serenelli}, {Shane},
  {Shen}, {Shetrone}, {Shu}, {Silva Aguirre}, {Sivarani}, {Skrutskie},
  {Slosar}, {Smith}, {Sobreira}, {Souto}, {Stassun}, {Steinmetz}, {Stello},
  {Strauss}, {Streblyanska}, {Suzuki}, {Swanson}, {Tan}, {Tayar}, {Terrien},
  {Thakar}, {Thomas}, {Thomas}, {Thompson}, {Tinker}, {Tojeiro}, {Troup},
  {Vargas-Maga{\~n}a}, {Vazquez}, {Verde}, {Viel}, {Vogt}, {Wake}, {Wang},
  {Weaver}, {Weinberg}, {Weiner}, {White}, {Wilson}, {Wisniewski},
  {Wood-Vasey}, {Ye`che}, {York}, {Zakamska}, {Zamora}, {Zasowski}, {Zehavi},
  {Zhao}, {Zheng}, {Zhou}, {Zhou}, {Zou}, \& {Zhu}}]{2015ApJS..219...12A}
{Alam}, S., {Albareti}, F.~D., {Allende Prieto}, C., {et~al.} 2015, \apjs, 219,
  12

\bibitem[{{Aric{\`o}} {et~al.}(2023){Aric{\`o}}, {Angulo}, {Zennaro},
  {Contreras}, {Chen}, \& {Hern{\'a}ndez-Monteagudo}}]{2023A&A...678A.109A}
{Aric{\`o}}, G., {Angulo}, R.~E., {Zennaro}, M., {et~al.} 2023, \aap, 678, A109

\bibitem[{{Asgari} {et~al.}(2021){Asgari}, {Lin}, {Joachimi}, {Giblin},
  {Heymans}, {Hildebrandt}, {Kannawadi}, {St{\"o}lzner}, {Tr{\"o}ster}, {van
  den Busch}, {Wright}, {Bilicki}, {Blake}, {de Jong}, {Dvornik}, {Erben},
  {Getman}, {Hoekstra}, {K{\"o}hlinger}, {Kuijken}, {Miller}, {Radovich},
  {Schneider}, {Shan}, \& {Valentijn}}]{2021A&A...645A.104A}
{Asgari}, M., {Lin}, C.-A., {Joachimi}, B., {et~al.} 2021, \aap, 645, A104

\bibitem[{{Bacon} {et~al.}(2005){Bacon}, {Taylor}, {Brown}, {Gray}, {Wolf},
  {Meisenheimer}, {Dye}, {Wisotzki}, {Borch}, \&
  {Kleinheinrich}}]{2005MNRAS.363..723B}
{Bacon}, D.~J., {Taylor}, A.~N., {Brown}, M.~L., {et~al.} 2005, \mnras, 363,
  723

\bibitem[{{Bartelmann} \& {Schneider}(2001)}]{2001PhR...340..291B}
{Bartelmann}, M. \& {Schneider}, P. 2001, \physrep, 340, 291

\bibitem[{{Begeman} {et~al.}(2013){Begeman}, {Belikov}, {Boxhoorn}, \&
  {Valentijn}}]{2013ExA....35....1B}
{Begeman}, K., {Belikov}, A.~N., {Boxhoorn}, D.~R., \& {Valentijn}, E.~A. 2013,
  Experimental Astronomy, 35, 1

\bibitem[{{Ben{\'\i}tez}(2000)}]{2000ApJ...536..571B}
{Ben{\'\i}tez}, N. 2000, \apj, 536, 571

\bibitem[{{Bigwood} {et~al.}(2024){Bigwood}, {Amon}, {Schneider}, {Salcido},
  {McCarthy}, {Preston}, {Sanchez}, {Sijacki}, {Schaan}, {Ferraro},
  {Battaglia}, {Chen}, {Dodelson}, {Roodman}, {Pieres}, {Fert{\'e}}, {Alarcon},
  {Drlica-Wagner}, {Choi}, {Navarro-Alsina}, {Campos}, {Ross}, {Carnero
  Rosell}, {Yin}, {Yanny}, {S{\'a}nchez}, {Chang}, {Davis}, {Doux}, {Gruen},
  {Rykoff}, {Huff}, {Sheldon}, {Tarsitano}, {Andrade-Oliveira}, {Bernstein},
  {Giannini}, {Diehl}, {Huang}, {Harrison}, {Sevilla-Noarbe}, {Tutusaus},
  {Elvin-Poole}, {McCullough}, {Zuntz}, {Blazek}, {DeRose}, {Cordero}, {Prat},
  {Myles}, {Eckert}, {Bechtol}, {Herner}, {Secco}, {Gatti}, {Raveri}, {Kind},
  {Becker}, {Troxel}, {Jarvis}, {MacCrann}, {Friedrich}, {Alves}, {Leget},
  {Chen}, {Rollins}, {Wechsler}, {Gruendl}, {Cawthon}, {Allam}, {Bridle},
  {Pandey}, {Everett}, {Shin}, {Hartley}, {Fang}, {Zhang}, {Aguena}, {Annis},
  {Bacon}, {Bertin}, {Bocquet}, {Brooks}, {Carretero}, {Castander}, {da Costa},
  {Pereira}, {De Vicente}, {Desai}, {Doel}, {Ferrero}, {Flaugher}, {Frieman},
  {Garc{\'\i}a-Bellido}, {Gaztanaga}, {Gutierrez}, {Hinton}, {Hollowood},
  {Honscheid}, {Huterer}, {James}, {Kuehn}, {Lahav}, {Lee}, {Marshall},
  {Mena-Fern{\'a}ndez}, {Miquel}, {Muir}, {Paterno}, {Plazas Malag{\'o}n},
  {Porredon}, {Romer}, {Samuroff}, {Sanchez}, {Sanchez Cid}, {Smith},
  {Soares-Santos}, {Suchyta}, {Swanson}, {Tarle}, {To}, {Weaverdyck}, {Weller},
  {Wiseman}, \& {Yamamoto}}]{2024MNRAS.534..655B}
{Bigwood}, L., {Amon}, A., {Schneider}, A., {et~al.} 2024, \mnras, 534, 655

\bibitem[{{Blake} {et~al.}(2016){Blake}, {Amon}, {Childress}, {Erben},
  {Glazebrook}, {Harnois-Deraps}, {Heymans}, {Hildebrandt}, {Hinton},
  {Janssens}, {Johnson}, {Joudaki}, {Klaes}, {Kuijken}, {Lidman}, {Marin},
  {Parkinson}, {Poole}, \& {Wolf}}]{2016MNRAS.462.4240B}
{Blake}, C., {Amon}, A., {Childress}, M., {et~al.} 2016, \mnras, 462, 4240

\bibitem[{{Bridle} \& {King}(2007)}]{2007NJPh....9..444B}
{Bridle}, S. \& {King}, L. 2007, New Journal of Physics, 9, 444

\bibitem[{Brooks {et~al.}(2011)Brooks, Gelman, Jones, \&
  Meng}]{rooks2011handbook}
Brooks, S., Gelman, A., Jones, G., \& Meng, X. 2011, Handbook of Markov Chain
  Monte Carlo, Chapman \& Hall/CRC Handbooks of Modern Statistical Methods (CRC
  Press)

\bibitem[{{Broxterman} \& {Kuijken}(2024)}]{2024arXiv240913404B}
{Broxterman}, J.~C. \& {Kuijken}, K. 2024, \aap, 692, A201

\bibitem[{{Bucko} {et~al.}(2024){Bucko}, {Giri}, {Peters}, \&
  {Schneider}}]{2024A&A...683A.152B}
{Bucko}, J., {Giri}, S.~K., {Peters}, F.~H., \& {Schneider}, A. 2024, \aap,
  683, A152

\bibitem[{{Burger} {et~al.}(2024){Burger}, {Porth}, {Heydenreich}, {Linke},
  {Wielders}, {Schneider}, {Asgari}, {Castro}, {Dolag}, {Harnois-D{\'e}raps},
  {Hildebrandt}, {Kuijken}, \& {Martinet}}]{2024A&A...683A.103B}
{Burger}, P.~A., {Porth}, L., {Heydenreich}, S., {et~al.} 2024, \aap, 683, A103

\bibitem[{{Crittenden} {et~al.}(2001){Crittenden}, {Natarajan}, {Pen}, \&
  {Theuns}}]{2001ApJ...559..552C}
{Crittenden}, R.~G., {Natarajan}, P., {Pen}, U.-L., \& {Theuns}, T. 2001, \apj,
  559, 552

\bibitem[{{Croft} \& {Metzler}(2000)}]{2000ApJ...545..561C}
{Croft}, R. A.~C. \& {Metzler}, C.~A. 2000, \apj, 545, 561

\bibitem[{{Edge} {et~al.}(2013){Edge}, {Sutherland}, {Kuijken}, {Driver},
  {McMahon}, {Eales}, \& {Emerson}}]{2013Msngr.154...32E}
{Edge}, A., {Sutherland}, W., {Kuijken}, K., {et~al.} 2013, The Messenger, 154,
  32

\bibitem[{{Erben} {et~al.}(2005){Erben}, {Schirmer}, {Dietrich}, {Cordes},
  {Haberzettl}, {Hetterscheidt}, {Hildebrandt}, {Schmithuesen}, {Schneider},
  {Simon}, {Deul}, {Hook}, {Kaiser}, {Radovich}, {Benoist}, {Nonino}, {Olsen},
  {Prandoni}, {Wichmann}, {Zaggia}, {Bomans}, {Dettmar}, \&
  {Miralles}}]{2005AN....326..432E}
{Erben}, T., {Schirmer}, M., {Dietrich}, J.~P., {et~al.} 2005, Astronomische
  Nachrichten, 326, 432

\bibitem[{{Euclid Collaboration: Mellier} {et~al.}(2025){Euclid Collaboration:
  Mellier}, {Abdurro'uf}, {Acevedo Barroso}, {Ach{\'u}carro}, {Adamek}, {Adam},
  {Addison}, {Aghanim}, {Aguena}, {Ajani}, {Akrami}, {Al-Bahlawan}, {Alavi},
  {Albuquerque}, {Alestas}, {Alguero}, {Allaoui}, {Allen}, {Allevato},
  {Alonso-Tetilla}, {Altieri}, {Alvarez-Candal}, {Amara}, {Amendola}, {Amiaux},
  {Andika}, {Andreon}, {Andrews}, {Angora}, {Angulo}, {Annibali}, {Anselmi},
  {Anselmi}, {Arcari}, {Archidiacono}, {Aric{\`o}}, {Arnaud}, {Arnouts},
  {Asgari}, {Asorey}, {Atayde}, {Atek}, {Atrio-Barandela}, {Aubert}, {Aubourg},
  {Auphan}, {Auricchio}, {Aussel}, {Aussel}, {Avelino}, {Avgoustidis}, {Avila},
  {Awan}, {Azzollini}, {Baccigalupi}, {Bachelet}, {Bacon}, {Baes}, {Bagley},
  {Bahr-Kalus}, {Balaguera-Antolinez}, {Balbinot}, {Balcells}, {Baldi},
  {Baldry}, {Balestra}, {Ballardini}, {Ballester}, {Balogh}, {Ba{\~n}ados},
  {Barbier}, {Bardelli}, {Barreiro}, {Barriere}, {Barros}, {Barthelemy},
  {Bartolo}, {Basset}, {Battaglia}, {Battisti}, {Baugh}, {Baumont},
  {Bazzanini}, {Beaulieu}, {Beckmann}, {Belikov}, {Bel}, {Bellagamba}, {Bella},
  {Bellini}, {Benabed}, {Bender}, {Benevento}, {Bennett}, {Benson},
  {Bergamini}, {Bermejo-Climent}, {Bernardeau}, {Bertacca}, {Berthe},
  {Berthier}, {Bethermin}, {Beutler}, {Bevillon}, {Bhargava}, {Bhatawdekar},
  {Bisigello}, {Biviano}, {Blake}, {Blanchard}, {Blazek}, {Blot}, {Bosco},
  {Bodendorf}, {Boenke}, {B{\"o}hringer}, {Bolzonella}, {Bonchi}, {Bonici},
  {Bonino}, {Bonino}, {Bonvin}, {Bon}, {Booth}, {Borgani}, {Borlaff},
  {Borsato}, {Bosco}, {Bose}, {Botticella}, {Boucaud}, {Bouche}, {Boucher},
  {Boutigny}, {Bouvard}, {Bouy}, {Bowler}, {Bozza}, {Bozzo}, {Branchini},
  {Brau-Nogue}, {Brekke}, {Bremer}, {Brescia}, {Breton}, {Brinchmann},
  {Brinckmann}, {Brockley-Blatt}, {Brodwin}, {Brouard}, {Brown}, {Bruton},
  {Bucko}, {Buddelmeijer}, {Buenadicha}, {Buitrago}, {Burger}, {Burigana},
  {Busillo}, {Busonero}, {Cabanac}, {Cabayol-Garcia}, {Cagliari}, {Caillat},
  {Caillat}, {Calabrese}, {Calabro}, {Calderone}, {Calura}, {Camacho Quevedo},
  {Camera}, {Campos}, {Canas-Herrera}, {Candini}, {Cantiello}, {Capobianco},
  {Cappellaro}, {Cappelluti}, {Cappi}, {Caputi}, {Cara}, {Carbone}, {Cardone},
  {Carella}, {Carlberg}, {Carle}, {Carminati}, {Caro}, {Carrasco}, {Carretero},
  {Carrilho}, {Carron Duque}, {Carry}, {Carvalho}, {Carvalho}, {Casas},
  {Casas}, {Casenove}, {Casey}, {Cassata}, {Castander}, {Castelao},
  {Castellano}, {Castiblanco}, {Castignani}, {Castro}, {Cavet}, {Cavuoti},
  {Chabaud}, {Chambers}, {Charles}, {Charlot}, {Chartab}, {Chary}, {Chaumeil},
  {Cho}, {Chon}, {Ciancetta}, {Ciliegi}, {Cimatti}, {Cimino}, {Cioni},
  {Claydon}, {Cleland}, {Cl{\'e}ment}, {Clements}, {Clerc}, {Clesse}, {Codis},
  {Cogato}, {Colbert}, {Cole}, {Coles}, {Collett}, {Collins}, {Colodro-Conde},
  {Colombo}, {Combes}, {Conforti}, {Congedo}, {Conseil}, {Conselice},
  {Contarini}, {Contini}, {Conversi}, {Cooray}, {Copin}, {Corasaniti},
  {Corcho-Caballero}, {Corcione}, {Cordes}, {Corpace}, {Correnti}, {Costanzi},
  {Costille}, {Courbin}, {Courcoult Mifsud}, {Courtois}, {Cousinou}, {Covone},
  {Cowell}, {Cragg}, {Cresci}, {Cristiani}, {Crocce}, {Cropper}, {E Crouzet},
  {Csizi}, {Cuby}, {Cucchetti}, {Cucciati}, {Cuillandre}, {Cunha}, {Cuozzo},
  {Daddi}, {D'Addona}, {Dafonte}, {Dagoneau}, {Dalessandro}, {Dalton},
  {D'Amico}, {Dannerbauer}, {Danto}, {Das}, {Da Silva}, {da Silva}, {Daste},
  {Davies}, {Davini}, {de Boer}, {Decarli}, {De Caro}, {Degaudenzi}, {Degni},
  {de Jong}, {de la Bella}, {de la Torre}, {Delhaise}, {Delley}, {Delucchi},
  {De Lucia}, {Denniston}, {De Paolis}, {De Petris}, {Derosa}, {Desai},
  {Desjacques}, {Despali}, {Desprez}, {De Vicente-Albendea}, {Deville}, {Dias},
  {D{\'\i}az-S{\'a}nchez}, {Diaz}, {Di Domizio}, {Diego}, {Di Ferdinando}, {Di
  Giorgio}, {Dimauro}, {Dinis}, {Dolag}, {Dolding}, {Dole}, {Dom{\'\i}nguez
  S{\'a}nchez}, {Dor{\'e}}, {Dournac}, {Douspis}, {Dreihahn}, {Droge}, {Dryer},
  {Dubath}, {Duc}, {Ducret}, {Duffy}, {Dufresne}, {Duncan}, {Dupac}, {Duret},
  {Durrer}, {Durret}, {Dusini}, {Ealet}, {Eggemeier}, {Eisenhardt}, {Elbaz},
  {Elkhashab}, {Ellien}, {Endicott}, {Enia}, {Erben}, {Escartin Vigo},
  {Escoffier}, {Escudero Sanz}, {Essert}, {Ettori}, {Ezziati}, {Fabbian},
  {Fabricius}, {Fang}, {Farina}, {Farina}, {Farinelli}, {Farrens}, {Faustini},
  {Feltre}, {Ferguson}, {Ferrando}, {Ferrari}, {Ferr{\'e}-Mateu}, {Ferreira},
  {Ferreras}, {Ferrero}, {Ferriol}, {Ferruit}, {Filleul}, {Finelli},
  {Finkelstein}, {Finoguenov}, {Fiorini}, {Flentge}, {Focardi}, {Fonseca},
  {Fontana}, {Fontanot}, {Fornari}, {Fosalba}, {Fossati}, {Fotopoulou},
  {Fouchez}, {Fourmanoit}, {Frailis}, {Fraix-Burnet}, {Franceschi}, {Franco},
  {Franzetti}, {Freihoefer}, {Frittoli}, {Frugier}, {Frusciante}, {Fumagalli},
  {Fumagalli}, {Fumana}, {Fu}, {Gabarra}, {Galeotta}, {Galluccio}, {Ganga},
  {Gao}, {Garc{\'\i}a-Bellido}, {Garcia}, {Gardner}, {Garilli},
  {Gaspar-Venancio}, {Gasparetto}, {Gautard}, {Gavazzi}, {Gaztanaga},
  {Genolet}, {Genova Santos}, {Gentile}, {George}, {Ghaffari}, {Giacomini},
  {Gianotti}, {Gibb}, {Gillard}, {Gillis}, {Ginolfi}, {Giocoli}, {Girardi},
  {Giri}, {Goh}, {G{\'o}mez-Alvarez}, {Gonzalez}, {Gonzalez}, {Gonzalez},
  {Gouyou Beauchamps}, {Gozaliasl}, {Gracia-Carpio}, {Grandis}, {Granett},
  {Granvik}, {Grazian}, {Gregorio}, {Grenet}, {Grillo}, {Grupp}, {Gruppioni},
  {Gruppuso}, {Guerbuez}, {Guerrini}, {Guidi}, {Guillard}, {Gutierrez},
  {Guttridge}, {Guzzo}, {Gwyn}, {Haapala}, {Haase}, {Haddow}, {Hailey}, {Hall},
  {Hall}, {Hamaus}, {Haridasu}, {Harnois-D{\'e}raps}, {Harper}, {Hartley},
  {Hasinger}, {Hassani}, {Hatch}, {Haugan}, {H{\"a}u{\ss}ler}, {Heavens},
  {Heisenberg}, {Helmi}, {Helou}, {Hemmati}, {Henares}, {Herent},
  {Hern{\'a}ndez-Monteagudo}, {Heuberger}, {Hewett}, {Heydenreich},
  {Hildebrandt}, {Hirschmann}, {Hjorth}, {Hoar}, {Hoekstra}, {Holland},
  {Holliman}, {Holmes}, {Hook}, {Horeau}, {Hormuth}, {Hornstrup}, {Hosseini},
  {Hu}, {Hudelot}, {Hudson}, {Huertas-Company}, {Huff}, {Hughes}, {Humphrey},
  {Hunt}, {Huynh}, {Ibata}, {Ichikawa}, {Iglesias-Groth}, {Ilbert}, {Ili{\'c}},
  {Ingoglia}, {Iodice}, {Israel}, {Israelsson}, {Izzo}, {Jablonka}, {Jackson},
  {Jacobson}, {Jafariyazani}, {Jahnke}, {Jansen}, {Jarvis}, {Jasche}, {Jauzac},
  {Jeffrey}, {Jhabvala}, {Jimenez-Teja}, {Jimenez Mu{\~n}oz}, {Joachimi},
  {Johansson}, {Joudaki}, {Jullo}, {Kajava}, {Kang}, {Kannawadi}, {Kansal},
  {Karagiannis}, {K{\"a}rcher}, {Kashlinsky}, {Kazandjian}, {Keck},
  {Keih{\"a}nen}, {Kerins}, {Kermiche}, {Khalil}, {Kiessling}, {Kiiveri},
  {Kilbinger}, {Kim}, {King}, {Kirkpatrick}, {Kitching}, {Kluge}, {Knabenhans},
  {Knapen}, {Knebe}, {Kneib}, {Kohley}, {Koopmans}, {Koskinen}, {Koulouridis},
  {Kou}, {Kov{\'a}cs}, {Kova\{{\v{c}}\}i{\'c}}, {Kowalczyk}, {Koyama},
  {Kraljic}, {Krause}, {Kruk}, {Kubik}, {Kuchner}, {Kuijken}, {K{\"u}mmel},
  {Kunz}, {Kurki-Suonio}, {Lacasa}, {Lacey}, {La Franca}, {Lagarde}, {Lahav},
  {Laigle}, {La Marca}, {La Marle}, {Lamine}, {Lam}, {Lan{\c{c}}on}, {Landt},
  {Langer}, {Lapi}, {Larcheveque}, {Larsen}, {Lattanzi}, {Laudisio}, {Laugier},
  {Laureijs}, {Lavaux}, {Lawrenson}, {Lazanu}, {Lazeyras}, {Le Boulc'h}, {Le
  Brun}, {Le Brun}, {Leclercq}, {Lee}, {Le Graet}, {Legrand}, {Leirvik}, {Le
  Jeune}, {Lembo}, {Le Mignant}, {Lepinzan}, {Lepori}, {Lesci}, {Lesgourgues},
  {Leuzzi}, {Levi}, {Liaudat}, {Libet}, {Liebing}, {Ligori}, {Lilje}, {Lin},
  {Linde}, {Linder}, {Lindholm}, {Linke}, {Li}, {Liu}, {Lloro}, {Lobo},
  {Lodieu}, {Lombardi}, {Lombriser}, {Lonare}, {Longo}, {L{\'o}pez-Caniego},
  {Lopez Lopez}, {Alvarez}, {Loureiro}, {Loveday}, {Lusso}, {Macias-Perez},
  {Maciaszek}, {Magliocchetti}, {Magnard}, {Magnier}, {Magro}, {Mahler},
  {Mainetti}, {Maino}, {Maiorano}, {Maiorano}, {Malavasi}, {Mamon}, {Mancini},
  {Mandelbaum}, {Manera}, {Manj{\'o}n-Garc{\'\i}a}, {Mannucci}, {Mansutti},
  {Manteiga Outeiro}, {Maoli}, {Maraston}, {Marcin}, {Marcos-Arenal},
  {Margalef-Bentabol}, {Marggraf}, {Marinucci}, {Marinucci}, {Markovic},
  {Marleau}, {Marpaud}, {Martignac}, {Mart{\'\i}n-Fleitas}, {Martin-Moruno},
  {Martin}, {Martinelli}, {Martinet}, {Martin}, {Martins}, {Marulli},
  {Massari}, {Massey}, {Masters}, {Matarrese}, {Matsuoka}, {Matthew},
  {Maughan}, {Mauri}, {Maurin}, {Maurogordato}, {McCarthy}, {McConnachie},
  {McCracken}, {McDonald}, {McEwen}, {McPartland}, {Medinaceli}, {Mehta},
  {Mei}, {Melchior}, {Melin}, {M{\'e}nard}, {Mendes}, {Mendez-Abreu},
  {Meneghetti}, {Mercurio}, {Merlin}, {Metcalf}, {Meylan}, {Migliaccio},
  {Mignoli}, {Miller}, {Miluzio}, {Milvang-Jensen}, {Mimoso}, {Miquel},
  {Miyatake}, {Mobasher}, {Mohr}, {Monaco}, {Mongui{\'o}}, {Montoro}, {Mora},
  {Moradinezhad Dizgah}, {Moresco}, {Moretti}, {Morgante}, {Morisset},
  {Moriya}, {Morris}, {Mortlock}, {Moscardini}, {Mota}, {Moustakas}, {Moutard},
  {M{\"u}ller}, {Munari}, {Murphree}, {Murray}, {Murray}, {Musi}, {Nadathur},
  {Nagam}, {Nagao}, {Naidoo}, {Nakajima}, {Nally}, {Natoli}, {Navarro-Alsina},
  {Navarro Girones}, {Neissner}, {Nersesian}, {Nesseris}, {Nguyen-Kim},
  {Nicastro}, {Nichol}, {Nielbock}, {Niemi}, {Nieto}, {Nilsson}, {Noller},
  {Norberg}, {Nourizonoz}, {Ntelis}, {Nucita}, {Nugent}, {Nunes}, {Nutma},
  {Ocampo}, {Odier}, {Oesch}, {Oguri}, {Magalhaes Oliveira}, {Onoue},
  {Oosterbroek}, {Oppizzi}, {Ordenovic}, {Osato}, {Pacaud}, {Pace}, {Padilla},
  {Paech}, {Pagano}, {Page}, {Palazzi}, {Paltani}, {Pamuk}, {Pandolfi},
  {Paoletti}, {Paolillo}, {Papaderos}, {Pardede}, {Parimbelli}, {Parmar},
  {Partmann}, {Pasian}, {Passalacqua}, {Paterson}, {Patrizii}, {Pattison},
  {Paulino-Afonso}, {Paviot}, {Peacock}, {Pearce}, {Pedersen}, {Peel},
  {Peletier}, {Pellejero Ibanez}, {Pello}, {Penny}, {Percival},
  {Perez-Garrido}, {Perotto}, {Pettorino}, {Pezzotta}, {Pezzuto}, {Philippon},
  {Piersanti}, {Pietroni}, {Piga}, {Pilo}, {Pires}, {Pisani}, {Pizzella},
  {Pizzuti}, {Plana}, {Polenta}, {Pollack}, {Poncet}, {P{\"o}ntinen}, {Pool},
  {Popa}, {Popa}, {Popp}, {Porciani}, {Porth}, {Potter}, {Poulain},
  {Pourtsidou}, {Pozzetti}, {Prandoni}, {Pratt}, {Prezelus}, {Prieto}, {Pugno},
  {Quai}, {Quilley}, {Racca}, {Raccanelli}, {R{\'a}cz}, {Radinovi{\'c}},
  {Radovich}, {Ragagnin}, {Ragnit}, {Raison}, {Ramos-Chernenko}, {Ranc},
  {Raylet}, {Rebolo}, {Refregier}, {Reimberg}, {Reiprich}, {Renk}, {Renzi},
  {Retre}, {Revaz}, {Reyl{\'e}}, {Reynolds}, {Rhodes}, {Ricci}, {Ricci},
  {Riccio}, {Ricken}, {Rissanen}, {Risso}, {Rix}, {Robin}, {Rocca-Volmerange},
  {Rocci}, {Rodenhuis}, {Rodighiero}, {Rodriguez Monroy}, {Rollins},
  {Romanello}, {Roman}, {Romelli}, {Romero-Gomez}, {Roncarelli}, {Rosati},
  {Rosset}, {Rossetti}, {Roster}, {Rottgering}, {Rozas-Fern{\'a}ndez}, {Ruane},
  {Rubino-Martin}, {Rudolph}, {Ruppin}, {Rusholme}, {Sacquegna},
  {S{\'a}ez-Casares}, {Saga}, {Saglia}, {Sahl{\'e}n}, {Saifollahi}, {Sakr},
  {Salvalaggio}, {Salvaterra}, {Salvati}, {Salvato}, {Salvignol},
  {S{\'a}nchez}, {Sanchez}, {Sanders}, {Sapone}, {Saponara}, {Sarpa}, {Sarron},
  {Sartori}, {Sassolas}, {Sauniere}, {Sauvage}, {Sawicki}, {Scaramella},
  {Scarlata}, {Scharr{\'e}}, {Schaye}, {Schewtschenko}, {Schindler},
  {Schinnerer}, {Schirmer}, {Schmidt}, {Schmidt}, {Schmidt}, {Schneider},
  {Schneider}, {Schneider}, {Sch{\"o}neberg}, {Schrabback}, {Schultheis},
  {Schulz}, {Schwartz}, {Sciotti}, {Scodeggio}, {Scognamiglio}, {Scott},
  {Scottez}, {Secroun}, {Sefusatti}, {Seidel}, {Seiffert}, {Sellentin},
  {Selwood}, {Semboloni}, {Sereno}, {Serjeant}, {Serrano}, {Shankar},
  {Sharples}, {Short}, {Shulevski}, {Shuntov}, {Sias}, {Sikkema}, {Silvestri},
  {Simon}, {Sirignano}, {Sirri}, {Skottfelt}, {Slezak}, {Sluse}, {Smith},
  {Smith}, {Smith}, {Smit}, {Soldano}, {Solheim}, {Sorce}, {Sorrenti},
  {Soubrie}, {Spinoglio}, {Spurio Mancini}, {Stadel}, {Stagnaro}, {Stanco},
  {Stanford}, {Starck}, {Stassi}, {Steinwagner}, {Stern}, {Stone}, {Strada},
  {Strafella}, {Stramaccioni}, {Surace}, {Sureau}, {Suyu}, {Swindells},
  {Szafraniec}, {Szapudi}, {Taamoli}, {Talia}, {Tallada-Cresp{\'\i}},
  {Tanidis}, {Tao}, {Tarr{\'\i}o}, {Tavagnacco}, {Taylor}, {Taylor}, {Taylor},
  {Teixeira}, {Tenti}, {Teodoro Idiago}, {Teplitz}, {Tereno}, {Tessore},
  {Testa}, {Testera}, {Tewes}, {Teyssier}, {Theret}, {Thizy}, {Thomas}, {Toba},
  {Toft}, {Toledo-Moreo}, {Tolstoy}, {Tommasi}, {Torbaniuk}, {Torradeflot},
  {Tortora}, {Tosi}, {Tosti}, {Trifoglio}, {Troja}, {Trombetti}, {Tronconi},
  {Tsedrik}, {Tsyganov}, {Tucci}, {Tutusaus}, {Uhlemann}, {Ulivi}, {Urbano},
  {Vacher}, {Vaillon}, {Valdes}, {Valentijn}, {Valenziano}, {Valieri},
  {Valiviita}, {Van den Broeck}, {Vassallo}, {Vavrek}, {Venemans}, {Venhola},
  {Ventura}, {Verdoes Kleijn}, {Vergani}, {Verma}, {Vernizzi}, {Veropalumbo},
  {Verza}, {Vescovi}, {Vibert}, {Viel}, {Vielzeuf}, {Viglione}, {Viitanen},
  {Villaescusa-Navarro}, {Vinciguerra}, {Visticot}, {Voggel}, {von
  Wietersheim-Kramsta}, {Vriend}, {Wachter}, {Walmsley}, {Walth}, {Walton},
  {Walton}, {Wander}, {Wang}, {Wang}, {Weaver}, {Weller}, {Whalen}, {Wiesmann},
  {Wilde}, {Williams}, {Winther}, {Wittje}, {Wong}, {Wright}, {Yankelevich},
  {Yeung}, {Youles}, {Yung}, {Zacchei}, {Zalesky}, {Zamorani}, {Zamorano
  Vitorelli}, {Zanoni Marc}, {Zennaro}, {Zerbi}, {Zinchenko}, {Zoubian},
  {Zucca}, \& {Zumalacarregui}}]{2024arXiv240513491E}
{Euclid Collaboration: Mellier}, Y., {Abdurro'uf}, {Acevedo Barroso}, J.~A.,
  {et~al.} 2025, \aap, 697, A1

\bibitem[{{Fenech Conti} {et~al.}(2017){Fenech Conti}, {Herbonnet}, {Hoekstra},
  {Merten}, {Miller}, \& {Viola}}]{2017MNRAS.467.1627F}
{Fenech Conti}, I., {Herbonnet}, R., {Hoekstra}, H., {et~al.} 2017, \mnras,
  467, 1627

\bibitem[{{Ferreira} {et~al.}(2024){Ferreira}, {Alonso}, {Garcia-Garcia}, \&
  {Chisari}}]{2023arXiv230911129F}
{Ferreira}, T., {Alonso}, D., {Garcia-Garcia}, C., \& {Chisari}, N.~E. 2024,
  \prl, 133, 051001

\bibitem[{{Fortuna} {et~al.}(2021){Fortuna}, {Hoekstra}, {Joachimi},
  {Johnston}, {Chisari}, {Georgiou}, \& {Mahony}}]{2021MNRAS.501.2983F}
{Fortuna}, M.~C., {Hoekstra}, H., {Joachimi}, B., {et~al.} 2021, \mnras, 501,
  2983

\bibitem[{{Garc{\'\i}a-Garc{\'\i}a} {et~al.}(2024){Garc{\'\i}a-Garc{\'\i}a},
  {Zennaro}, {Aric{\`o}}, {Alonso}, \& {Angulo}}]{2024arXiv240313794G}
{Garc{\'\i}a-Garc{\'\i}a}, C., {Zennaro}, M., {Aric{\`o}}, G., {Alonso}, D., \&
  {Angulo}, R.~E. 2024, JCAP, 2024, 024

\bibitem[{Gelman {et~al.}(2003)Gelman, Carlin, Stern, \&
  Rubin}]{gelman2003bayesian}
Gelman, A., Carlin, J., Stern, H., \& Rubin, D. 2003, Bayesian Data Analysis,
  Chapman \& Hall/CRC Texts in Statistical Science (Chapman \& Hall/CRC)

\bibitem[{{Gelman} \& {Rubin}(1992)}]{1992StaSc...7..457G}
{Gelman}, A. \& {Rubin}, D.~B. 1992, Statistical Science, 7, 457

\bibitem[{{Giblin} {et~al.}(2021){Giblin}, {Heymans}, {Asgari}, {Hildebrandt},
  {Hoekstra}, {Joachimi}, {Kannawadi}, {Kuijken}, {Lin}, {Miller},
  {Tr{\"o}ster}, {van den Busch}, {Wright}, {Bilicki}, {Blake}, {de Jong},
  {Dvornik}, {Erben}, {Getman}, {Napolitano}, {Schneider}, {Shan}, \&
  {Valentijn}}]{2021A&A...645A.105G}
{Giblin}, B., {Heymans}, C., {Asgari}, M., {et~al.} 2021, \aap, 645, A105

\bibitem[{{Harnois-D{\'e}raps} {et~al.}(2022){Harnois-D{\'e}raps}, {Martinet},
  \& {Reischke}}]{2022MNRAS.509.3868H}
{Harnois-D{\'e}raps}, J., {Martinet}, N., \& {Reischke}, R. 2022, \mnras, 509,
  3868

\bibitem[{{Harnois-D{\'e}raps} {et~al.}(2015){Harnois-D{\'e}raps}, {van
  Waerbeke}, {Viola}, \& {Heymans}}]{2015MNRAS.450.1212H}
{Harnois-D{\'e}raps}, J., {van Waerbeke}, L., {Viola}, M., \& {Heymans}, C.
  2015, \mnras, 450, 1212

\bibitem[{{Hartlap} {et~al.}(2007){Hartlap}, {Simon}, \&
  {Schneider}}]{2007A&A...464..399H}
{Hartlap}, J., {Simon}, P., \& {Schneider}, P. 2007, \aap, 464, 399

\bibitem[{{Heymans} {et~al.}(2013){Heymans}, {Grocutt}, {Heavens}, {Kilbinger},
  {Kitching}, {Simpson}, {Benjamin}, {Erben}, {Hildebrandt}, {Hoekstra},
  {Mellier}, {Miller}, {Van Waerbeke}, {Brown}, {Coupon}, {Fu},
  {Harnois-D{\'e}raps}, {Hudson}, {Kuijken}, {Rowe}, {Schrabback}, {Semboloni},
  {Vafaei}, \& {Velander}}]{2013MNRAS.432.2433H}
{Heymans}, C., {Grocutt}, E., {Heavens}, A., {et~al.} 2013, \mnras, 432, 2433

\bibitem[{{Heymans} {et~al.}(2021){Heymans}, {Tr{\"o}ster}, {Asgari}, {Blake},
  {Hildebrandt}, {Joachimi}, {Kuijken}, {Lin}, {S{\'a}nchez}, {van den Busch},
  {Wright}, {Amon}, {Bilicki}, {de Jong}, {Crocce}, {Dvornik}, {Erben},
  {Fortuna}, {Getman}, {Giblin}, {Glazebrook}, {Hoekstra}, {Joudaki},
  {Kannawadi}, {K{\"o}hlinger}, {Lidman}, {Miller}, {Napolitano}, {Parkinson},
  {Schneider}, {Shan}, {Valentijn}, {Verdoes Kleijn}, \&
  {Wolf}}]{2021A&A...646A.140H}
{Heymans}, C., {Tr{\"o}ster}, T., {Asgari}, M., {et~al.} 2021, \aap, 646, A140

\bibitem[{{Hilbert} {et~al.}(2009){Hilbert}, {Hartlap}, {White}, \&
  {Schneider}}]{2009A&A...499...31H}
{Hilbert}, S., {Hartlap}, J., {White}, S.~D.~M., \& {Schneider}, P. 2009, \aap,
  499, 31

\bibitem[{{Hildebrandt} {et~al.}(2021){Hildebrandt}, {van den Busch}, {Wright},
  {Blake}, {Joachimi}, {Kuijken}, {Tr{\"o}ster}, {Asgari}, {Bilicki}, {de
  Jong}, {Dvornik}, {Erben}, {Getman}, {Giblin}, {Heymans}, {Kannawadi}, {Lin},
  \& {Shan}}]{2021A&A...647A.124H}
{Hildebrandt}, H., {van den Busch}, J.~L., {Wright}, A.~H., {et~al.} 2021,
  \aap, 647, A124

\bibitem[{{Hinshaw} {et~al.}(2013){Hinshaw}, {Larson}, {Komatsu}, {Spergel},
  {Bennett}, {Dunkley}, {Nolta}, {Halpern}, {Hill}, {Odegard}, {Page}, {Smith},
  {Weiland}, {Gold}, {Jarosik}, {Kogut}, {Limon}, {Meyer}, {Tucker}, {Wollack},
  \& {Wright}}]{2013ApJS..208...19H}
{Hinshaw}, G., {Larson}, D., {Komatsu}, E., {et~al.} 2013, \apjs, 208, 19

\bibitem[{{Hirata} \& {Seljak}(2004)}]{2004PhRvD..70f3526H}
{Hirata}, C.~M. \& {Seljak}, U. 2004, \prd, 70, 063526

\bibitem[{{Hu} \& {Keeton}(2002)}]{2002PhRvD..66f3506H}
{Hu}, W. \& {Keeton}, C.~R. 2002, \prd, 66, 063506

\bibitem[{{Jarvis}(2015)}]{2015ascl.soft08007J}
{Jarvis}, M. 2015, {TreeCorr: Two-point correlation functions}, Astrophysics
  Source Code Library, record ascl:1508.007

\bibitem[{{Jarvis} {et~al.}(2004){Jarvis}, {Bernstein}, \&
  {Jain}}]{2004MNRAS.352..338J}
{Jarvis}, M., {Bernstein}, G., \& {Jain}, B. 2004, \mnras, 352, 338

\bibitem[{{Jing} {et~al.}(2006){Jing}, {Zhang}, {Lin}, {Gao}, \&
  {Springel}}]{2006ApJ...640L.119J}
{Jing}, Y.~P., {Zhang}, P., {Lin}, W.~P., {Gao}, L., \& {Springel}, V. 2006,
  \apjl, 640, L119

\bibitem[{{Joachimi} {et~al.}(2021){Joachimi}, {Lin}, {Asgari}, {Tr{\"o}ster},
  {Heymans}, {Hildebrandt}, {K{\"o}hlinger}, {S{\'a}nchez}, {Wright},
  {Bilicki}, {Blake}, {van den Busch}, {Crocce}, {Dvornik}, {Erben}, {Getman},
  {Giblin}, {Hoekstra}, {Kannawadi}, {Kuijken}, {Napolitano}, {Schneider},
  {Scoccimarro}, {Sellentin}, {Shan}, {von Wietersheim-Kramsta}, \&
  {Zuntz}}]{2021A&A...646A.129J}
{Joachimi}, B., {Lin}, C.~A., {Asgari}, M., {et~al.} 2021, \aap, 646, A129

\bibitem[{{Joachimi} {et~al.}(2011){Joachimi}, {Mandelbaum}, {Abdalla}, \&
  {Bridle}}]{2011A&A...527A..26J}
{Joachimi}, B., {Mandelbaum}, R., {Abdalla}, F.~B., \& {Bridle}, S.~L. 2011,
  \aap, 527, A26

\bibitem[{{Joachimi} \& {Schneider}(2008)}]{2008A&A...488..829J}
{Joachimi}, B. \& {Schneider}, P. 2008, \aap, 488, 829

\bibitem[{{Kaiser}(1992)}]{1992ApJ...388..272K}
{Kaiser}, N. 1992, \apj, 388, 272

\bibitem[{{Kamada} {et~al.}(2016){Kamada}, {Inoue}, \&
  {Takahashi}}]{2016PhRvD..94b3522K}
{Kamada}, A., {Inoue}, K.~T., \& {Takahashi}, T. 2016, \prd, 94, 023522

\bibitem[{{Kilbinger}(2015)}]{2015RPPh...78h6901K}
{Kilbinger}, M. 2015, Reports on Progress in Physics, 78, 086901

\bibitem[{{Kilbinger} {et~al.}(2017){Kilbinger}, {Heymans}, {Asgari},
  {Joudaki}, {Schneider}, {Simon}, {Van Waerbeke}, {Harnois-D{\'e}raps},
  {Hildebrandt}, {K{\"o}hlinger}, {Kuijken}, \& {Viola}}]{2017MNRAS.472.2126K}
{Kilbinger}, M., {Heymans}, C., {Asgari}, M., {et~al.} 2017, \mnras, 472, 2126

\bibitem[{{Kuijken}(2011)}]{2011Msngr.146....8K}
{Kuijken}, K. 2011, The Messenger, 146, 8

\bibitem[{{Kuijken} {et~al.}(2019){Kuijken}, {Heymans}, {Dvornik},
  {Hildebrandt}, {de Jong}, {Wright}, {Erben}, {Bilicki}, {Giblin}, {Shan},
  {Getman}, {Grado}, {Hoekstra}, {Miller}, {Napolitano}, {Paolilo}, {Radovich},
  {Schneider}, {Sutherland}, {Tewes}, {Tortora}, {Valentijn}, \& {Verdoes
  Kleijn}}]{2019A&A...625A...2K}
{Kuijken}, K., {Heymans}, C., {Dvornik}, A., {et~al.} 2019, \aap, 625, A2

\bibitem[{{Kuijken} {et~al.}(2015){Kuijken}, {Heymans}, {Hildebrandt},
  {Nakajima}, {Erben}, {de Jong}, {Viola}, {Choi}, {Hoekstra}, {Miller}, {van
  Uitert}, {Amon}, {Blake}, {Brouwer}, {Buddendiek}, {Conti}, {Eriksen},
  {Grado}, {Harnois-D{\'e}raps}, {Helmich}, {Herbonnet}, {Irisarri},
  {Kitching}, {Klaes}, {La Barbera}, {Napolitano}, {Radovich}, {Schneider},
  {Sif{\'o}n}, {Sikkema}, {Simon}, {Tudorica}, {Valentijn}, {Verdoes Kleijn},
  \& {van Waerbeke}}]{2015MNRAS.454.3500K}
{Kuijken}, K., {Heymans}, C., {Hildebrandt}, H., {et~al.} 2015, \mnras, 454,
  3500

\bibitem[{{Lagu{\"e}} {et~al.}(2024){Lagu{\"e}}, {Schwabe}, {Hlo{\v{z}}ek},
  {Marsh}, \& {Rogers}}]{2024PhRvD.109d3507L}
{Lagu{\"e}}, A., {Schwabe}, B., {Hlo{\v{z}}ek}, R., {Marsh}, D. J.~E., \&
  {Rogers}, K.~K. 2024, \prd, 109, 043507

\bibitem[{{Lamman} {et~al.}(2024){Lamman}, {Tsaprazi}, {Shi},
  {{\v{S}}ar{\v{c}}evi{\'c}}, {Pyne}, {Legnani}, \&
  {Ferreira}}]{2023arXiv230908605L}
{Lamman}, C., {Tsaprazi}, E., {Shi}, J., {et~al.} 2024, The Open Journal of
  Astrophysics, 7, 14

\bibitem[{{Li} {et~al.}(2023){Li}, {Zhang}, {Sugiyama}, {Dalal}, {Terasawa},
  {Rau}, {Mandelbaum}, {Takada}, {More}, {Strauss}, {Miyatake}, {Shirasaki},
  {Hamana}, {Oguri}, {Luo}, {Nishizawa}, {Takahashi}, {Nicola}, {Osato},
  {Kannawadi}, {Sunayama}, {Armstrong}, {Bosch}, {Komiyama}, {Lupton}, {Lust},
  {MacArthur}, {Miyazaki}, {Murayama}, {Nishimichi}, {Okura}, {Price}, {Tait},
  {Tanaka}, \& {Wang}}]{2023PhRvD.108l3518L}
{Li}, X., {Zhang}, T., {Sugiyama}, S., {et~al.} 2023, \prd, 108, 123518

\bibitem[{{Mauland} {et~al.}(2024){Mauland}, {Winther}, \&
  {Ruan}}]{2024A&A...685A.156M}
{Mauland}, R., {Winther}, H.~A., \& {Ruan}, C.-Z. 2024, \aap, 685, A156

\bibitem[{{Mead} {et~al.}(2021){Mead}, {Brieden}, {Tr{\"o}ster}, \&
  {Heymans}}]{2021MNRAS.502.1401M}
{Mead}, A.~J., {Brieden}, S., {Tr{\"o}ster}, T., \& {Heymans}, C. 2021, \mnras,
  502, 1401

\bibitem[{{Mead} {et~al.}(2015){Mead}, {Peacock}, {Heymans}, {Joudaki}, \&
  {Heavens}}]{2015MNRAS.454.1958M}
{Mead}, A.~J., {Peacock}, J.~A., {Heymans}, C., {Joudaki}, S., \& {Heavens},
  A.~F. 2015, \mnras, 454, 1958

\bibitem[{{Mezzetti} {et~al.}(2012){Mezzetti}, {Bonometto}, {Casarini}, \&
  {Murante}}]{2012JCAP...06..005M}
{Mezzetti}, M., {Bonometto}, S.~A., {Casarini}, L., \& {Murante}, G. 2012,
  JCAP, 2012, 005

\bibitem[{{Miller} {et~al.}(2013){Miller}, {Heymans}, {Kitching}, {van
  Waerbeke}, {Erben}, {Hildebrandt}, {Hoekstra}, {Mellier}, {Rowe}, {Coupon},
  {Dietrich}, {Fu}, {Harnois-D{\'e}raps}, {Hudson}, {Kilbinger}, {Kuijken},
  {Schrabback}, {Semboloni}, {Vafaei}, \& {Velander}}]{2013MNRAS.429.2858M}
{Miller}, L., {Heymans}, C., {Kitching}, T.~D., {et~al.} 2013, \mnras, 429,
  2858

\bibitem[{{Murata} \& {Takeuchi}(2022)}]{2022PASJ...74.1329M}
{Murata}, K. \& {Takeuchi}, T.~T. 2022, \pasj, 74, 1329

\bibitem[{{Pen} {et~al.}(2003){Pen}, {Lu}, {van Waerbeke}, \&
  {Mellier}}]{2003MNRAS.346..994P}
{Pen}, U.-L., {Lu}, T., {van Waerbeke}, L., \& {Mellier}, Y. 2003, \mnras, 346,
  994

\bibitem[{{Perez Sarmiento} {et~al.}(2025){Perez Sarmiento}, {Lagu{\"e}},
  {Madhavacheril}, {Jain}, \& {Sherwin}}]{2025arXiv250206687P}
{Perez Sarmiento}, K., {Lagu{\"e}}, A., {Madhavacheril}, M., {Jain}, B., \&
  {Sherwin}, B. 2025, arXiv:2502.06687

\bibitem[{{Planck Collaboration} {et~al.}(2020){Planck Collaboration},
  {Aghanim}, {Akrami}, {Ashdown}, {Aumont}, {Baccigalupi}, {Ballardini},
  {Banday}, {Barreiro}, {Bartolo}, {Basak}, {Battye}, {Benabed}, {Bernard},
  {Bersanelli}, {Bielewicz}, {Bock}, {Bond}, {Borrill}, {Bouchet}, {Boulanger},
  {Bucher}, {Burigana}, {Butler}, {Calabrese}, {Cardoso}, {Carron},
  {Challinor}, {Chiang}, {Chluba}, {Colombo}, {Combet}, {Contreras}, {Crill},
  {Cuttaia}, {de Bernardis}, {de Zotti}, {Delabrouille}, {Delouis}, {Di
  Valentino}, {Diego}, {Dor{\'e}}, {Douspis}, {Ducout}, {Dupac}, {Dusini},
  {Efstathiou}, {Elsner}, {En{\ss}lin}, {Eriksen}, {Fantaye}, {Farhang},
  {Fergusson}, {Fernandez-Cobos}, {Finelli}, {Forastieri}, {Frailis},
  {Fraisse}, {Franceschi}, {Frolov}, {Galeotta}, {Galli}, {Ganga},
  {G{\'e}nova-Santos}, {Gerbino}, {Ghosh}, {Gonz{\'a}lez-Nuevo}, {G{\'o}rski},
  {Gratton}, {Gruppuso}, {Gudmundsson}, {Hamann}, {Handley}, {Hansen},
  {Herranz}, {Hildebrandt}, {Hivon}, {Huang}, {Jaffe}, {Jones}, {Karakci},
  {Keih{\"a}nen}, {Keskitalo}, {Kiiveri}, {Kim}, {Kisner}, {Knox},
  {Krachmalnicoff}, {Kunz}, {Kurki-Suonio}, {Lagache}, {Lamarre}, {Lasenby},
  {Lattanzi}, {Lawrence}, {Le Jeune}, {Lemos}, {Lesgourgues}, {Levrier},
  {Lewis}, {Liguori}, {Lilje}, {Lilley}, {Lindholm}, {L{\'o}pez-Caniego},
  {Lubin}, {Ma}, {Mac{\'\i}as-P{\'e}rez}, {Maggio}, {Maino}, {Mandolesi},
  {Mangilli}, {Marcos-Caballero}, {Maris}, {Martin}, {Martinelli},
  {Mart{\'\i}nez-Gonz{\'a}lez}, {Matarrese}, {Mauri}, {McEwen}, {Meinhold},
  {Melchiorri}, {Mennella}, {Migliaccio}, {Millea}, {Mitra},
  {Miville-Desch{\^e}nes}, {Molinari}, {Montier}, {Morgante}, {Moss}, {Natoli},
  {N{\o}rgaard-Nielsen}, {Pagano}, {Paoletti}, {Partridge}, {Patanchon},
  {Peiris}, {Perrotta}, {Pettorino}, {Piacentini}, {Polastri}, {Polenta},
  {Puget}, {Rachen}, {Reinecke}, {Remazeilles}, {Renzi}, {Rocha}, {Rosset},
  {Roudier}, {Rubi{\~n}o-Mart{\'\i}n}, {Ruiz-Granados}, {Salvati}, {Sandri},
  {Savelainen}, {Scott}, {Shellard}, {Sirignano}, {Sirri}, {Spencer},
  {Sunyaev}, {Suur-Uski}, {Tauber}, {Tavagnacco}, {Tenti}, {Toffolatti},
  {Tomasi}, {Trombetti}, {Valenziano}, {Valiviita}, {Van Tent}, {Vibert},
  {Vielva}, {Villa}, {Vittorio}, {Wandelt}, {Wehus}, {White}, {White},
  {Zacchei}, \& {Zonca}}]{2020A&A...641A...6P}
{Planck Collaboration}, {Aghanim}, N., {Akrami}, Y., {et~al.} 2020, \aap, 641,
  A6

\bibitem[{Pranjal {et~al.}(2025)Pranjal, {Krause}, {Dolag}, {Benabed},
  {Eifler}, {Ay{\c{c}}oberry}, \& {Dubois}}]{2024arXiv241021980P}
Pranjal, R., {Krause}, E., {Dolag}, K., {et~al.} 2025, \jcap, 2025, 041

\bibitem[{{Preston} {et~al.}(2023){Preston}, {Amon}, \&
  {Efstathiou}}]{2023MNRAS.525.5554P}
{Preston}, C., {Amon}, A., \& {Efstathiou}, G. 2023, \mnras, 525, 5554

\bibitem[{{Preston} {et~al.}(2024){Preston}, {Amon}, \&
  {Efstathiou}}]{2024arXiv240418240P}
{Preston}, C., {Amon}, A., \& {Efstathiou}, G. 2024, \mnras, 533, 621

\bibitem[{{Salcido} {et~al.}(2023){Salcido}, {McCarthy}, {Kwan}, {Upadhye}, \&
  {Font}}]{2023MNRAS.523.2247S}
{Salcido}, J., {McCarthy}, I.~G., {Kwan}, J., {Upadhye}, A., \& {Font}, A.~S.
  2023, \mnras, 523, 2247

\bibitem[{{Schaller} {et~al.}(2025){Schaller}, {Schaye}, {Kugel}, {Broxterman},
  \& {van Daalen}}]{2024arXiv241017109S}
{Schaller}, M., {Schaye}, J., {Kugel}, R., {Broxterman}, J.~C., \& {van
  Daalen}, M.~P. 2025, \mnras, in press, arXiv:2410.17109

\bibitem[{{Schneider} {et~al.}(2022){Schneider}, {Giri}, {Amodeo}, \&
  {Refregier}}]{2022MNRAS.514.3802S}
{Schneider}, A., {Giri}, S.~K., {Amodeo}, S., \& {Refregier}, A. 2022, \mnras,
  514, 3802

\bibitem[{{Schneider} \& {Teyssier}(2015)}]{2015JCAP...12..049S}
{Schneider}, A. \& {Teyssier}, R. 2015, JCAP, 2015, 049

\bibitem[{{Schneider}(2006)}]{2006glsw.conf..269S}
{Schneider}, P. 2006, in Saas-Fee Advanced Course 33: Gravitational Lensing:
  Strong, Weak and Micro, ed. G.~{Meylan}, P.~{Jetzer}, P.~{North},
  P.~{Schneider}, C.~S. {Kochanek}, \& J.~{Wambsganss}, 269--451

\bibitem[{{Schneider} {et~al.}(2010){Schneider}, {Eifler}, \&
  {Krause}}]{2010A&A...520A.116S}
{Schneider}, P., {Eifler}, T., \& {Krause}, E. 2010, \aap, 520, A116

\bibitem[{{Schneider} {et~al.}(2002){Schneider}, {van Waerbeke}, {Kilbinger},
  \& {Mellier}}]{2002A&A...396....1S}
{Schneider}, P., {van Waerbeke}, L., {Kilbinger}, M., \& {Mellier}, Y. 2002,
  \aap, 396, 1

\bibitem[{Secco {et~al.}(2022)Secco, Samuroff, Krause, Jain, Blazek, Raveri,
  Campos, Amon, Chen, Doux, Choi, Gruen, Bernstein, Chang, DeRose, Myles,
  Fert\'e, Lemos, Huterer, Prat, Troxel, MacCrann, Liddle, Kacprzak, Fang,
  S\'anchez, Pandey, Dodelson, Chintalapati, Hoffmann, Alarcon, Alves,
  Andrade-Oliveira, Baxter, Bechtol, Becker, Brandao-Souza, Camacho,
  Carnero~Rosell, Carrasco~Kind, Cawthon, Cordero, Crocce, Davis, Di~Valentino,
  Drlica-Wagner, Eckert, Eifler, Elidaiana, Elsner, Elvin-Poole, Everett,
  Fosalba, Friedrich, Gatti, Giannini, Gruendl, Harrison, Hartley, Herner,
  Huang, Huff, Jarvis, Jeffrey, Kuropatkin, Leget, Muir, Mccullough,
  Navarro~Alsina, Omori, Park, Porredon, Rollins, Roodman, Rosenfeld, Ross,
  Rykoff, Sanchez, Sevilla-Noarbe, Sheldon, Shin, Troja, Tutusaus, Varga,
  Weaverdyck, Wechsler, Yanny, Yin, Zhang, Zuntz, Abbott, Aguena, Allam, Annis,
  Bacon, Bertin, Bhargava, Bridle, Brooks, Buckley-Geer, Burke, Carretero,
  Costanzi, da~Costa, De~Vicente, Diehl, Dietrich, Doel, Ferrero, Flaugher,
  Frieman, Garc\'{\i}a-Bellido, Gaztanaga, Gerdes, Giannantonio, Gschwend,
  Gutierrez, Hinton, Hollowood, Honscheid, Hoyle, James, Jeltema, Kuehn, Lahav,
  Lima, Lin, Maia, Marshall, Martini, Melchior, Menanteau, Miquel, Mohr,
  Morgan, Ogando, Palmese, Paz-Chinch\'on, Petravick, Pieres, Plazas~Malag\'on,
  Rodriguez-Monroy, Romer, Sanchez, Scarpine, Schubnell, Scolnic, Serrano,
  Smith, Soares-Santos, Suchyta, Swanson, Tarle, Thomas, \&
  To}]{PhysRevD.105.023515}
Secco, L.~F., Samuroff, S., Krause, E., {et~al.} 2022, Phys. Rev. D, 105,
  023515

\bibitem[{{Seljak}(1998)}]{1998ApJ...506...64S}
{Seljak}, U. 1998, \apj, 506, 64

\bibitem[{{Simon}(2012)}]{2012A&A...543A...2S}
{Simon}, P. 2012, \aap, 543, A2

\bibitem[{Simon \& Hilbert(2018)}]{Simon_2018}
Simon, P. \& Hilbert, S. 2018, \aap, 613, A15

\bibitem[{{Smith} \& {Markovic}(2011)}]{2011PhRvD..84f3507S}
{Smith}, R.~E. \& {Markovic}, K. 2011, \prd, 84, 063507

\bibitem[{{Smith} {et~al.}(2003){Smith}, {Peacock}, {Jenkins}, {White},
  {Frenk}, {Pearce}, {Thomas}, {Efstathiou}, \&
  {Couchman}}]{2003MNRAS.341.1311S}
{Smith}, R.~E., {Peacock}, J.~A., {Jenkins}, A., {et~al.} 2003, \mnras, 341,
  1311

\bibitem[{{Spergel} {et~al.}(2015){Spergel}, {Gehrels}, {Baltay}, {Bennett},
  {Breckinridge}, {Donahue}, {Dressler}, {Gaudi}, {Greene}, {Guyon}, {Hirata},
  {Kalirai}, {Kasdin}, {Macintosh}, {Moos}, {Perlmutter}, {Postman},
  {Rauscher}, {Rhodes}, {Wang}, {Weinberg}, {Benford}, {Hudson}, {Jeong},
  {Mellier}, {Traub}, {Yamada}, {Capak}, {Colbert}, {Masters}, {Penny},
  {Savransky}, {Stern}, {Zimmerman}, {Barry}, {Bartusek}, {Carpenter}, {Cheng},
  {Content}, {Dekens}, {Demers}, {Grady}, {Jackson}, {Kuan}, {Kruk}, {Melton},
  {Nemati}, {Parvin}, {Poberezhskiy}, {Peddie}, {Ruffa}, {Wallace}, {Whipple},
  {Wollack}, \& {Zhao}}]{2015arXiv150303757S}
{Spergel}, D., {Gehrels}, N., {Baltay}, C., {et~al.} 2015, arXiv:1503.03757

\bibitem[{{Sugiyama}(1995)}]{1995ApJS..100..281S}
{Sugiyama}, N. 1995, \apjs, 100, 281

\bibitem[{{Takahashi} {et~al.}(2017){Takahashi}, {Hamana}, {Shirasaki},
  {Namikawa}, {Nishimichi}, {Osato}, \& {Shiroyama}}]{2017ApJ...850...24T}
{Takahashi}, R., {Hamana}, T., {Shirasaki}, M., {et~al.} 2017, \apj, 850, 24

\bibitem[{{Takahashi} {et~al.}(2012){Takahashi}, {Sato}, {Nishimichi},
  {Taruya}, \& {Oguri}}]{2012ApJ...761..152T}
{Takahashi}, R., {Sato}, M., {Nishimichi}, T., {Taruya}, A., \& {Oguri}, M.
  2012, \apj, 761, 152

\bibitem[{{Tegmark} \& {Zaldarriaga}(2002)}]{2002PhRvD..66j3508T}
{Tegmark}, M. \& {Zaldarriaga}, M. 2002, \prd, 66, 103508

\bibitem[{{The LSST Dark Energy Science Collaboration} {et~al.}(2018){The LSST
  Dark Energy Science Collaboration}, {Mandelbaum}, {Eifler}, {Hlo{\v{z}}ek},
  {Collett}, {Gawiser}, {Scolnic}, {Alonso}, {Awan}, {Biswas}, {Blazek},
  {Burchat}, {Chisari}, {Dell'Antonio}, {Digel}, {Frieman}, {Goldstein},
  {Hook}, {Ivezi{\'c}}, {Kahn}, {Kamath}, {Kirkby}, {Kitching}, {Krause},
  {Leget}, {Marshall}, {Meyers}, {Miyatake}, {Newman}, {Nichol}, {Rykoff},
  {Sanchez}, {Slosar}, {Sullivan}, \& {Troxel}}]{2018arXiv180901669T}
{The LSST Dark Energy Science Collaboration}, {Mandelbaum}, R., {Eifler}, T.,
  {et~al.} 2018, arXiv:1809.01669

\bibitem[{{Tr{\"o}ster} {et~al.}(2021){Tr{\"o}ster}, {Asgari}, {Blake},
  {Cataneo}, {Heymans}, {Hildebrandt}, {Joachimi}, {Lin}, {S{\'a}nchez},
  {Wright}, {Bilicki}, {Bose}, {Crocce}, {Dvornik}, {Erben}, {Giblin},
  {Glazebrook}, {Hoekstra}, {Joudaki}, {Kannawadi}, {K{\"o}hlinger}, {Kuijken},
  {Lidman}, {Lombriser}, {Mead}, {Parkinson}, {Shan}, {Wolf}, \&
  {Xia}}]{2021A&A...649A..88T}
{Tr{\"o}ster}, T., {Asgari}, M., {Blake}, C., {et~al.} 2021, \aap, 649, A88

\bibitem[{{Truttero} {et~al.}(2025){Truttero}, {Zuntz}, {Pourtsidou}, \&
  {Robertson}}]{2024arXiv241018191T}
{Truttero}, O., {Zuntz}, J., {Pourtsidou}, A., \& {Robertson}, N. 2025, The
  Open Journal of Astrophysics, 8, 19

\bibitem[{{Wright} {et~al.}(2020{\natexlab{a}}){Wright}, {Hildebrandt}, {van
  den Busch}, \& {Heymans}}]{2020A&A...637A.100W}
{Wright}, A.~H., {Hildebrandt}, H., {van den Busch}, J.~L., \& {Heymans}, C.
  2020{\natexlab{a}}, \aap, 637, A100

\bibitem[{{Wright} {et~al.}(2020{\natexlab{b}}){Wright}, {Hildebrandt}, {van
  den Busch}, {Heymans}, {Joachimi}, {Kannawadi}, \&
  {Kuijken}}]{2020A&A...640L..14W}
{Wright}, A.~H., {Hildebrandt}, H., {van den Busch}, J.~L., {et~al.}
  2020{\natexlab{b}}, \aap, 640, L14

\bibitem[{{Ye} {et~al.}(2024){Ye}, {Jiang}, \&
  {Silvestri}}]{2024arXiv241107082Y}
{Ye}, G., {Jiang}, J.-Q., \& {Silvestri}, A. 2024, arXiv:2411.07082

\bibitem[{Yoshida(1990)}]{YOSHIDA1990262}
Yoshida, H. 1990, Physics Letters A, 150, 262

\end{thebibliography}

\begin{appendix}

\section{Hamiltonian Monte Carlo algorithm and code verification}
\label{ap:mcmc}

The Bayesian reconstruction of the 3D matter power spectrum, expressed
by $\vec{\pi}=(\ldots,f_{\delta,mn},\ldots)$, is numerically
challenging due to the high number of variables, up to
\mbox{$N_z\times N_k=60$} or more in future Stage IV applications. As
practical solution, the following describes our implementation details
of a Hamiltonian Monte Carlo sampler of the posterior PDF
$P(\vec{\pi}|\vec{d})$, as well as our verification and convergence
tests. Compared to a Metropolis-Hastings algorithm, this MCMC approach
has an improved convergence and quicker decorrelation -- but requires
first-order derivatives of the posterior PDF. The derivatives were
here easily obtained analytically, contrary to previous KiDS-1000
cosmology analyses where other MCMC algorithms are used. We focus on
the specifics of our implementation but refer to
\cite{rooks2011handbook} for more background information.

\subsection{Basic concept}

In brief, the Hamiltonian MCMC algorithm is based on the insight from
statistical mechanics that a canonical ensemble of particles
\mbox{$j=1\ldots N_z\times N_k$} with generalised coordinates $\pi_j$,
masses $m_j$, canonical moments $p_j$, the Hamiltonian
\begin{equation}
  H(\vec{\pi},\vec{p})=
  U(\vec{\pi})+\sum_j\frac{p_j^2}{2m_j}\;,
\end{equation}
and temperature parameter $\beta$ randomly occupies states in phase
space according to the PDF
\begin{multline}
  f(\vec{\pi},\vec{p})=
  \frac{1}{Z}\exp{\Big(-\beta\,H[\vec{\pi},\vec{p}]\Big)}\\
  =\frac{1}{Z}\, \exp{\Big(-\beta\,U[\vec{\pi}]\Big)}
  \,\exp{\left(-\beta\,\sum_j\frac{p_j^2}{2m_j}\right)}\;,
\end{multline}
where the partition function, $Z$, is a normalisation
constant. Turning this insight around, our sampler simulates a
canonical ensemble following individual particles along trajectories
under the influence of the potential
\mbox{$U(\vec{\pi})=-\ln{P(\vec{\pi}|\vec{d})}$} to trace the
posterior PDF, $P(\vec{\pi}|\vec{d})$. The trajectory of each particle
is determined by the symplectic Hamilton equations
\begin{eqnarray}
 \nonumber
  \frac{\d\pi_j}{\d t}&=&+\frac{\partial H(\vec{\pi},\vec{p})}{\partial p_j}
                          =\frac{\beta\,p_j}{m_j}\;;
  \\  \label{eq:hameq}
  \frac{\d p_j}{\d t}&=&-\frac{\partial H(\vec{\pi},\vec{p})}{\partial\pi_j}
  =-\beta\,\frac{\partial U(\vec{\pi})}{\partial\pi_j} \;,
\end{eqnarray}
interrupted by random jumps between hyper-planes of constant energy,
\mbox{$E(\vec{\pi},\vec{p})=H(\vec{\pi},\vec{p})=\rm const$}, such
that energy levels change from $E$ to $E^\prime$ with probability
$\min\{1,\e^{\beta (E-E^\prime)}\}$. These random events simulate the
transfer of energy between particles and a heat bath of temperature
\mbox{$T\propto\beta^{-1}$}. Ignoring the moments $p_j$, the
distribution of states $\vec{\pi}_i$ of an ensemble of particles then
samples the marginalised PDF
\begin{equation}
  \int_{V_{\vec{p}}}\d\vec{p}\;f(\vec{\pi},\vec{p})\propto
  \exp{\Big(-\beta\,U[\vec{\pi}]\Big)}=
  \left[P(\vec{\pi}|\vec{d})\right]^\beta\;.
\end{equation}
This does not mean, however, moments can be chosen
arbitrarily. Instead, the $p_j$ obey the Gaussian distribution of a
maximum-entropy ensemble in $f(\vec{\pi},\vec{p})$ and, therefore, are
randomly set to $p_j\sim{\cal N}(0,\sigma_p)$, where
$\sigma_p^2=m_j/\beta$, in the algorithm below. For convenience, we
set the heat bath temperature to \mbox{$\beta\equiv1$}, thus the
ensemble $\vec{\pi}_i$ samples the posterior $P(\vec{\pi}|\vec{d})$,
as needed.

\subsection{Implementation details}

The art of optimising the Hamiltonian MCMC is in the way we evolve
states in the canonical ensemble, how we switch between energy levels,
which sampling points are kept, and how particle masses, $m_j$, are
chosen. For our application, we found sufficient convergence already
when using a simplistic model of equal mass weights,
\mbox{$m_j\equiv1$}. Furthermore, to produce a sequence of sampling
points $\vec{\pi}_i$, we followed the MCMC standard procedure.
\begin{enumerate}

\item We pick a random starting point, $\vec{\pi}$, $i=0$.
  
\item The elements of the momentum in $\vec{p}$, conjugate to
  $\vec{\pi}$, are randomised independently,
  \mbox{$p_j\sim {\cal N}(0,\sigma_p)$}.

\item The algorithm then numerically integrates the Hamilton
  Eqs. \Ref{eq:hameq} with a symplectic integrator to follow the
  trajectory, starting from $(\vec{\pi},\vec{p})$, for $n$ equal time
  steps $\Delta t$, proposing the end point as a new MCMC point,
  $(\vec{\pi}_{n\Delta t},\vec{p}_{n\Delta t})$.

\item This proposal $\vec{\pi}_{n\Delta t}$ is accepted with
  probability \mbox{$\min\{1,\e^{\beta (E-E_{n\Delta t})}\}$}, where
  \mbox{$E:=H(\vec{\pi},\vec{p})$} and
  \mbox{$E_{n\Delta t}:=H(\vec{\pi}_{n\Delta t},\vec{p}_{n\Delta t})$}
  are the energies of the initial and the proposed new state. If
  rejected, we go back to the initial trajectory point, $\vec{\pi}$,
  and start again from step 2.

\item If accepted, we keep the proposed state as MCMC data point,
  $\vec{\pi}_i=\vec{\pi}_{n\Delta t}$. Since rejections mean we have
  to go back and reuse a sampling point, the stored $\vec{\pi}_i$ has
  weight $w_i=1+n_{{\rm r},i}$, where $n_{{\rm r},i}$ denotes the
  number of rejections needed for $\vec{\pi}_i$.

\item We stop after $n_{\rm mcmc}$ accepted sampling points
  $(\vec{\pi}_i,w_i)$, or start over again from step 2 with
  $\vec{\pi}=\vec{\pi}_i$ and $i\to i+1$ otherwise.

\end{enumerate}

In contrast to Metropolis-Hastings, the Hamiltonian method proposes a
new MCMC point by (approximately) transporting the previous point
along the iso-contours of the posterior density, keeping the rejection
rate low even if the previous and proposed point are well separated,
thereby decorrelating sampling points. Notably, this way of proposing
new points is symmetric due to the time symmetry of the Hamilton
equations and the isotropy of the randomised momenta, satisfying the
conditions of a detailed balance. In addition, as long as the time
symmetry is preserved, unavoidable inaccuracies when numerically
integrating Eqs. \Ref{eq:hameq} do not bias the sampling, although
they usually increase rejection rates when moving too far away from
the posterior density iso-contours. As discussed in
\cite{rooks2011handbook}, the leapfrog method is a suitable choice as
numerical integrator of \Ref{eq:hameq} for a MCMC sampler due to its
symplectic symmetry. To increase the acceptance rate for larger time
steps $\Delta t$, we use the more accurate fourth-order, also
symplectic, integrator. The various integrator steps are not spelt out
here in detail to save space but can be found in
\cite{YOSHIDA1990262}.

To control the rejection rate of the MCMC sampler, we adjusted the
size of a time step, $\Delta t$. Decreasing the time step, $\Delta t$,
typically lowers the rejection rate, but boosts correlations between
the sampling points. Starting from $\Delta t=10^{-3}$, we followed the
heuristic to keep the rejection rate somewhere around $30\%$ by (i)
decreasing \mbox{$\Delta t\to\frac{\Delta t}{1.2}$}, if the rate
exceeds $50\%$, or (ii) by increasing
\mbox{$\Delta t\to1.2\,\Delta t$}, if it falls below $16\%$, after
$20$ consecutive accepted sampling points. At the same time, the
number of trajectory steps was fixed to \mbox{$n=100$}. With these
settings, the time steps stabilise typically around
\mbox{$\Delta t=10^{-2}$}. As final data product, we kept
\mbox{$n_{\rm mcmc}=5\times10^4$} sampling points after a burn-in
phase of $2\times10^4$ steps during which the step size was
continuously adjusted. For all chains, we started the burn-in with the
initial state drawn from a normal distribution centred around
\mbox{$f_{\delta,mn}=1$}, \mbox{$f_{\delta,mn}\sim {\cal N}(1,0.5)$}.

To compute the movement along a trajectory by the Hamilton equations,
we employed an analytic expression for the gradient
\begin{multline}
  \nabla_{\vec{\pi}}U(\vec{\pi})\\
  =-\nabla_{\vec{\pi}}\,\ln{{\cal L}_{\vec{q}}(\vec{d}|\vec{\pi})}
  -\nabla_{\vec{\pi}}\,\ln{P_{\rm hat}(\vec{\pi})}
  -\nabla_{\vec{\pi}}\,\ln{P_\tau(\vec{\pi})}
\end{multline}
which, for Eq. \Ref{eqn:posterior}, is split into three summands: one
for the likelihood,
\begin{equation}
  -\nabla_{\vec{\pi}}\ln{{\cal L}_{\vec{q}}(\vec{d}|\vec{\pi})}=
  \left(\mat{C}^{-1}\mat{X}_{\vec{q}}\right)^{\rm T}\,
  \left(\mat{X}_{\vec{q}}\vec{\pi}+\vec{\xi}^{\rm fid}_{\vec{q}}-\vec{d}\right)
  \,\;,
\end{equation}
one for the top-hat priors,
\begin{multline}
  -\frac{\partial\ln{P_{\rm hat}(\vec{\pi})}}{\partial f_{\delta,mn}}\\
 = \left\{\begin{array}{ll}
           2\sigma_{\rm f}^{-2}\,f_{\delta,mn} & {,\rm if~}f_{\delta,mn}\le0\\
           2\sigma_{\rm f}^{-2}\,\left(f_{\delta,mn}-f_{\delta,\rm max}\right) & {,\rm if~}f_{\delta,mn}>f_{\delta,\rm max}
         \end{array}\right.\;,
\end{multline}
and one for the Tikhonov regularisation,
\begin{multline}
  -\frac{\partial\ln{P_\tau(\vec{\pi})}}{\partial f_{\delta,mn}}\\
 = \left\{\begin{array}{ll}
           2\tau\,\left(f_{\delta,1n}-f_{\delta,2n}\right) & {,\rm if~}m=1\\
           2\tau\,\left(f_{\delta,N_kn}-f_{\delta,(N_k-1)n}\right) & {,\rm if~}m=N_k\\
           2\tau\,\left(2f_{\delta,mn}-f_{\delta,(m-1)n}-f_{\delta,(m+1)n}\right) & {,\rm if~}1<m<N_k
         \end{array}\right.\;.
\end{multline}
The computation time of the sampler was reduced by computing the
matrix $(\mat{C}^{-1}\mat{X}_{\vec{q}})^{\rm T}$ just once and then
reusing it for fixed projection parameters $\vec{q}$.

\begin{figure}
  \begin{center}
    \includegraphics[width=95mm]{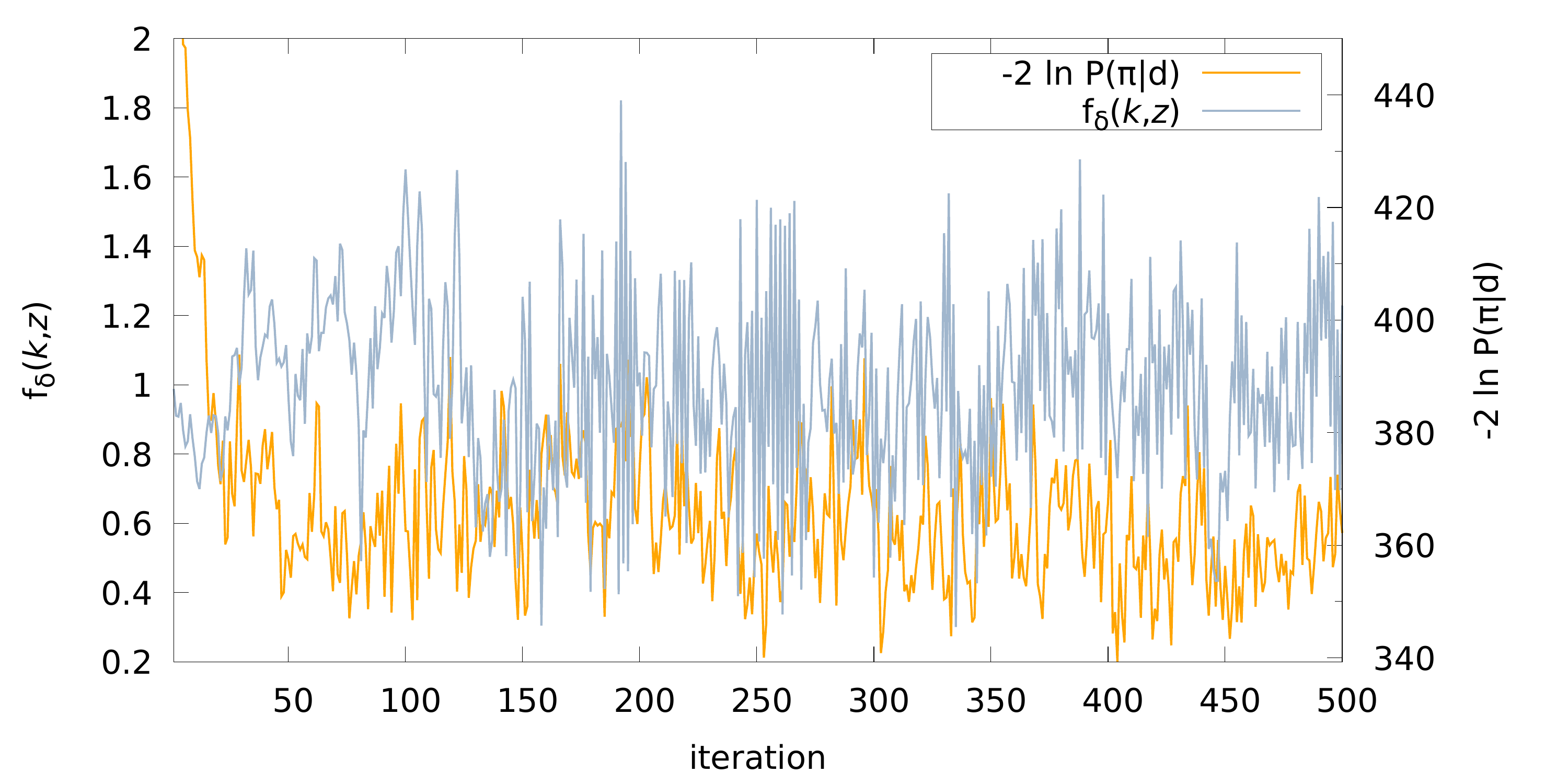}    
    \includegraphics[width=95mm]{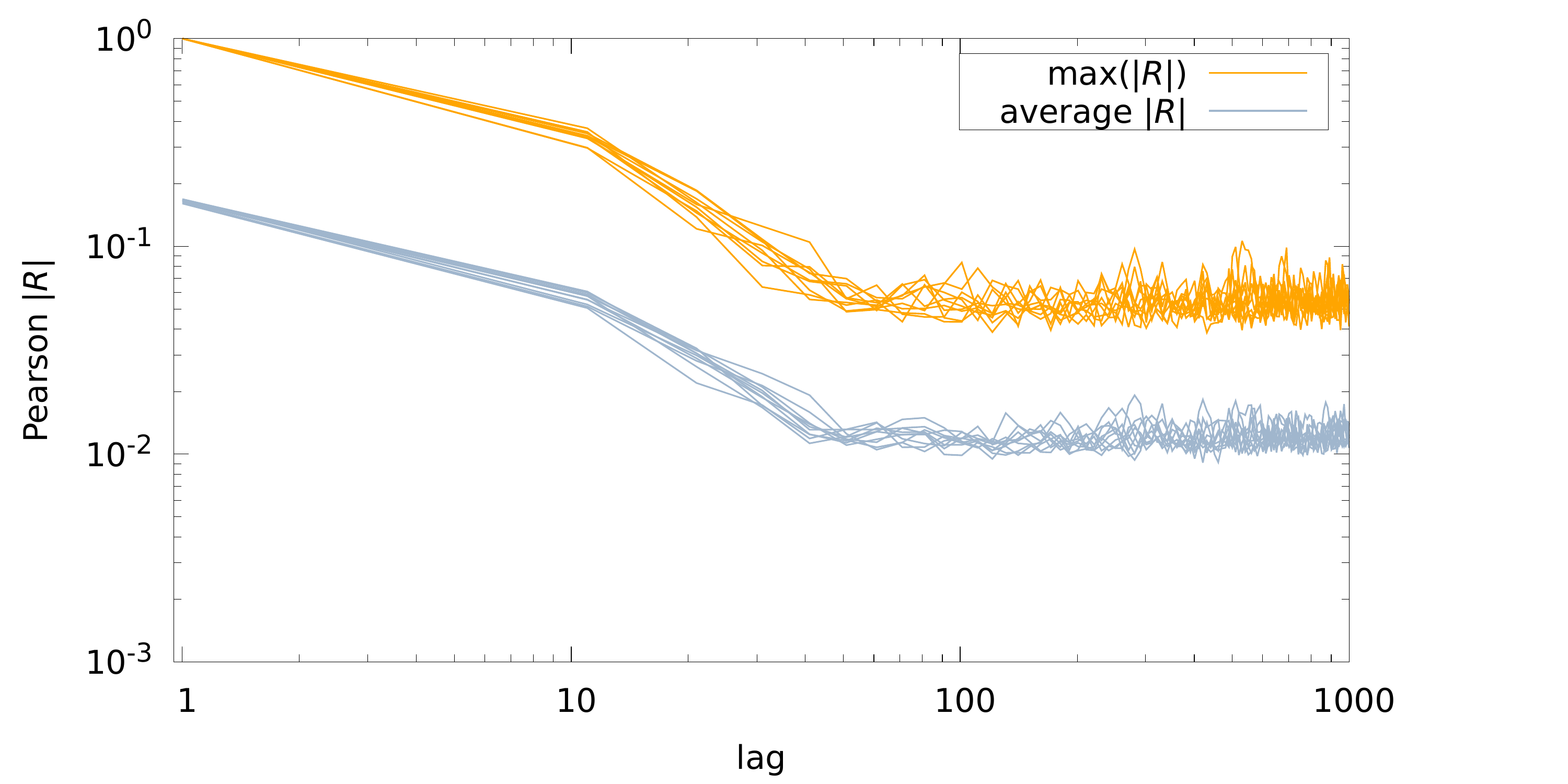}
  \end{center}
  \caption{\label{fig:sampleval} Diagnostic plots of the Hamiltonian
    MCMC sampler with KiDS-1000 data and our final analysis set-up for
    fixed projection parameters as in Table
    \ref{tab:fiducialmodel}. \emph{Top panel:} Example of
    $f_{\delta,mn}$ variations along one random chain at the beginning
    of the burn-in, starting from the random initial point $i=0$ (blue
    line). The orange line depicts the (negative log-) posterior
    probability $-2\ln{P(\vec{\pi}_i|\vec{d})}$. \emph{Bottom panel:}
    Decrease in the correlation of $f_{\delta,mn}$ and $f_{\delta,kl}$
    with $\delta i$ iterations lag as estimated in ten randomly
    chosen, independent chains. The orange lines follows, one line per
    chain, the maximum absolute-value Pearson correlation coefficient
    $|R_{mn,kl}|$, while the blue lines follows the average
    $|R_{mn,kl}|$ over all variable combinations $(mn,kl)$. The
    sampling points quickly decorrelate after a few hundred
    iterations; the flattening after lag $\delta i\sim100$ is due to
    the statistical error in the estimated $|R|$ for chains of finite
    length $n_{\rm mcmc}=5\times10^4$.}
\end{figure}

\subsection{Verification and convergence tests}

To validate the convergence of the Monte Carlo sampler, we carried out
three different tests using the KiDS-1000 data vector for the
projection parameters $\vec{q}$ as in Table
\ref{tab:fiducialmodel}. The set-up is identical to that in our
analysis in Sect. \ref{sect:threezbins} for the most complex model
with three redshift bins, \mbox{$N_z=3$}. The first test is a sanity
check to verify by eye the transition towards a stationary process
during the burn-in phase. For this, we inspected as function of
iteration $i$ the variation of $f_{\delta,mn}$ and the (negative log-)
probability $-2\ln{P(\vec{\pi}|d)}$ along the chain. The top panel of
Fig. \ref{fig:sampleval} is one random example; other chains and
variables $f_{\delta,mn}$ in $\vec{\pi}_i$ look qualitatively similar
and are not shown here. Apart from the deep drop of $-2\ln{P}$ at the
very beginning of the chain, where the sampler takes the points away
from the initial random $\vec{\pi}$, there are no more systematic
drifts present for $f_{\delta,mn}$ and $-2\ln{P}$. This indicates a
stationary sampling is quickly reached.

The second test, bottom panel of Fig. \ref{fig:sampleval}, quantifies
the decorrelation of sampling points, $\vec{\pi}_i$, along the
chain. For a fair sampling of the posterior density, the sampling
points have to become independent quickly. To quantify this, we looked
at the correlation coefficients $R_{mn,kl}$ in the covariance matrix
$\ave{\vec{\pi}_i\,\vec{\pi}^{\rm T}_{i+\delta i}}$ of two sampling
points $\vec{\pi}_i=(\ldots,f_{\delta,mn},\ldots)$ and
$\vec{\pi}_{i+\delta i}=(\ldots,f_{\delta,kl},\ldots)$ separated by a
lag of $\delta i$ iterations along the same chain. The expectation is
that the amplitude of $R_{mn,kl}$ between a pair of model coefficients
$f_{\delta,mn}$ and $f_{\delta,kl}$ decreases towards zero with
increasing lag $\delta i$. This is indeed the case, as shown in the
figure bottom panel for several independent chain runs. The orange
curves report for a series of ten MCMCs the trend of the maximum
$|R_{mn,kl}|$ for all coefficient combinations $(m,n)$ and $(k,l)$,
the blue curves are the average of all $|R_{mn,kl}|$. All chains
exhibit the same trend: the maximum $|R_{mn,kl}|$ quickly falls below
a few per cent already after $\delta i\sim100$, whereas the average
$|R_{mn,kl}|$ is even lower, reaching sub-per cent levels. After that,
all curves tend to flat out, probably due to statistical noise in our
$|R_{mn,kl}|$ estimates from the finite number of pairs of sampling
points in a given chain.

For the third convergence check, we applied the Gelman--Rubin
diagnostics to all coefficients $f_{\delta,mn}$ individually
\citep{1992StaSc...7..457G}. This test probes if a single chain has
sufficiently converged, by comparing our \mbox{$n_{\rm chain}=10$}
independent MCMC runs. Similar to the marginalisation process, Sect.
\ref{sect:errormodel}, we were not using the full chains but, instead,
randomly draw \mbox{$n_{\rm merge}=10^3$} points for a thinned each
chain. The Gelman--Rubin statistic
\mbox{$t_{\rm gr}:=\sqrt{\hat{V}/W}$} is based on the average
$\bar{f}_{\delta,mnk}$ and variance of $\sigma^2_{\delta,mnk}$ of
$f_{\delta,mn}$ within a (thinned) chain $k$, where
\mbox{$W:=\sum_k\sigma^2_{\delta,mnk}/n_{\rm chain}$} is the average
of all within-chain variances,
\mbox{$B/n_{\rm
    merge}:=\sum_k(\bar{f}_{\delta,mnk}-\bar{f}_{\delta,mn})^2/(n_{\rm
    chain}-1)$} for
\mbox{$\bar{f}_{\delta,mn}:=\sum_k\bar{f}_{\delta,mnk}/n_{\rm chain}$}
is the variance between chains, and
\mbox{$\hat{V}:=\frac{n_{\rm merge}-1}{n_{\rm merge}}\,W+\frac{n_{\rm
      chain}+1}{n_{\rm chain}}\,B/n_{\rm merge}$}. The ratio $B/W$
follows a $F$-distribution in the null hypothesis of independent,
stationary chains, and normally distributed $f_{\delta,mn}$; and
\mbox{$t_{\rm gr}\sim1$} is to high confidence
\mbox{$t_{\rm gr}<1.1$}. We repeatedly find \mbox{$t_{\rm gr}<1.001$}
for all our free model variables, $f_{\delta,mn}$, consistent with the
null hypothesis.

\begin{figure*}
  \begin{center}
    \includegraphics[width=60mm]{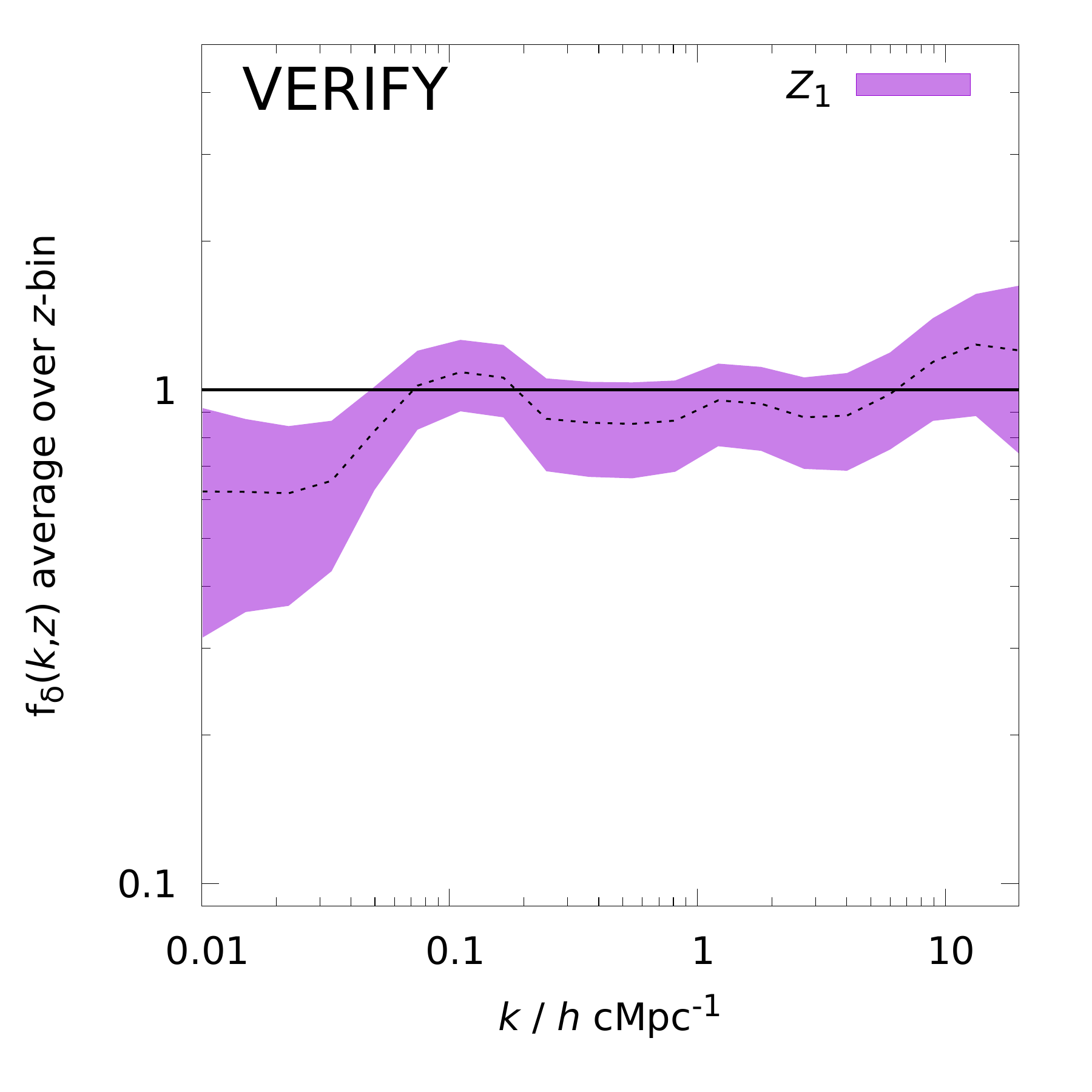}
    \includegraphics[width=60mm]{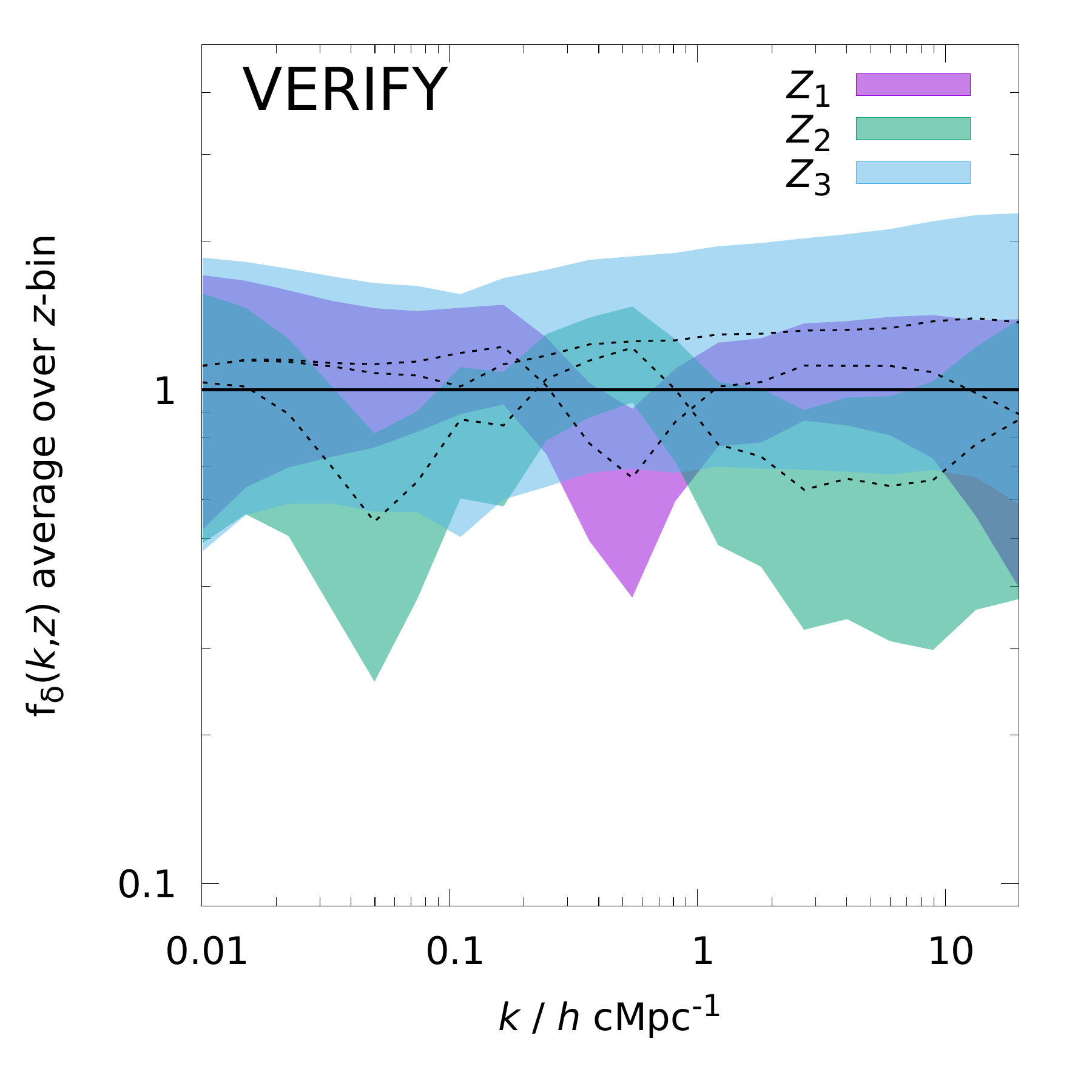}
    \includegraphics[width=60mm]{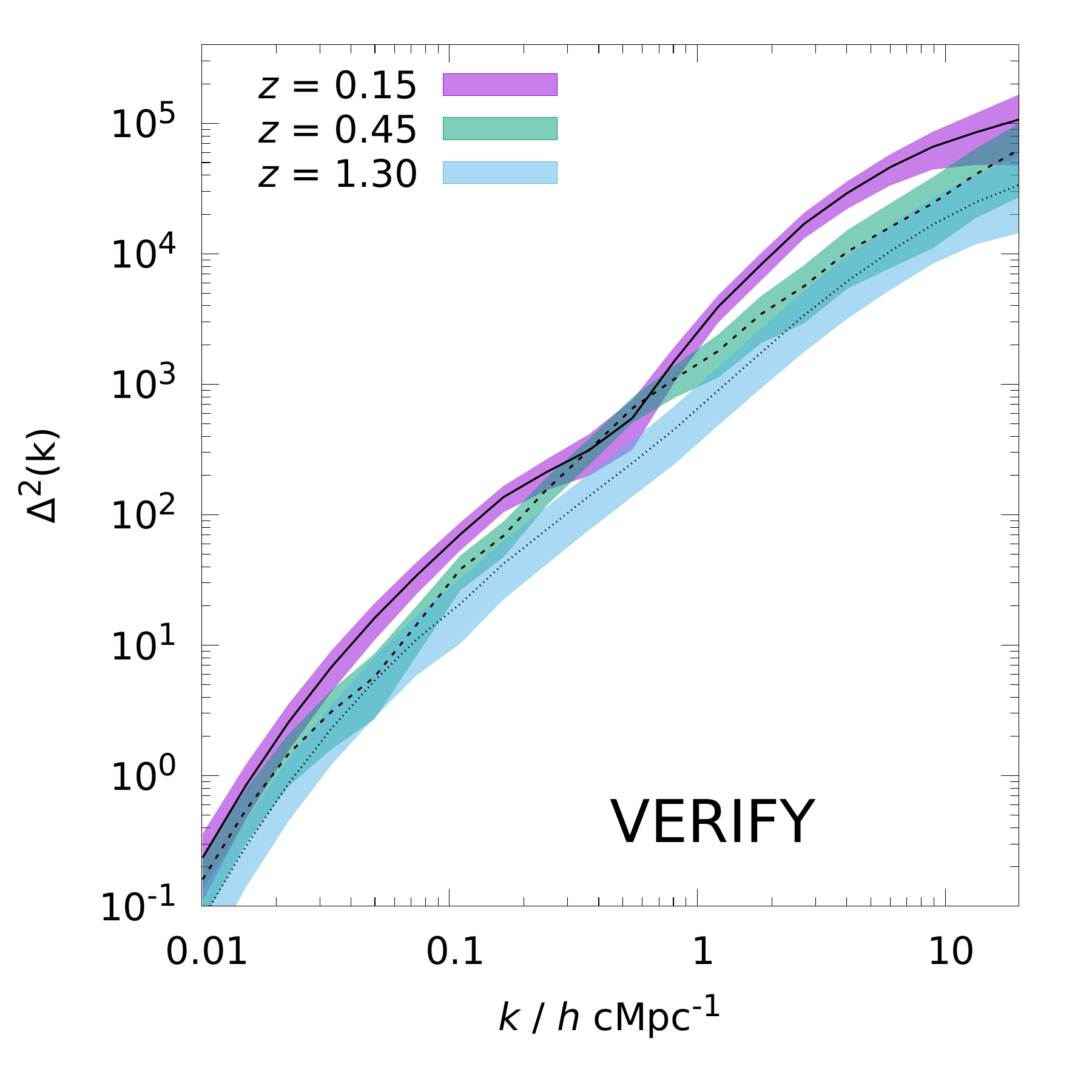}    
  \end{center}
  \caption{\label{fig:relpowerVer} Reconstructed relative power
    spectrum, $f_\delta(k,z)$, in redshift range \mbox{$0\le z<2$}
    using a verification data vector mimicking KiDS-1000 data (noise
    added, not marginalising IA or lensing kernel errors). The true
    relative power has $f_\delta\equiv1$ for all scales and
    redshift. The data vector is computed from an independent code,
    not based on the projection matrix $X^{(ij)}(\theta; m,n)$. Shown
    is the posterior median within a $68\%$ CI. The reconstructions
    use a Tikhonov parameter of $\tau=5.0$ and $N_k=20$
    $k$-bins. \emph{Left panel:} The relative power is averaged over
    the entire redshift range $Z_1=[0,2]$.  \emph{Middle panel:} For a
    relative power that is averaged independently for three redshift
    bins $Z_1=[0,0.3]$, $Z_2=[0.3,0.6]$, and
    $Z_3=[0.6,2]$. \emph{Right panel:} Dimensionless average power
    spectrum,
    $\Delta^2(k,z)=4\pi\,k^3P_\delta^{\rm fid}(k,z)\,f_\delta(k,z)$,
    interpolated to the centres of the redshift bins $Z_1$ to $Z_3$.}
\end{figure*}

For test reconstructions of the 3D matter power spectrum, Fig.
\ref{fig:relpowerVer} shows two successful verification runs of the
sampler, but now with simulated data, a model correlation vector,
$\vec{d}$, with added random noise according to the Gaussian
likelihood model of the KiDS-1000 data. The verification data assumed
a matter power spectrum identical to $P_\delta^{\rm fid}(k,z)$, thus
\mbox{$f_{\delta,mn}=1$} for all $m$ and $n$. In order to not rely on
our MCMC code and the $X^{(ij)}_\pm(\theta;m,n)$ expansion in
particular, we calculated the verification data vector
$\xi^{(ij)}_\pm(\theta)$, Eq. \Ref{eq:xipm}, with a separate computer
code from a past study \citep{Simon_2018}. The left panel in
Fig. \ref{fig:relpowerVer}, plots the results for $f_{\delta,mn}$
averaged over one redshift bin, \mbox{$N_z=1$}.  The middle panel,
uses another noise realisation and splits, as before, $f_{\delta,mn}$
into \mbox{$N_z=3$} broad redshift bins to probe the $z$-evolution. In
both panels, we find results consistent with the true values
\mbox{$f_{\delta,mn}=1$} on a $68\%$ CI level.

\begin{figure}
  \begin{center}
    \includegraphics[width=75mm]{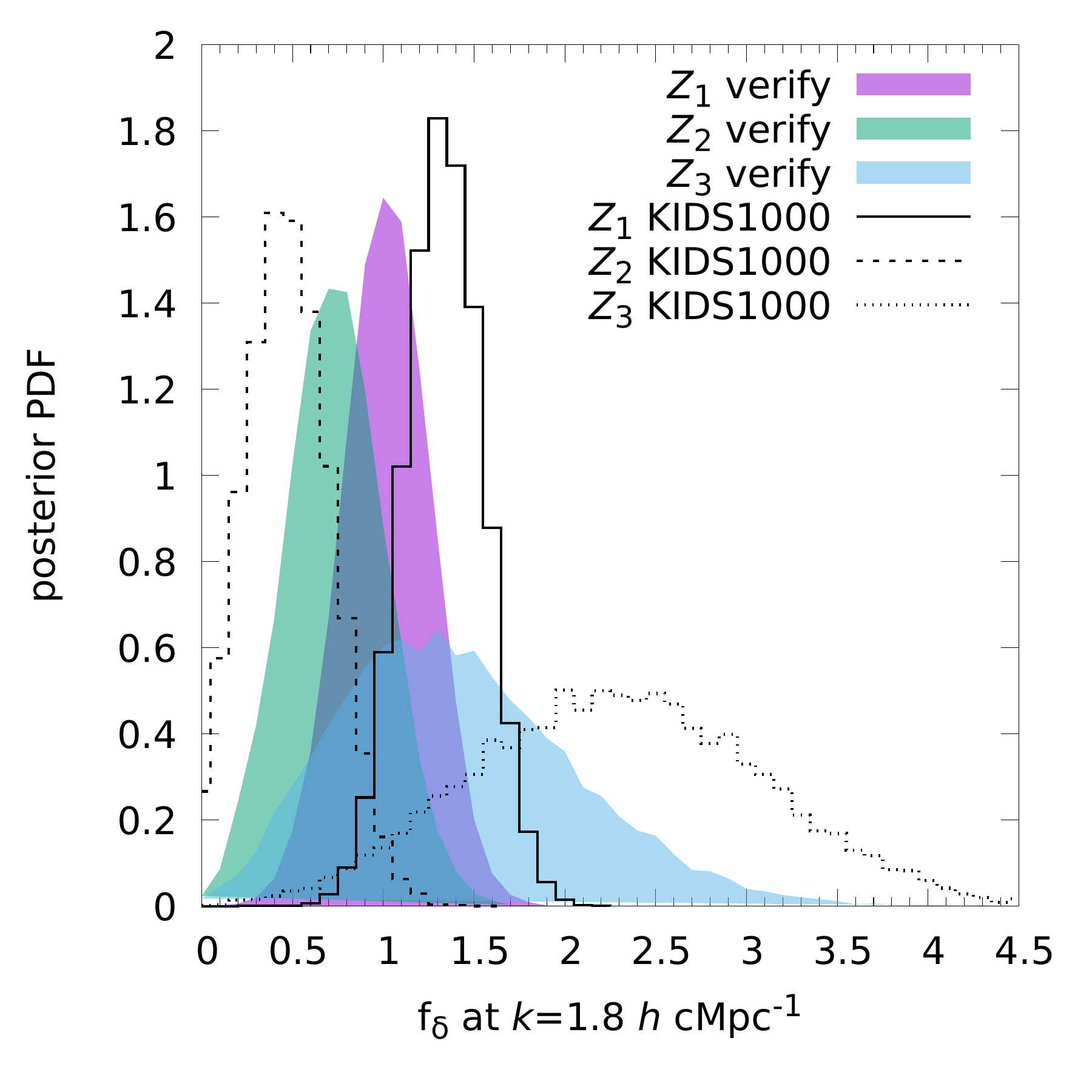}    
  \end{center}
  \caption{\label{fig:posteriorShape} Posterior PDF of $f_{\delta,mn}$
    at $k=1.8\,h\,\rm Mpc^{-1}$ in three separate redshift bins
    $Z_1=[0,0.3]$, $Z_2=[0.3,0.5]$, and $Z_3=[0.6,2]$. The filled
    regions are PDFs for a noisy verification run with
    $f_\delta(k,z)\equiv1$, as in Fig. \ref{fig:relpowerVer}, and the
    lines use the KiDS-1000 tomographic data (without marginalisation
    of projection parameters).}
\end{figure}

Increasing the number of $z$-bins decreases the constraining power for
each bin. This adds skewness to the posterior PDF of $f_{\delta,mn}$,
especially for $Z_3$, as illustrated by
Fig. \ref{fig:posteriorShape}. The figure shows the posterior PDF of
$f_{\delta,mn}$ in three $z$-bins at fixed comoving scale
\mbox{$k=1.8\,h\,\rm Mpc^{-1}$} for the $N_Z=3$ verification data
(shaded $68\%$ CI regions). Other scales look qualitatively
similar. Notably, the skewness systematically shifts our credible
regions towards the long tails in direction of higher $f_\delta$
values because we report $68\%$ intervals about the median posterior
value. This shift for the cyan $Z_3$ regions can also be seen in
random realisations of KiDS-1000 mock data, obtained by ray-tracing
$N$-body simulations, in Fig. \ref{fig:ths17relpower3}.

\begin{figure*}
  \begin{center}
    \includegraphics[width=185mm]{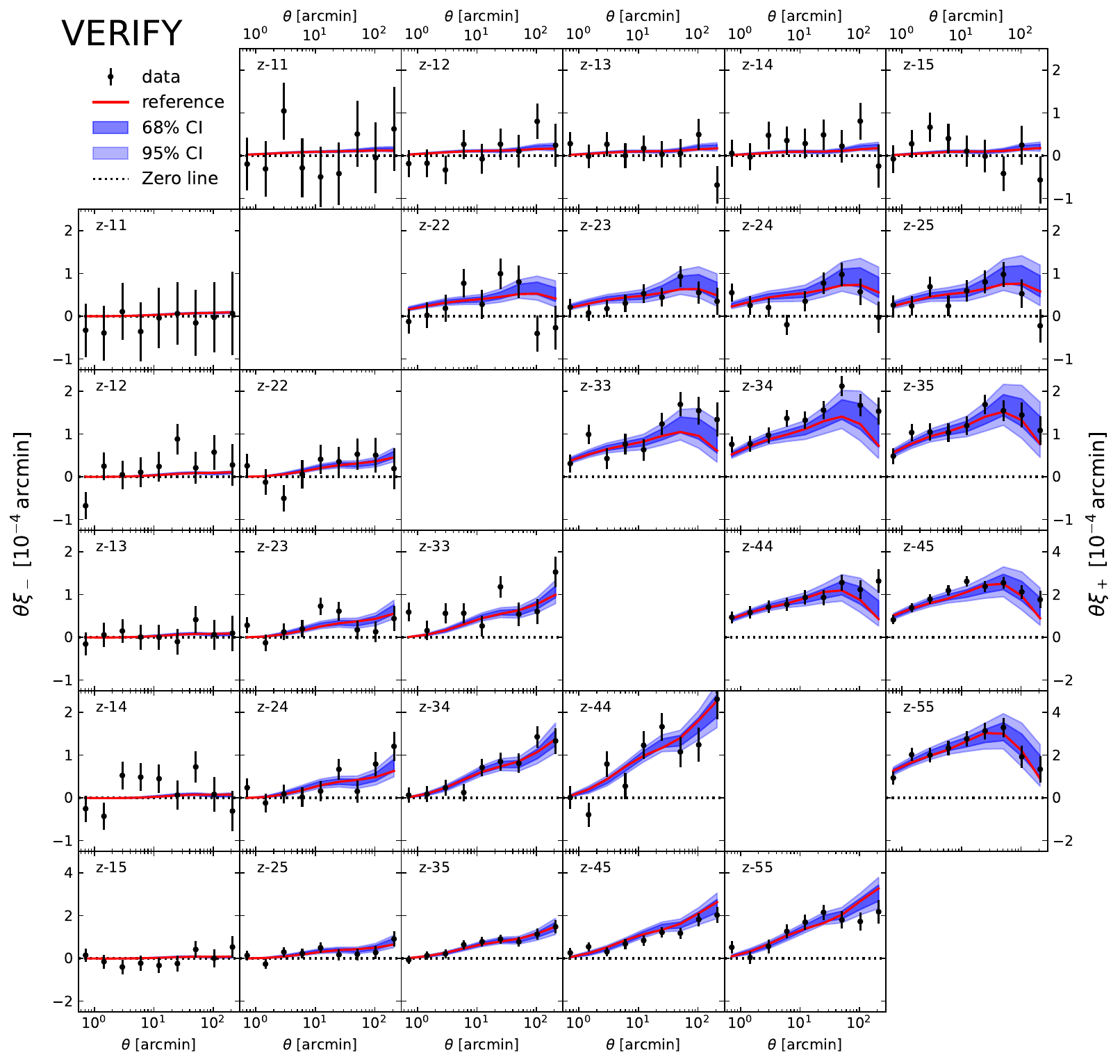}
  \end{center}
  \caption{\label{fig:postpredVer} Same as in
    Fig. \ref{fig:kidspostpred} but here for a verification data
    vector (black points with errors bars), this means
    \mbox{$f_{\delta,mn}\equiv1$} by construction, with added random
    noise based on the error covariance matrix for KiDS-1000.}
\end{figure*}

\begin{figure*}
  \begin{center}
    \includegraphics[width=45mm]{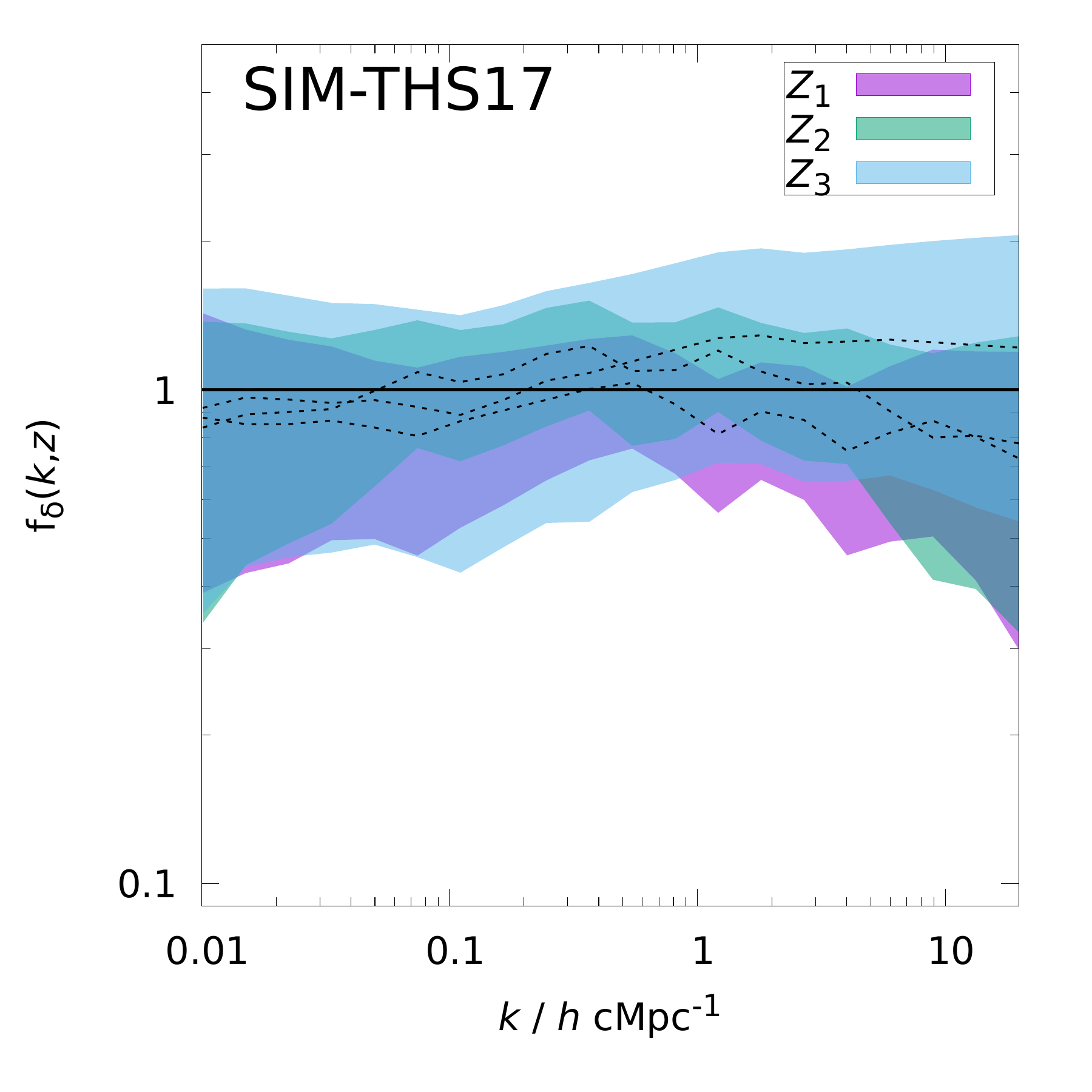}
    \includegraphics[width=45mm]{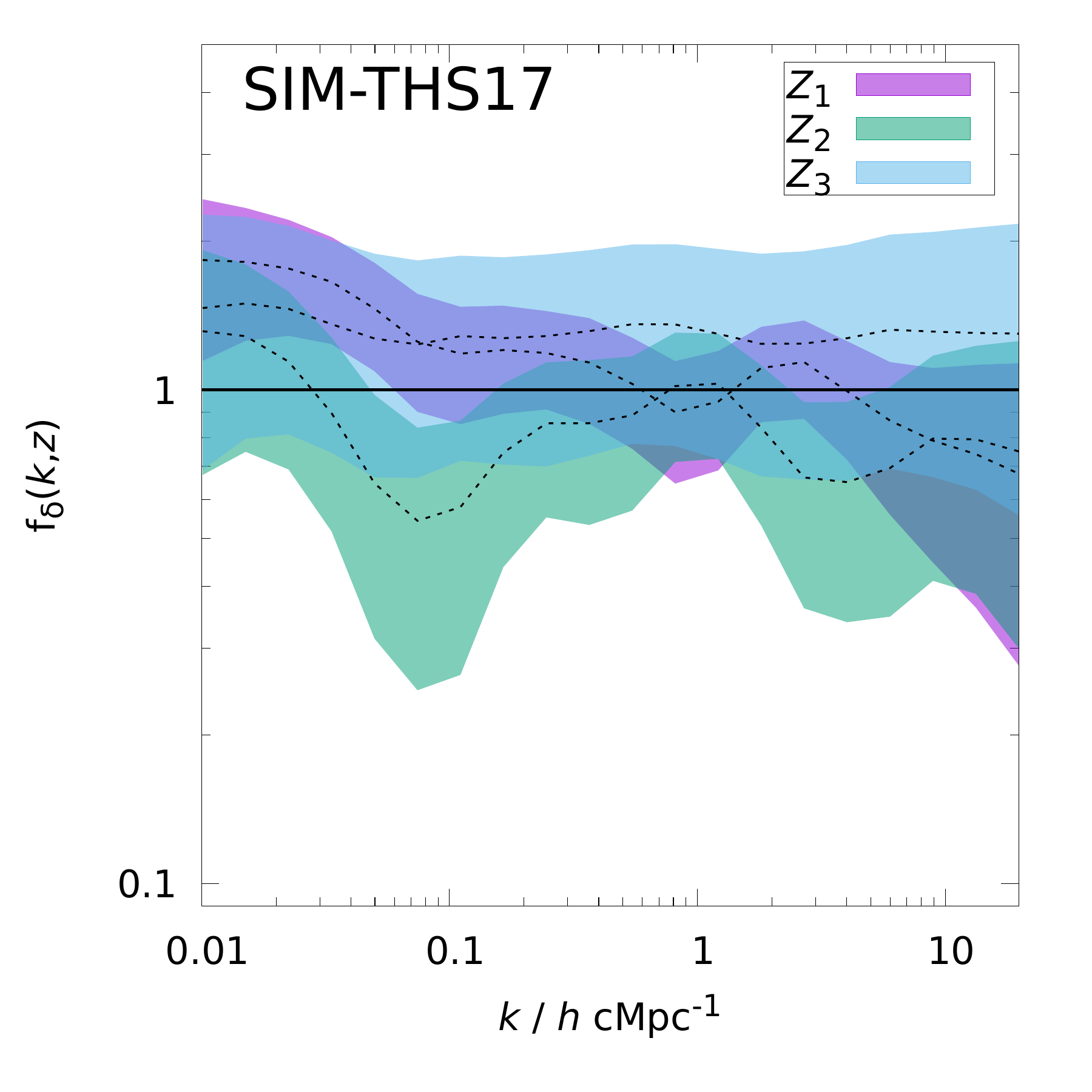}
    \includegraphics[width=45mm]{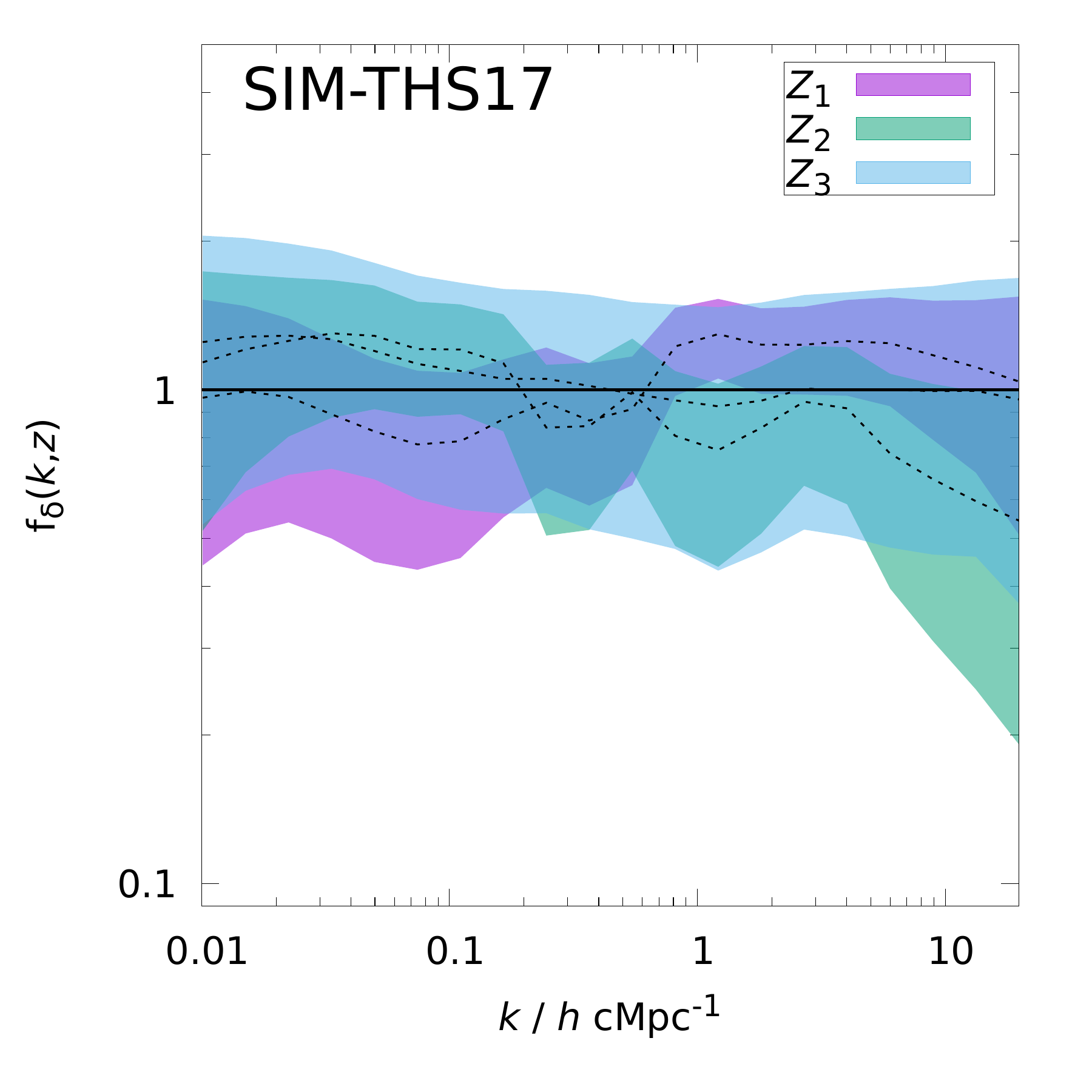}
    \includegraphics[width=45mm]{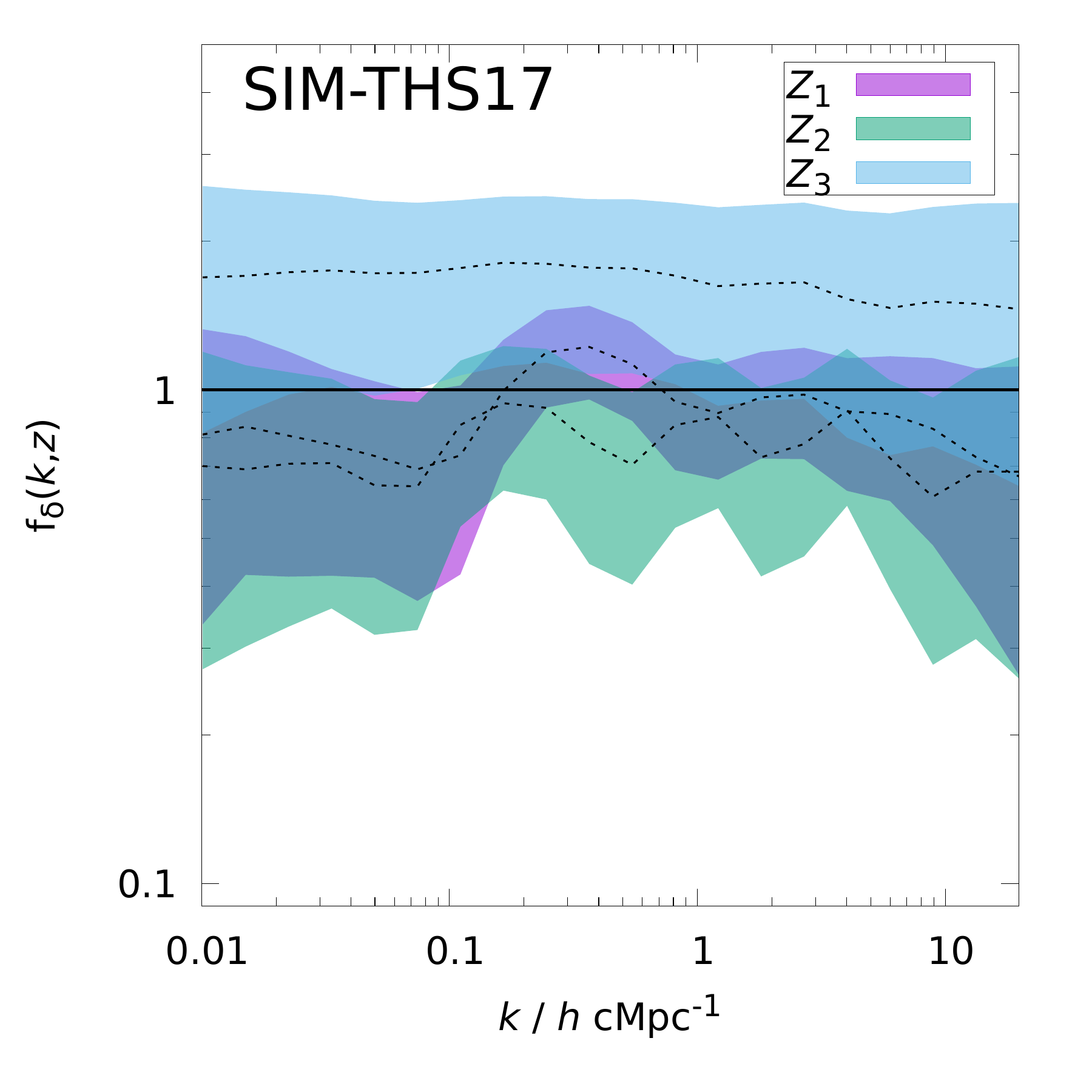}
    \\
    \includegraphics[width=45mm]{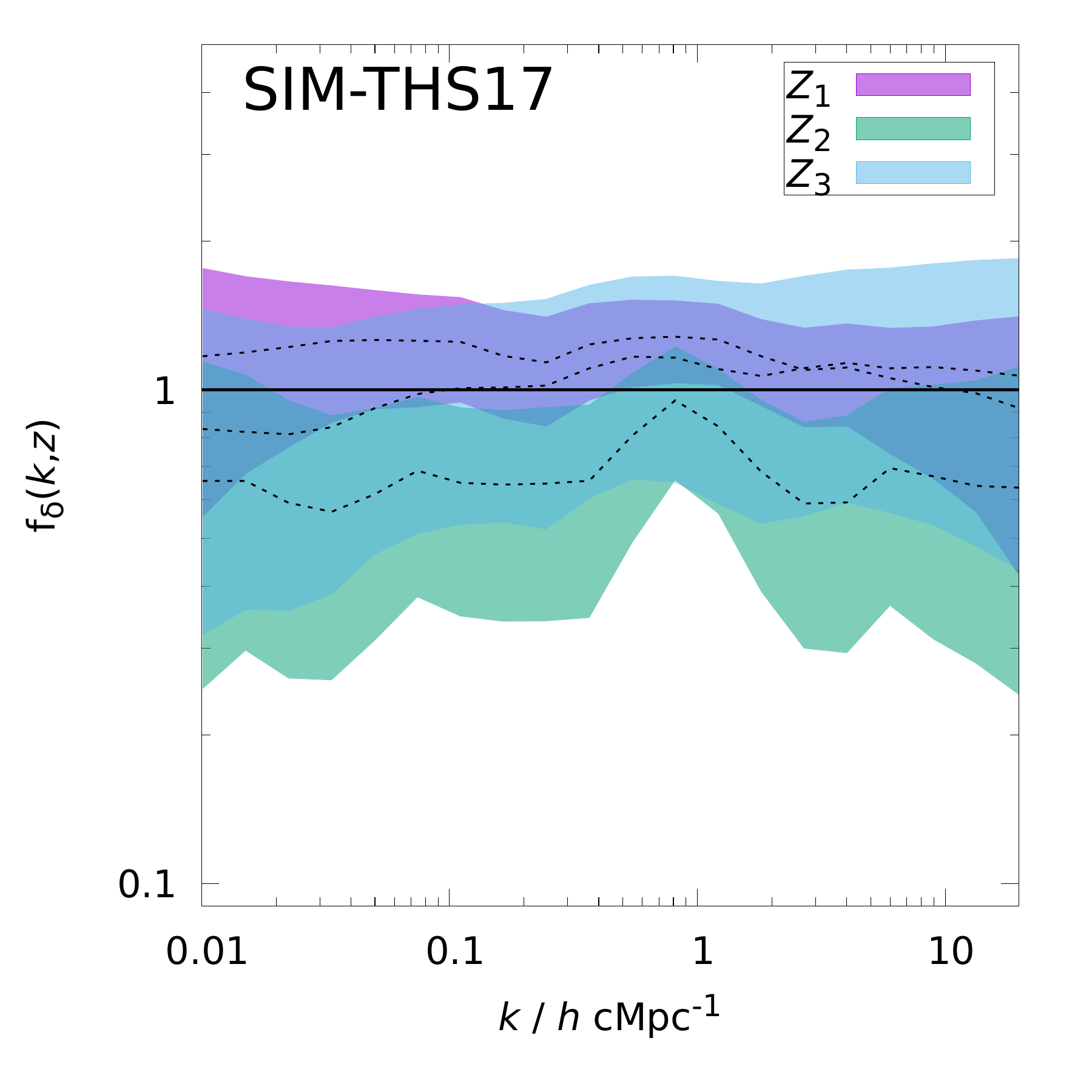}
    \includegraphics[width=45mm]{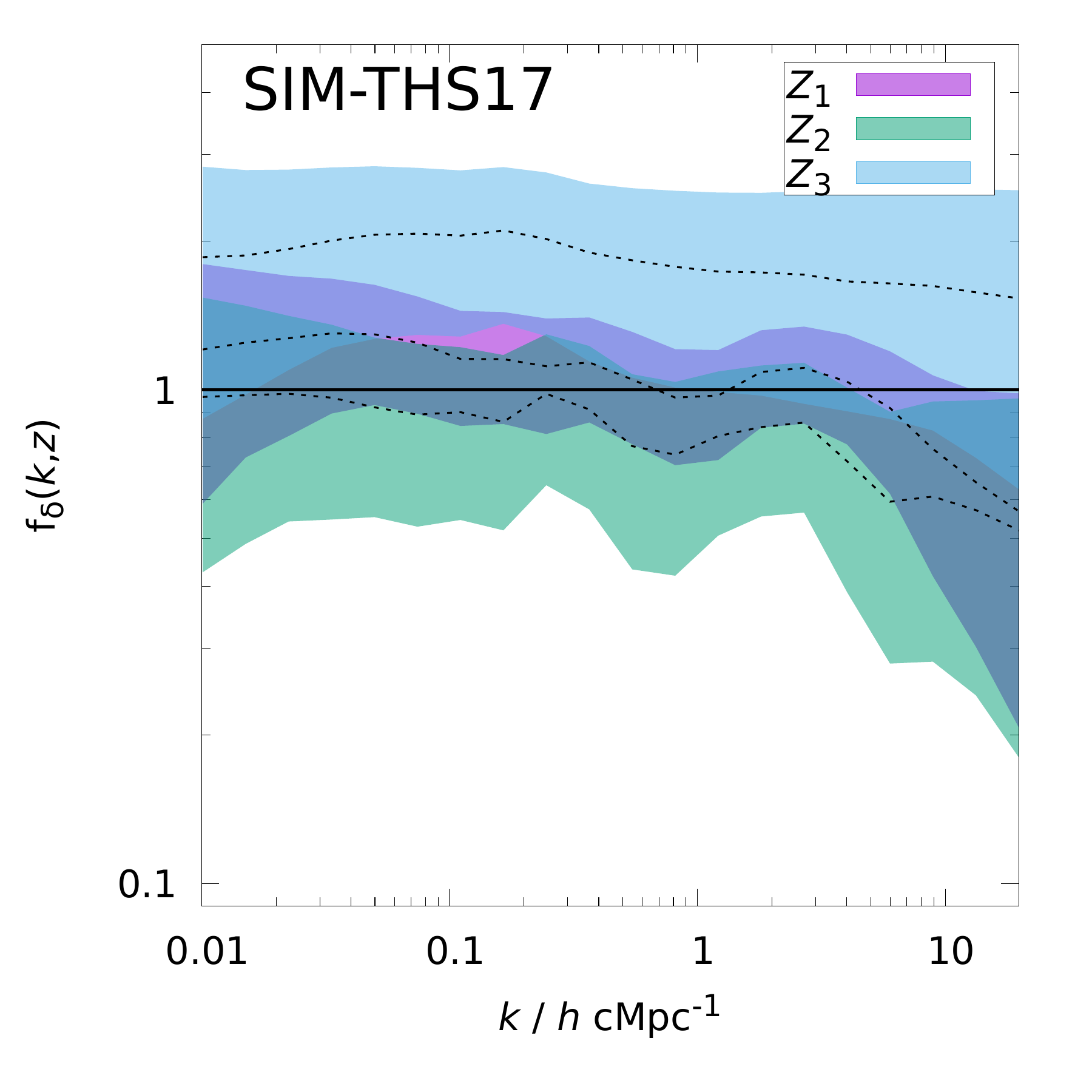}
    \includegraphics[width=45mm]{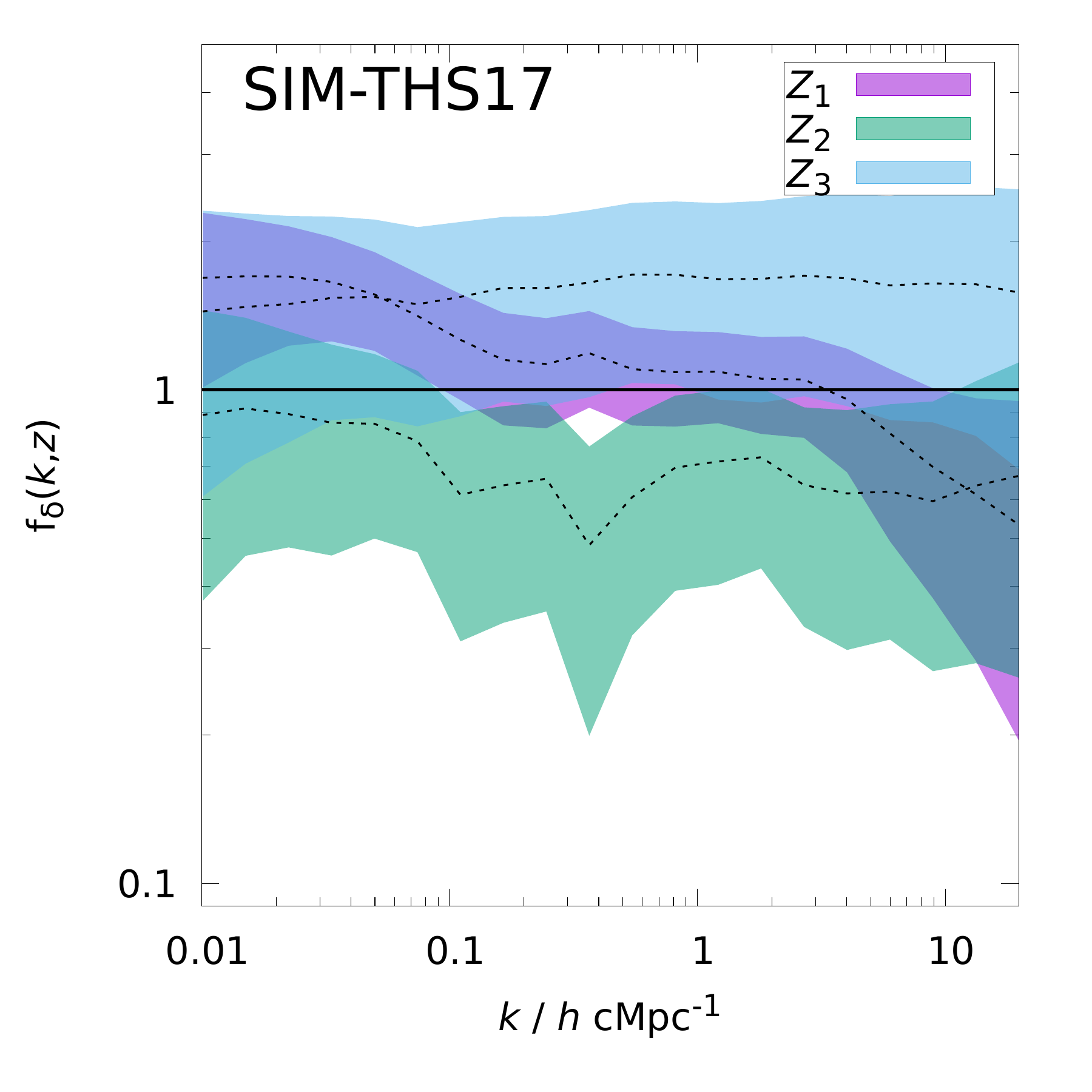}
    \includegraphics[width=45mm]{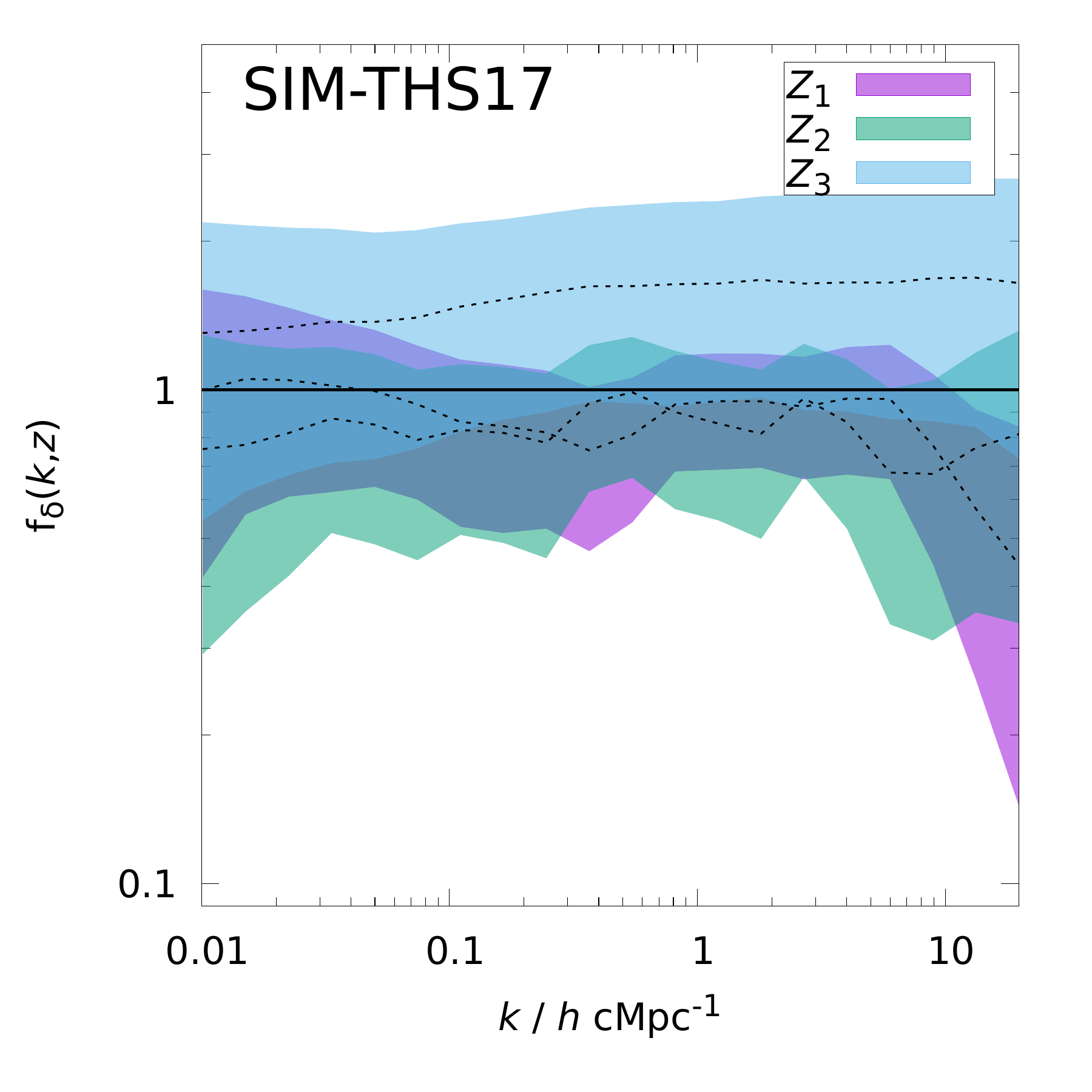}
    \\
    \includegraphics[width=45mm]{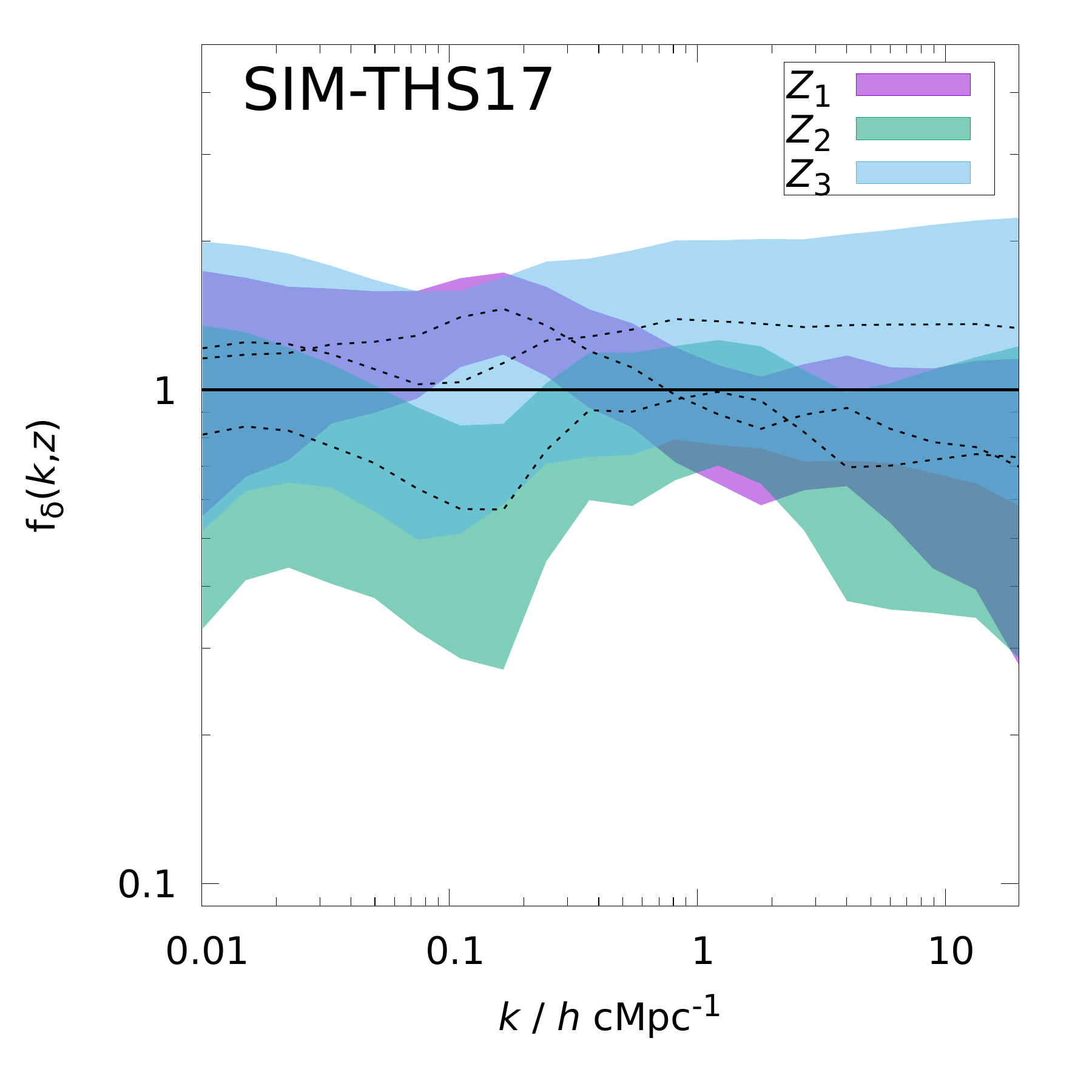}
    \includegraphics[width=45mm]{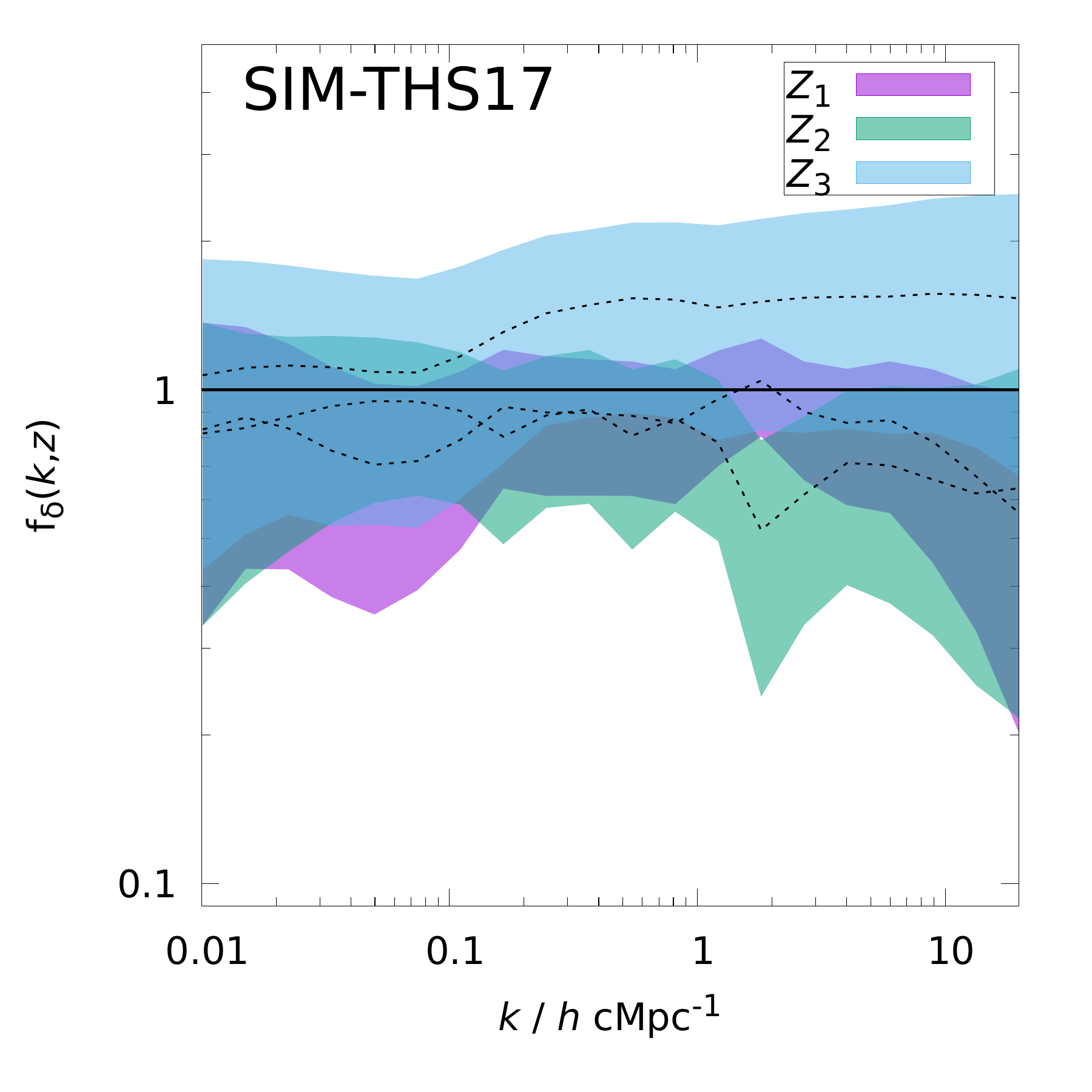}
    \includegraphics[width=45mm]{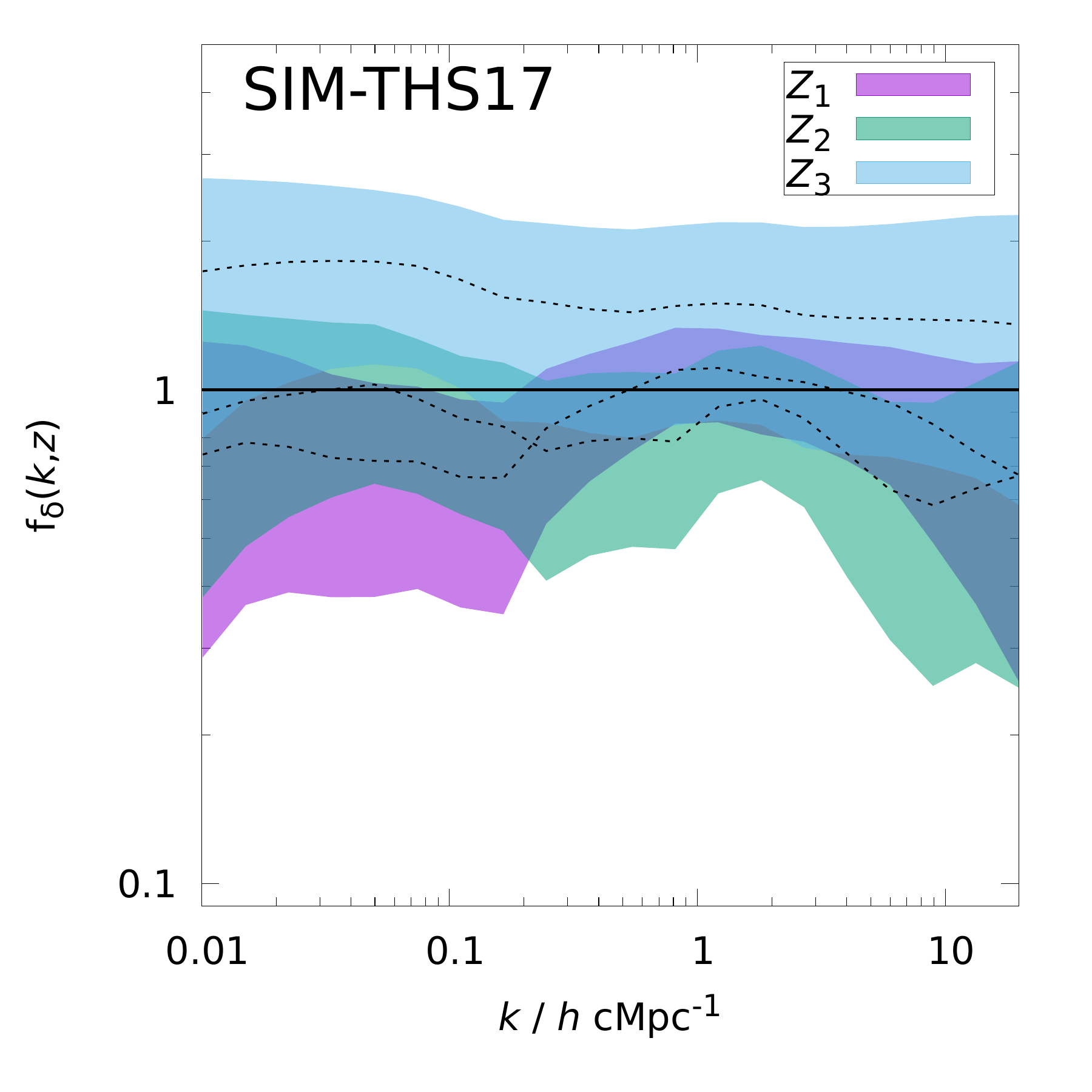}
    \includegraphics[width=45mm]{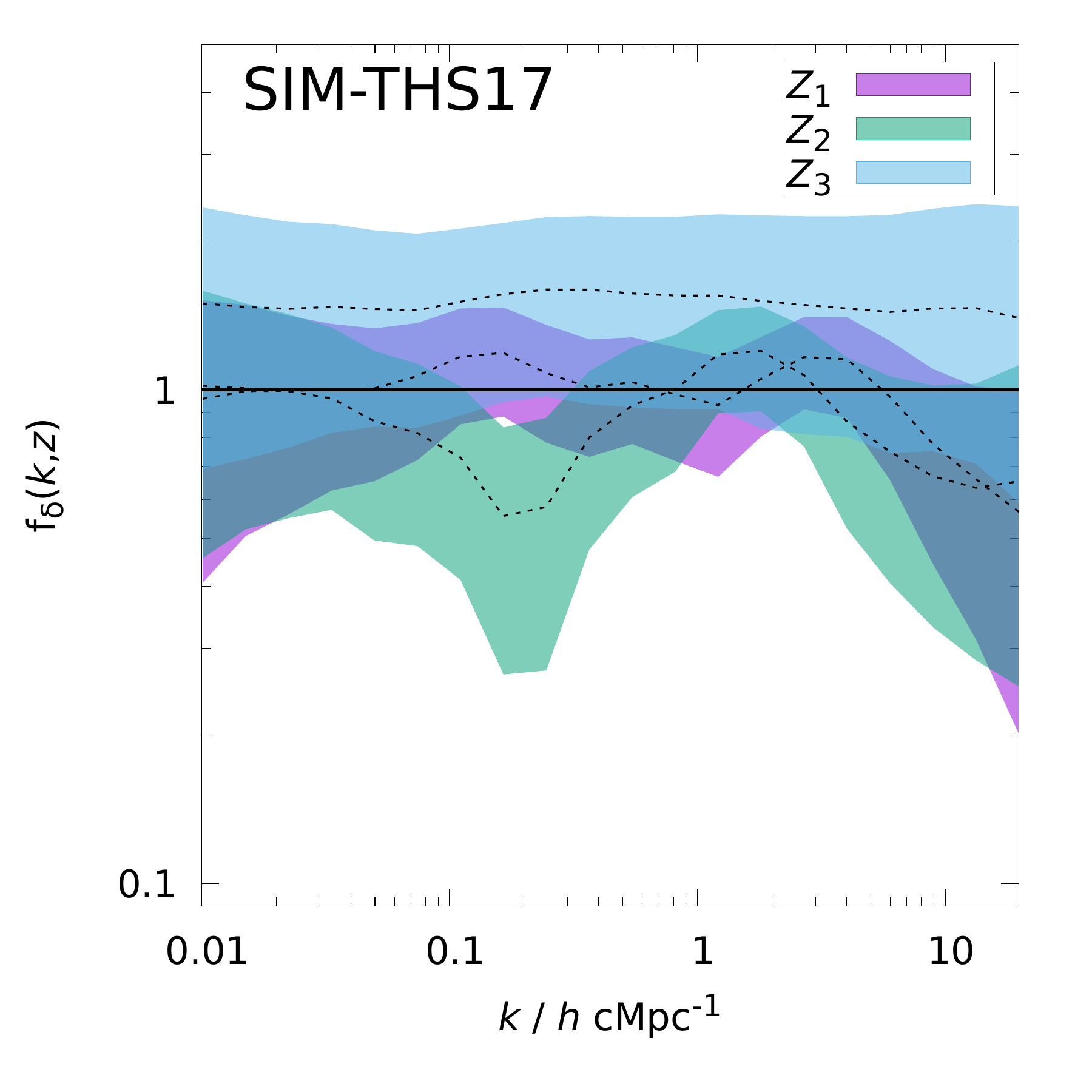}
    \\
    \includegraphics[width=45mm]{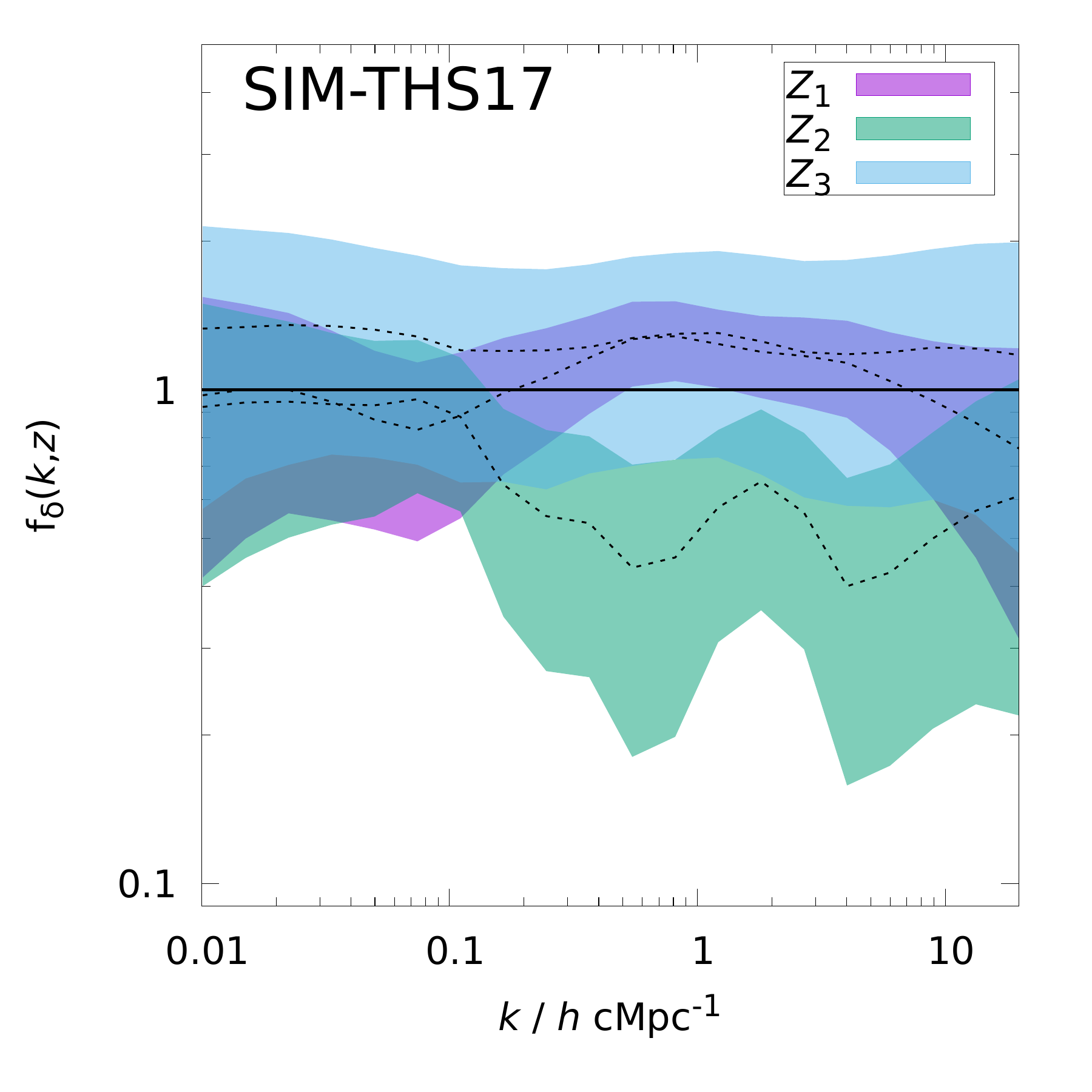}
    \includegraphics[width=45mm]{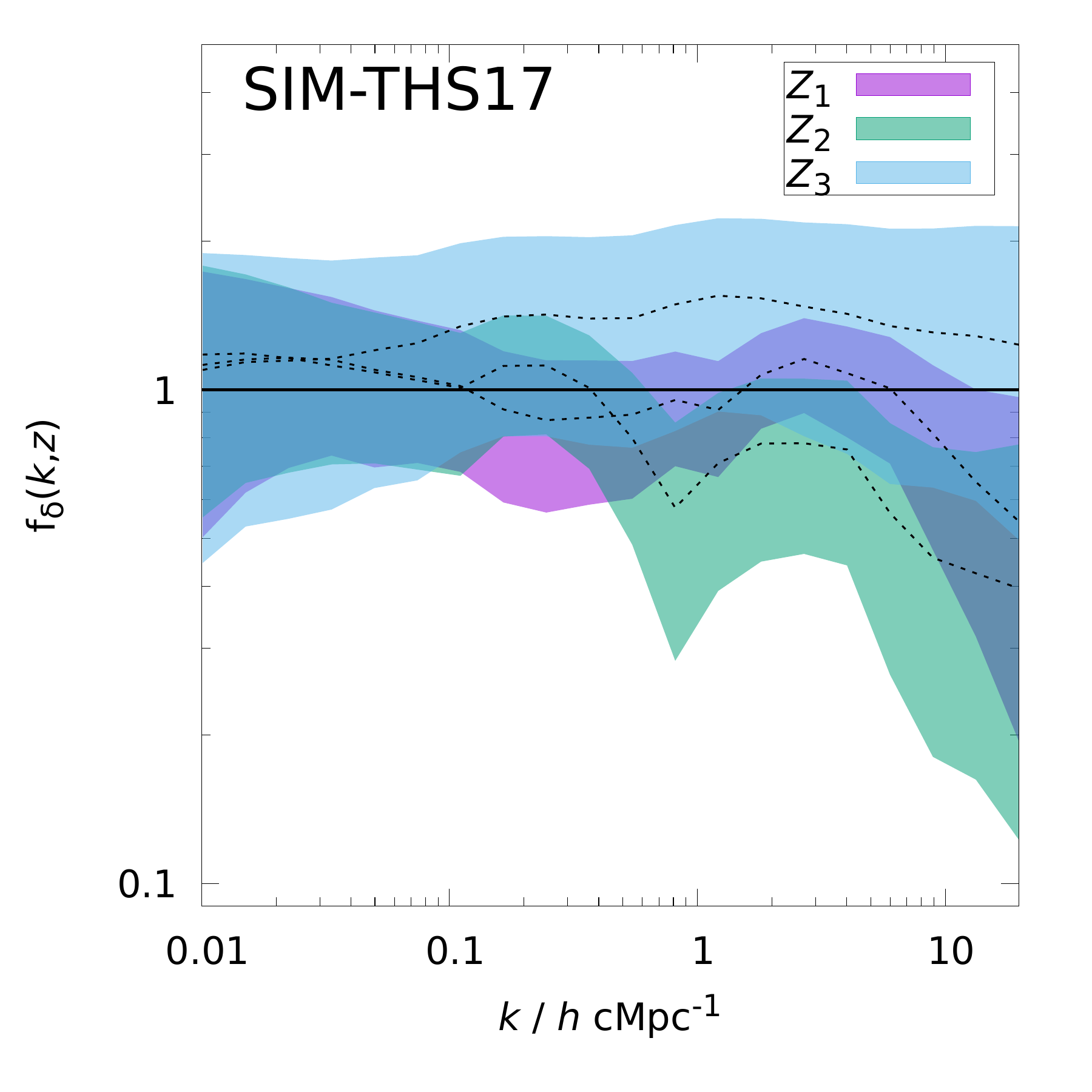}
    \includegraphics[width=45mm]{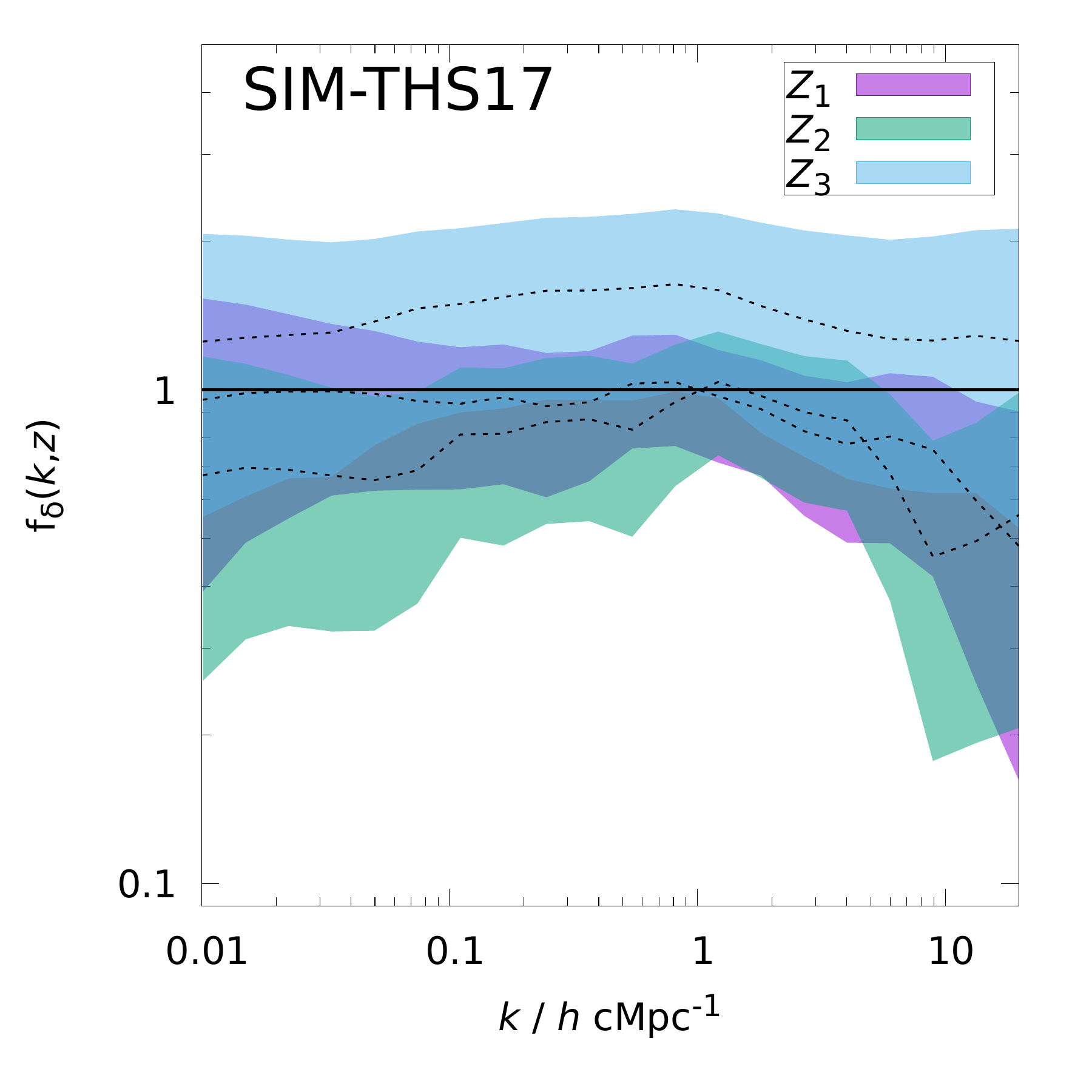}
    \includegraphics[width=45mm]{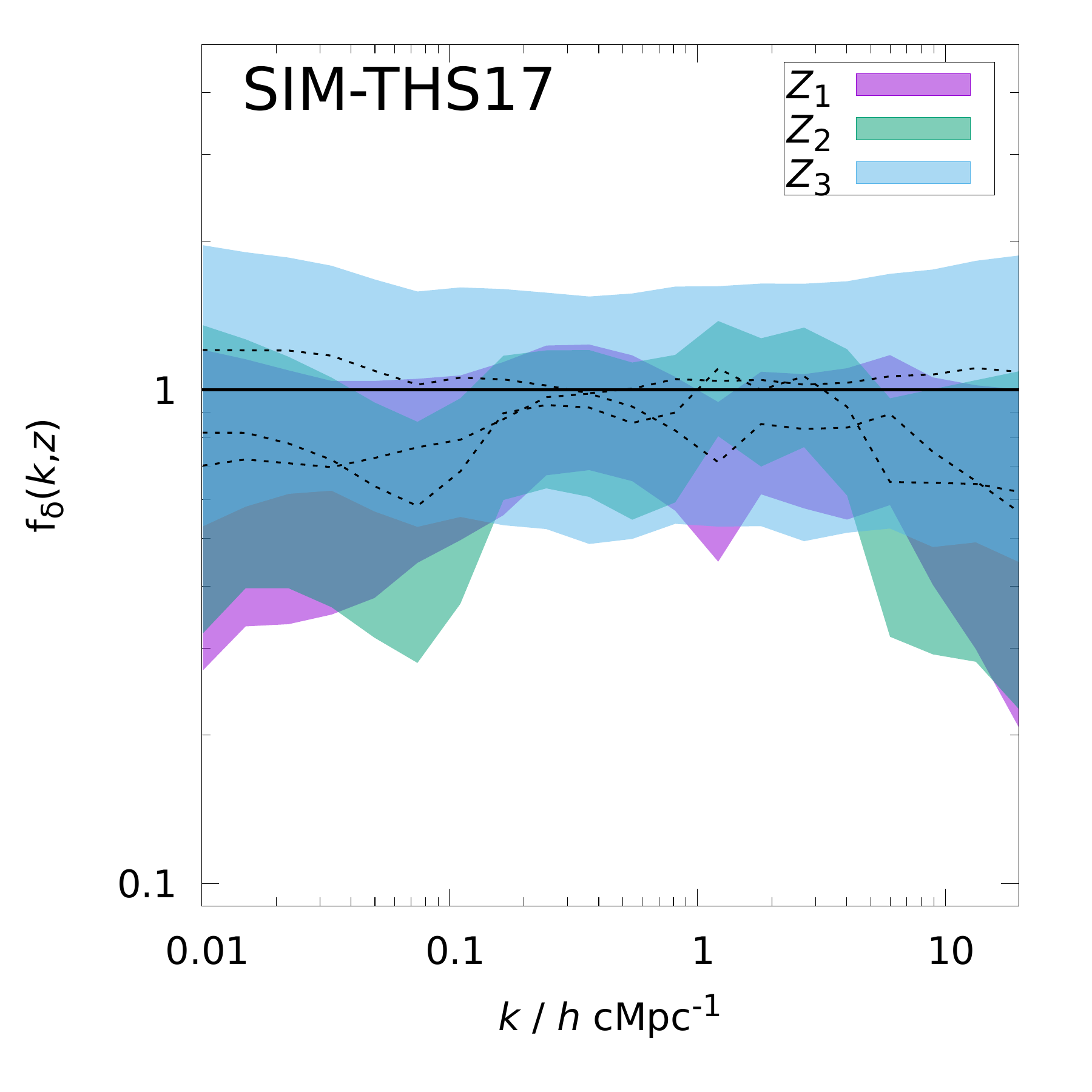}
  \end{center}
  \caption{\label{fig:ths17relpower3} Illustration of possible
    variations in $f_\delta(k,z)$ reconstructions ($68\%$ CIs about
    the median) using $\tau=5.0$ and $N_k=20$ on noisy data vectors
    from the ray-traced KiDS-1000 mock-data which has
    $f_\delta\approx1$ (Sect. \ref{sect:tks17}). There is a tendency
    for $f_\delta>1$ in $Z_3=[0.6,2]$ (cyan), and a tendency for
    $f_\delta<1$ in $Z_2=[0.3,0.6]$ (green) due to skewed PDFs and
    correlated errors. The magenta region shows the constraints for
    $Z_1=[0,0.3]$ which are the tightest compared to $Z_2$ and
    $Z_3$. An average over many reconstructions is shown in the bottom
    panel of Fig. \ref{fig:ths17relpower}.}
\end{figure*}

\begin{figure*}
  \begin{center}
    \includegraphics[width=185mm]{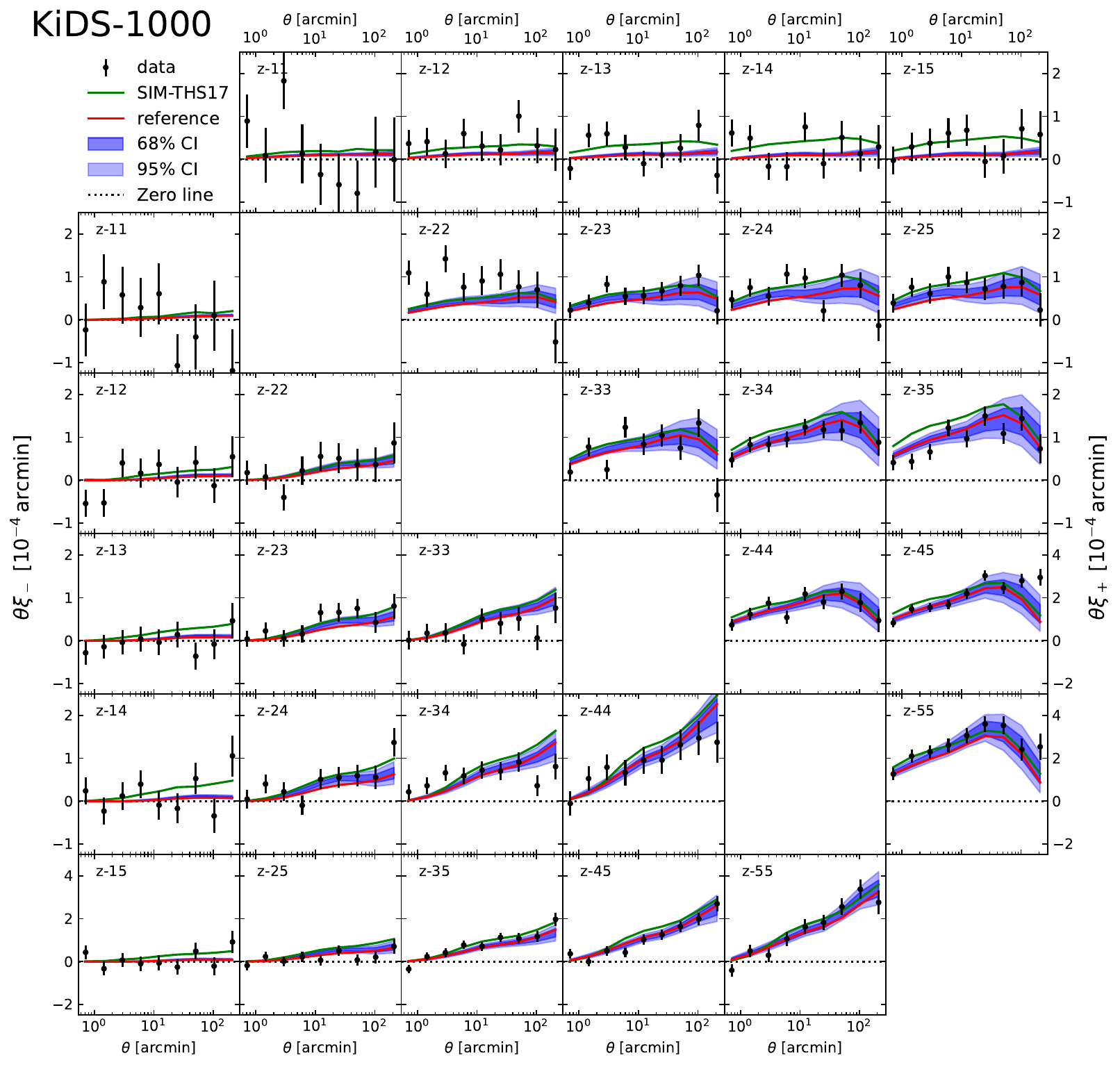}
  \end{center}
  \caption{\label{fig:kidspostpred} Posterior predictive of the
    Bayesian analysis using KiDS-1000 tomographic shear correlations
    and a relative power, $f_\delta(k,z)$, averaged inside the three
    redshift bins $Z_1=[0,0.3]$, $Z_2=[0.3,0.6]$, and
    $Z_3=[0.6,2]$. For each tomographic bin combination $(ij)$, the
    panels with labels `$z$--$ij$' show the posterior model
    constraints as light blue ($95\%$ CI) regions and dark blue
    ($68\%$ CI) regions about the median for either
    $\theta\,\xi^{(ij)}_-(\theta)$ (lower left triangle) or
    $\theta\,\xi^{(ij)}_+(\theta)$ (upper right triangle), both in
    units of $10^{-4}\,\rm arcmin$ and as function of lag
    $\theta$. Black points with error bars ($1\sigma$) are the
    KiDS-1000 data points. The red lines correspond to the
    $\Lambda\rm CDM$ reference power spectrum with $S_8\approx0.72$,
    the solid green lines `SIM-THS17' correspond to the prediction by
    \cite{2017ApJ...850...24T}, see Sect. \ref{sect:tks17}, for
    $S_8\approx0.79$. Errors of $\xi^{(ij)}_\pm(\theta)$ are
    correlated between $\theta$-bins and tomographic bins, marginal
    errors due to lensing kernel and IA uncertainties are not included
    here (adding another $\sim10\%$ to CIs). Conflicts with the data
    are visible for $\theta\,\xi^{(ij)}_+$ in $z$--$22$ and to a
    lesser degree for $z$--$12$ to $z$--$15$. Figure
    \ref{fig:postpredVer} shows a random realisation of the reference
    model.}
\end{figure*}

The excellent fit of the model to the verification input data vector,
$\vec{d}$, is best viewed in the posterior predictive plot,
Fig. \ref{fig:postpredVer}. This plot compares the posterior model
(blue regions for the two CIs $68\%$ and $95\%$ about the median) to
the $\xi^{(ij)}_\pm(\theta)$ data points with $1\sigma$ error bars
(black) as function of $\theta$. The lower triangle of panels
`$z$--$ij$' shows $\theta\,\xi^{(ij)}_-$ for the tomographic bin pair
$(ij)$, and likewise for the upper triangle and
$\theta\,\xi^{(ij)}_+$.  The data points scatter symmetrically about
the median model at the centre of the CIs.  Sometimes the residuals
exceed $1\sigma$ or more, expected statistically for hundreds of
(correlated) points. On the whole, we cannot see clear model
shortcomings to match the verification data at any angular scale or
redshift. This is underscored by the red solid lines for the true
(fiducial) $\xi_\pm^{(ij)}(\theta)$: the red lines are well within the
$68\%$ CI, in dark blue, of the posterior model constraints. Figure
\ref{fig:kidspostpred} is the same plot but for the actual KiDS-1000
data (Sect. \ref{sect:threezbins} provides more details).

In summary, the foregoing tests demonstrate the excellent convergence
of the Hamiltonian Monte Carlo sampler used in the KiDS-1000 analysis
and its ability to infer the 3D matter power spectrum with up to
\mbox{$N_z\times N_k=60$} variables. More tests with
\mbox{$f_{\delta,mn}\ne1$} verification data or mock data based on a
CDM $N$-body simulation are presented in Sect. \ref{sect:tikhonov} and
Sect. \ref{sect:tks17}.

\newpage
\onecolumn

\section{Posterior predictive for KiDS-1000 data}

\end{appendix}
\end{document}